\documentclass{aa}  
\usepackage{comment}
\usepackage{siunitx}
\usepackage{subcaption}
\usepackage{nicefrac}
\usepackage{graphicx}
\usepackage{color}
\usepackage{hyperref}
\usepackage{csvsimple}
\usepackage{ulem}
\usepackage[switch]{lineno}
\usepackage{booktabs,wrapfig}
\hypersetup{colorlinks,linkcolor={blue},citecolor={blue}}

\usepackage{textcomp,gensymb}
\usepackage{caption} 
\usepackage{tabularx}
\usepackage[bottom]{footmisc}
\usepackage[varg]{txfonts}
\begin{document} 

    \title{HIP~15429: A newborn Be star on an eccentric binary orbit}

   \author{Johanna M\"uller-Horn \inst{1,2} \and
          Kareem El-Badry \inst{3,1} \and
          Hans-Walter Rix \inst{1} \and
          Tomer Shenar \inst{4} \and
          Rhys Seeburger \inst{1} \and
          Jaime I. Villase\~nor \inst{1} \and
          Julia Bodensteiner \inst{5} \and 
          David W. Latham 
          \inst{6} \and 
          Allyson Bieryla 
          \inst{6} \and 
          Lars A. Buchhave 
          \inst{7} \and 
          Howard Isaacson 
          \inst{8} \and 
           Andrew W. Howard 
           \inst{3}}

   \institute{
    Max-Planck-Institut für Astronomie, Königstuhl 17, 69117 Heidelberg, Germany \\
    \email{mueller-horn@mpia.de} \and Fakultät für Physik und Astronomie, Universität Heidelberg, Im Neuenheimer Feld 226, 69120 Heidelberg, Germany \and
    Department of Astronomy, California Institute of Technology, Pasadena, CA 91125, USA \and
    The School of Physics and Astronomy, Tel Aviv University, Tel Aviv 6997801, Israel \and
    European Southern Observatory (ESO), Karl-Schwarzschild-Str. 2, 85748 Garching bei München, Germany \and
    Center for Astrophysics | Harvard \& Smithsonian, 60 Garden Street, Cambridge, MA 02138, USA \and
    DTU Space, National Space Institute, Technical University of Denmark, Elektrovej, DK-2800 Kgs. Lyngby, Denmark \and
    Department of Astronomy, University of California, Berkeley, CA 94720, USA
}
    
   \date{Received 7 October 2024 / Accepted 4 July 2025}

  \abstract
    {Interaction in close binary systems is common in massive stars. Typically, the mass donor is stripped of its hydrogen envelope and evolves to become a hot helium star, while the accretor gains mass and angular momentum, spinning up in the process. However, the small number of well-constrained post-interaction binary systems currently limits detailed comparisons with binary evolution models. 

    We have identified a new post-interaction binary, HIP~15429, consisting of a stripped star and a recently formed rapidly rotating Be star companion ($v_\mathrm{rot} \sin i \approx 270\,$km/s) that shares many similarities with recently identified bloated stripped stars. 
    
    Based on the orbital fitting of multi-epoch radial velocities, we find a 221-day binary period. We also find an eccentricity of $e=0.52$, which is unexpectedly high, as tides are expected to have circularised the orbit efficiently during the presumed recent mass transfer. The formation of a circumbinary disc during the mass-transfer phase or the presence of an unseen tertiary companion might explain the orbit's high eccentricity. 
    
    We determined the physical parameters for both stars in HIP~15429 by fitting the spectra of the disentangled binary components and multi-band photometry. The stripped nature of the donor star is affirmed by its high luminosity at a low inferred mass ($\lesssim 1\,\mathrm{M}_\odot$) and the imprints of CNO-processed material on the surface abundances. The donor's large radius and cool temperature ($T_{\rm eff} = 13.5 \pm 0.5\,$kK) suggest that it has only recently ceased mass transfer. 
    Evolutionary models assuming a 5-6$\,\mathrm{M}_\odot$ progenitor can reproduce these parameters, and they imply that the binary is currently evolving towards a stage where the donor becomes a subdwarf orbiting a Be star. 

    The remarkably high eccentricity of HIP~15429 challenges standard tidal evolution models, suggesting either inefficient tidal dissipation or external influences, such as a tertiary companion or circumbinary disc. This underscores the need to identify and characterise more post-mass transfer binaries to benchmark and refine theoretical models of binary evolution.}


   \maketitle
%
%-------------------------------------------------------------------
  
\section{Introduction}
Binary interactions play a crucial role in the evolution of massive stars. Most massive stars are found in binary systems \citep[e.g.][]{Kobulnicky+2007,Sana+2008,Mason+2009,Moe_DiStefano2017,Offner+2023}, and a substantial fraction will interact at some point during their lifetimes \citep[e.g.][]{Sana+2012,deMink+2013,Marchant_Bodensteiner2023}. One common interaction channel is the exchange of mass and angular momentum through a phase of stable mass transfer \citep[e.g.][]{Kippenhahn_Weigert1967}. During this exchange, the initially more massive donor star is stripped of its hydrogen-rich envelope, and the companion star can accrete mass and angular momentum, spinning up in the process \citep[e.g.][]{Paczynski1971, Pols+1991, Marchant_Bodensteiner2023}.

The envelope-stripped stars exhibit a wide range of spectral characteristics depending on their mass and the extent of their remaining hydrogen envelope. The collective term `stripped star' historically refers to helium white dwarfs \citep{Kolb_Ritter1993} and hot subdwarf stars \citep[e.g.][]{Heber2009} at the low-mass end and to Wolf-Rayet stars \citep[e.g.][]{Paczynski1967} in the high-mass regime. More recently, intermediate-mass stripped stars were identified with masses in the range $\sim2 - 8\,\mathrm{M}_\odot$, thus bridging the helium star mass gap between subdwarfs and Wolf-Rayet stars \citep{Gotberg+2018,Drout+2023,Gotberg+2023}. Due to their high temperatures as exposed stellar cores, stripped stars can contribute substantially to the ionising radiation of stellar populations \citep[e.g.][]{Gotberg+2019}. As the likely progenitors of hydrogen-poor core-collapse supernovae \citep[e.g.][]{Smith+2011,Laplace+2021}, the stripped cores of massive stars play an important role in the formation of black holes and neutron stars.

The classical picture of envelope-stripped stars is that of hot compact stripped stars. However, recent studies have also identified cooler stripped stars in a transitional phase shortly after mass transfer. These stars, which are still in an inflated state and contracting into hot helium stars, have been described as `puffed up' or `bloated' in the recent literature \citep[e.g.][]{Villasenor+2023,Bodensteiner+2020,Dutta_Klencki2023}. The first such systems reported were LB-1 \citep{Irrgang+2020,Shenar+2020} and HR~6819 \citep{Bodensteiner+2020,ElBadry_Quataert2021}, and both feature low-mass stripped stars ($\lesssim 2\,\mathrm{M}_\odot$). More recently, bloated stripped stars of intermediate masses have been found in systems such as VFTS~291 \citep{Villasenor+2023}, SMCSGS-FS~69 \citep{Ramachandran+2023}, and AzV~476 \citep{Pauli+2022} in the Magellanic Clouds.
Bloated stripped stars exhibit cooler effective temperatures and larger radii than compact hot stripped stars, making them brighter in optical wavelengths and often indistinguishable from regular stars (i.e. single B-type stars) based only on their positions in a Hertzsprung–Russell diagram \citep[HRD; e.g.][]{Bodensteiner+2020, Villasenor+2023}. 

The envelope stripping of an initially more massive donor star can invert the binary mass ratio, making the accretor star the more massive but (seemingly) less evolved component in the binary, a process typically known as the Algol phenomenon. This mass and angular momentum transfer is believed to induce rapid rotation in the accreting star, potentially leading to the formation of Be-type stars \citep{Pols+1991,deMink+2013}. These are rapidly rotating B-type stars with Balmer-line emission from circumstellar decretion discs \citep[see e.g.][for extensive reviews on Be stars]{Rivinius+2013,Rivinius_Klement2024}. The variability of line strength and shape is a common feature of the emission lines formed in Be star-decretion discs. Emission can fully disappear and reappear on timescales of years to decades.
Variability on shorter timescales of a few days can be caused by phenomena in the close circumstellar environment or on the stellar surface \citep{Porter_Rivinius2003}.
The discovery of subdwarf companions to several Be stars supports the idea that binary interaction is the primary mechanism behind the formation of Be stars \citep[e.g.][]{Gies+1998,Peters+2008,Chojnowski+2018,Wang+2023,Klement+2024}. Additionally, the relative scarcity of main-sequence companions to Be stars, compared to regular B-type stars, strengthens the binary origin hypothesis \citep{Bodensteiner+2020b}.

Observational constraints on the stripped star population are essential to addressing the uncertainties that persist in theoretical models, particularly those concerning the efficiency and stability of mass transfer, modes of angular momentum loss, and the structural responses of donor and accretor stars to mass loss and gain. Although binary population synthesis models predict that Be + subdwarf binaries should be abundant \citep{Shao_Li2021}, to date only a few dozen have been identified, and even fewer have been studied in detail. The low number of identified systems is a consequence of challenges in detecting post-interaction binaries. When the stripped star evolves into a hot but faint subdwarf star, it is typically outshone by the Be star, making it difficult to identify in the optical part of the spectrum. Furthermore, the lower mass of the stripped star results in a low radial velocity (RV) amplitude for the Be star, which is only marginally detectable in the oftentimes variable and rotationally broadened spectra. 

Most of the known subdwarf companions in Be star binaries have been discovered using (far-)ultraviolet (UV) spectroscopy \citep[e.g.][]{Peters+2008, Peters+2013, Wang+2021, Wang+2023}. Although subdwarf companions are faint in the optical, their higher temperatures compared to the Be companions imply a greater flux contribution at shorter wavelengths (on the order of a few percent). Therefore, detections are more favourable in the UV \citep{Gotberg+2018, Wang+2018}. 

In addition to UV spectroscopy, several stripped stars have been discovered or confirmed through near-infrared interferometric observations \citep{Mourard+2015, Klement+2022, Klement+2024}. Long-baseline interferometry can detect stripped star companions with low flux contributions (as small as $\sim 1\%$; \citealt{Klement+2022}) and spatially resolve binary orbits. However, current telescope setups, particularly their limiting magnitudes, restrict these observations to bright and typically nearby targets.

Optical multi-epoch spectroscopy offers a complementary strategy for identifying post-mass transfer binaries, especially for binaries where the stripped star still is in the bloated phase shortly after detaching from mass transfer. Binaries such as LB-1, HR~6819, and VFTS~291 are examples of such systems, where optical spectroscopy revealed the stripped nature of the companions. In these systems, the former donor stars are in the process of contracting and evolving towards the subdwarf or helium white dwarf stage but retain their inflated radii, making them comparably or even more luminous than their Be star companions in the optical. The stripped nature of the stars could be revealed spectroscopically by anomalous surface abundances (particularly enhanced N and deficient C and O abundances) and low surface gravities, leading to lower inferred spectroscopic masses compared to regular B-type stars.

In this study, we present the binary system HIP~15429, which was highlighted in the third data release of the Gaia catalogue as a promising candidate for a dormant black hole companion due to its high binary mass function \citep{Gaia+2023}. Instead, our multi-epoch high-resolution spectroscopic and photometric analysis suggests that the system comprises a bloated stripped star and a recently spun-up Be star companion. This system bears similarities to the bloated stripped star binaries HR~6819 and LB-1, which were also initially suspected to harbour stellar-mass black hole companions \citep{Liu+2019,Rivinius+2020}. However, HIP~15429 stands out among these systems in that it has a highly eccentric orbit ($e = 0.52$) with a moderately long period ($P=221$ days). Eccentric orbits are unexpected for post-mass transfer binaries, as tidal interaction is expected to have circularised the orbit. Understanding the peculiarities of HIP~15429's orbit and confirming its status as a post-interaction binary will contribute to our understanding of binary evolution and add to the landscape of post-interaction binaries.

The remainder of this paper is organised as follows. Sect.~\ref{sec:observations} gives an overview of the observations and subsequent reduction of the multi-epoch high-resolution spectra obtained for HIP~15429. In Sect.~\ref{sec:variability} we describe the observed spectral variability, and in Sect.~\ref{sec:orbit_analysis} we describe the orbital analysis of the binary. In Sect.~\ref{sec:disentangling}, the results from the spectral disentangling of the components are presented followed by the spectral analysis of the disentangled single-star spectra in Sect.~\ref{sec:spectral_analysis}. A complementary photometric analysis of the binary components is presented in Sect.~\ref{sec:sed}. In Sect.~\ref{sec:evolution}, we construct binary evolution models to study the system’s formation history. We discuss the implications of our results and, in particular, the orbital eccentricity of the system in Sect.~\ref{sec:discussion}. Our main results are summarised in Sect.~\ref{sec:conclusion}.

\section{Observations and data reduction}
\label{sec:observations}
We analysed 24 optical spectra obtained with the Tillinghast Reflector Echelle Spectrograph (TRES; \citealt{Furesz2008}) mounted on the 1.5\,m Tillinghast Reflector telescope at the Fred Lawrence Whipple Observatory on Mount Hopkins, Arizona. The extracted TRES spectra cover a wavelength range $\simeq 3900$ to 7000\,Å with a spectral resolving power of $R = \nicefrac{\lambda}{\Delta \lambda} \simeq 44000$. The spectra were extracted and reduced as described in \cite{Buchhave+2010} including bias and flat-field corrections and wavelength calibration. We combined the spectra by merging the individual orders with linearly decreasing/increasing weights in overlapping wavelength regions. The data span 498\,d (2.25 orbital periods) from MJD 59890 to 60388 (11/07/2022 to 03/19/2024).
    
Furthermore, we obtained five spectra with the HIRES instrument \citep{Vogt+1994} on the Keck I telescope at the W. M. Keck Observatory on Mauna Kea. Data were obtained and reduced using the standard California Planet Survey (CPS) set up \citep{Howard+2010}, with resolving power $R \simeq 55000$ and in three wavelength bands; approximately 3700–4800, 5000-6400, and 6550-8000\,Å. 

We further analysed eight spectra with $R \simeq 85000$ and wavelength coverage 3800–9000\,Å taken with the High-Efficiency and High-Resolution Mercator Echelle Spectrograph (HERMES) instrument, which is mounted on the 1.2-m Mercator Telescope at the Roque de los Muchachos observatory in La Palma, Spain \citep{Raskin+2011}. Standard calibrations for this data set were performed with the \href{http://www.mercator.iac.es/instruments/hermes/drs/}{HERMES data reduction pipeline} \citep{Raskin+2011}. Spectra were rebinned to a lower resolution ($R \simeq 64000$) to increase signal-to-noise ratios (S/N). We combined three of the observations that were obtained during the same night to further improve S/N, using a S/N-weighted average.

We performed continuum normalisation individually for all 37 spectra. To do so, we used the Python package SciPy \citep{2020SciPy-NMeth} and fitted cubic splines to wavelength pixels in continuum regions. We followed the approach outlined by \cite{ElBadry_Quataert2021} to determine suitable spline anchor points in spectral regions without significant emission or absorption. A barycentric correction was applied to all spectra. The wavelength calibration between the data sets from the three instruments was verified using the interstellar Na absorption lines at 5890 and 5896\,Å. We estimated S/N empirically from the pixel-to-pixel scatter in regions without strong absorption or emission lines. The dates of all observations, typical S/N at $4500\,\AA$, and RVs, measured as described in Sect.~\ref{sec:orbit_analysis}, are listed in Table \ref{tab:obs_overview} in the Appendix.

\section{Spectral variability}
\label{sec:variability}

\begin{figure}[t]
    \centering
    \begin{minipage}[t]{\columnwidth}
        \centering
        \includegraphics[width=\columnwidth]{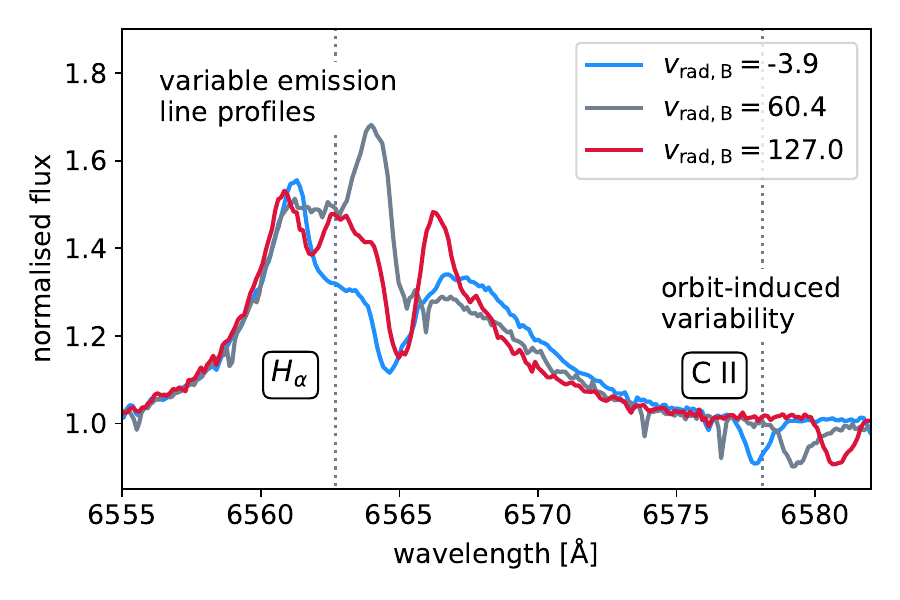}
    \end{minipage}
    \begin{minipage}[t]{\columnwidth}
        \centering
        \includegraphics[width=\columnwidth]{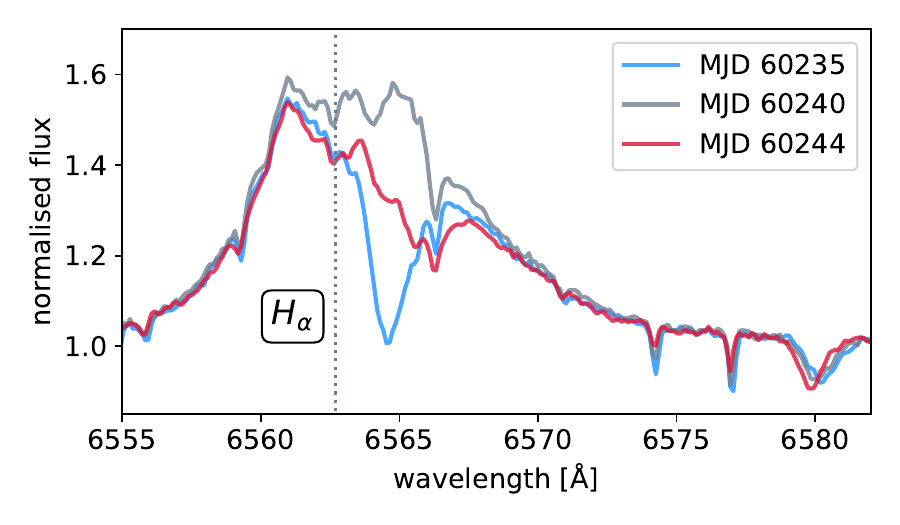}
    \end{minipage}
    \caption{\textit{Top:} {Balmer emission lines at quadrature.} Two TRES spectra taken close to quadrature (red and blue lines) and one spectrum close to the system's barycentric velocity (grey line). The grey dotted lines indicate the rest-frame wavelengths of the Balmer H$_\alpha$ and carbon C~II 6572 lines. Orbital motion of the narrow-lined star is apparent from wavelength shifts in the C~II lines, whereas the H$_\alpha$ line shows variable emission line profiles.\\
    \textit{Bottom:} {Short-term variability of emission line profiles.} Three TRES spectra observed at $t = \mathrm{MJD} \, 60235$, $t + 5\,$d, and $t+ 9\,$d, that is, spanning $<5\%$ of the orbital period. Solid grey, red, and blue lines show the variable H$_\alpha$ emission line profiles. The rest-frame wavelength of H$_\alpha$ is shown as in the top panel. Narrow lines at 6574, 6577\,\AA \, originate from variable telluric absorption.}
    \label{fig:spectral_variability}
\end{figure}

\begin{figure}[t]
    \centering
    \includegraphics[width=\columnwidth]{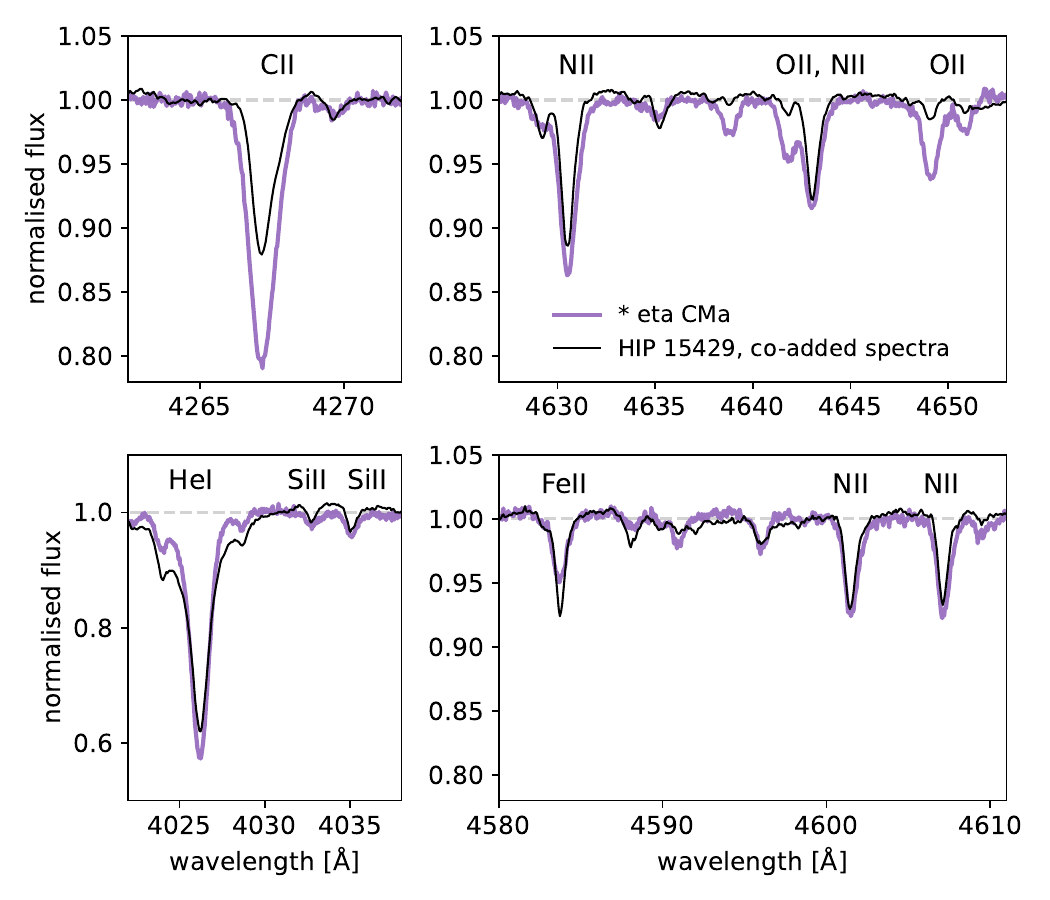}
    \caption{Comparison of CNO lines in the co-added spectra of HIP~15429 (black) and the B5 supergiant standard star Eta Canis Majoris (purple) presumed to be of similar temperature, surface gravity, and metallicity. Carbon and oxygen appear to be depleted in HIP~15429.}
    \label{fig:CNO_comparison_reference_star}
\end{figure}

HIP~15429 \citep[$V$=9.75\,mag,][]{Hiltner1956} was observed as part of the low-resolution spectral catalogue by \cite{Jacoby_Hunter1984} and identified as a B5Ib supergiant, a classification later confirmed by \cite{Navarro+2012}. Its binary nature was reported with a spectroscopic solution (SB1) published in the Gaia DR3 non-single-star catalogue \citep{Gaia+2023}. We point out two peculiarities about the multi-epoch spectra of HIP~15429 considering the suggested classification. 

First, the spectra exhibit disc emission features in the form of broad, double, and triple-peaked emission in the Balmer lines. The emission is most prominent in H$_\alpha$, where it was also observed by \cite{Jacoby_Hunter1984} and \cite{Navarro+2012}. Although there is clear orbit-induced variability in the narrow metal lines in the spectra, the Balmer emission appears to be nearly stationary, with potentially a small antiphase motion in the wings of the lines. This is shown in the upper panel of Fig.~\ref{fig:spectral_variability} where two TRES spectra are plotted close to quadrature, that is, close to the phase of maximum velocity separation.

Secondly, the strengths of the metal absorption lines seem to differ from those of known blue supergiants. In particular, the carbon and oxygen lines in HIP~15429 appear much weaker than expected. This is indicated in Fig.~\ref{fig:CNO_comparison_reference_star} where we compare HIP~15429 to spectra of the supergiant Eta Canis Majoris\footnote{The reference spectrum was downloaded from the ESO archive. It was observed by N. Przybilla with the FEROS spectrograph at the MPG/2.2m telescope with archive ID ADP.2016-09-27T09:50:39.855.}, listed as a B5Ia standard star in \cite{Gray_Corbally2009} and assumed to be of similar temperature, surface gravity, and comparable metallicity as HIP~15429.

Together, these spectral features already hint at a post-interaction binary nature of the system \citep[see e.g.][]{Marchant_Bodensteiner2023,Ramachandran+2024}. In this scenario, the observed spectra would be superpositions of the spectra of the binary components, featuring a narrow-lined, stripped star resembling a B5 supergiant and a fast-rotating, broad-lined companion with disc emission. 

The strength of Balmer emission and the shape of the line wings remain approximately constant throughout the observed period, but the central emission line profiles show substantial variability on short timescales (approximately a few days). The lower panel of Fig.~\ref{fig:spectral_variability} shows three epoch spectra observed within a 10-day period, illustrating the high variability of the central absorption feature. The short variability timescale, corresponding to $\lesssim5\%$ of the orbital period (see Sect.~\ref{sec:orbit_analysis}), suggests that these profile changes are not due to the superposition of the narrow-lined star's spectrum, nor does the absorption feature trace the companion star.

\section{Orbital analysis}
\label{sec:orbit_analysis}

We measured RVs of the narrow-lined star by cross-correlating the data with a synthetic template spectrum. We used a spectral template from the Potsdam Wolf-Rayet (PoWR) model atmospheres grids for OB-type stars \citep[][see Sect.~\ref{sec:spectral_analysis} for a more detailed description of the code]{Hainich+2019}  with temperature $T_\mathrm{eff} = 15\,\mathrm{kK}$, surface gravity $\log g = 2.6$, and solar metallicity. Cross-correlation was performed in eight wavelength regions of equal size, each 250\,Å wide, spanning the range of 4000 to 6000\,Å. RVs were determined by averaging measurements from all wavelength regions, with uncertainties derived as the standard deviation of the RVs across these regions. The most important spectral lines contributing to the RV measurements include He~I absorption lines (e.g. He~I $\lambda$4026, $\lambda$4144, $\lambda$4388, $\lambda$4922, $\lambda$5116, and $\lambda$5876) and narrow metal lines such as Mg~II, Ca~II, Si~II/III, and Fe~II/III lines. We masked wavelength regions with known telluric or interstellar features and the Balmer lines, where there is substantial contribution from the companion star. We cross-checked the measured RVs by fitting Voigt profiles to individual spectral lines, including multiple He~I and metal lines (e.g. Mg~II). Voigt profiles were chosen because they provided better fits to the slightly broader line wings compared to Gaussian profiles. The fitting was performed using the \texttt{funcFit} module of the \href{https://github.com/sczesla/PyAstronomy}{\texttt{PyAstronomy}} Python package.
To determine RVs and their uncertainties, we averaged the RVs obtained from all fitted lines and calculated the standard deviation among these measurements as the RV uncertainty. By comparing the RVs from the Voigt profile fitting and template cross-correlation, we found that the results were consistent within 2-$\sigma$ uncertainties.

We did not attempt to measure RVs for the companion star because its only discernible feature is the emission component in the Balmer lines, which is highly variable on short timescales (see Fig.~\ref{fig:spectral_variability}) and therefore is not suitable for RV measurements.

\begin{figure*}[t]
    \centering
    \includegraphics[width=0.9\textwidth]{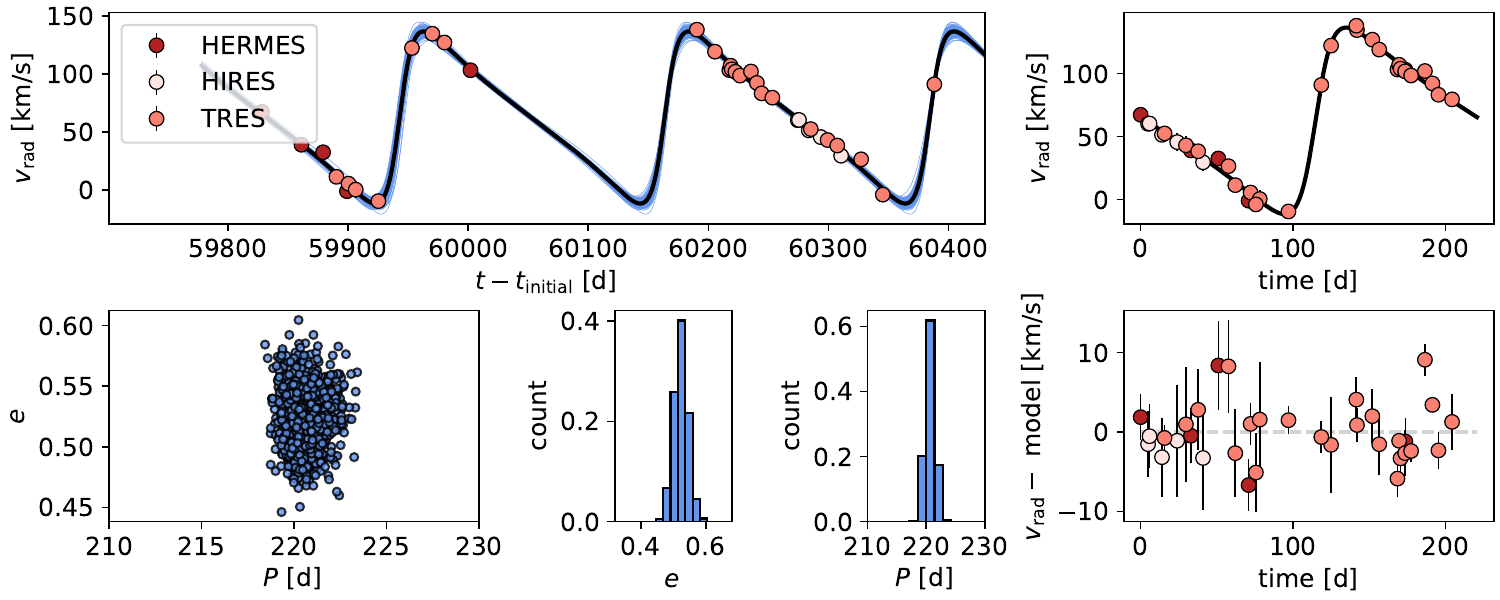}
    \caption{{Orbital analysis of HIP~15429}. The top panel shows the RV time series, with data points colour-coded by the instrument used for observation, and RV orbit models computed from posterior samples (blue lines). The lower-left and middle panels show the posterior distributions for the orbital period, $P$, and eccentricity, $e$. The MAP model is indicated by the solid black line. The phase-folded RVs and residuals from the MAP model are shown in the right-hand panels.} 
    \label{fig:orbit_fit}
\end{figure*} 

We used a nested sampling framework to determine the binary orbit. In particular, we used the algorithm MLFriends \citep{Buchner2016,Buchner2019}  as part of the \href{https://johannesbuchner.github.io/UltraNest/}{UltraNest} package \citep{Buchner2021} to infer posterior probability distribution functions for orbital parameters. The RV curve is parameterised by six orbital parameters $(K_{\mathrm{B}}, P, M_0, e, \omega, v_z)$, which we fitted together with an additional jitter term $s$ to account for potentially underestimated RV uncertainties. Here, $K_{\mathrm{B}}$ is the RV semi-amplitude of the narrow-lined star, $P$ denotes the orbital period, and $M_0 = \nicefrac{2\pi t_0}{P}$ is the mean anomaly at a reference time $t_0$. $e$ and $\omega$ are the orbital eccentricity and the argument of the pericentre, and $v_z$ denotes the system's barycentric velocity. The assumed prior distributions are uninformative and uniform for all parameters with prior ranges $K_B \in [0,200]\,\text{km/s},\,P \in [200,250]\,\text{d},\, M_0,\omega \in [0,2\pi], \, e \in [0.0,0.99], \, s \in [0.01,10]\,\text{km/s}, \, v_z \in [-100,100]\,\text{km/s}$. 

The assumed log-likelihood function has the form 
\begin{equation}
    \log \mathcal{L} = \sum_i -\frac{1}{2} \left(\frac{(v_\mathrm{rad, obs}(t_i)-v_\mathrm{rad, model}(t_i))^2}{\sigma_{v_{\mathrm{rad},i}}^2 + s^2}   - \log{\left(\frac{1}{\sigma_{v_{\mathrm{rad},i}}^2 + s^2}\right)}\right)
\end{equation}
for observed RVs of the narrow-lined star, $v_\mathrm{rad, obs}$, with uncertainties, $\sigma_{v_{\mathrm{rad},i}}$, and predicted RVs, $v_\mathrm{rad, model}$, summed over all available epochs, $t_i$. Given the different wavelength regimes and higher resolution of the obtained spectra with respect to Gaia RVS spectra as well as the good orbital coverage, we opted not to include the DR3 orbital solution in the prior or likelihood of the parameter inference so that we would get an independent estimate of the orbital parameters.

In Fig.~\ref{fig:orbit_fit}, the results of the orbital analysis are presented, showing the posterior samples of period and eccentricity and the maximum a posteriori (MAP) phase folded orbit model. The posterior samples are clearly unimodal and clustered around an orbital solution with a period of $P=221\pm1$\,d and eccentricity $e=0.52\pm0.03$. The mean values and uncertainties for all orbital parameters are listed in Table~\ref{tab:binary_parameters}. The derived orbital solution is broadly consistent with the spectroscopic solution published in Gaia DR3. However, we infer slightly higher period and higher eccentricity compared to the DR3 values; $P_\mathrm{DR3} = 217 \pm 2\,$d, $e_\mathrm{DR3} = 0.41 \pm 0.05$ \citep{Gaia+2023}.

\begin{table}[t]
\setlength{\extrarowheight}{2.5pt}
\centering
\caption{Orbital and physical parameters and uncertainties of the binary system and its component stars.}
\begin{tabular}{l l c}
\hline
\multicolumn{3}{l}{\textbf{Parameters of the binary system}} \\
Orbital period & $P$ [d] & $221 \pm 1$ \\
Eccentricity & $e$ & $0.52\pm0.03$ \\
Mass ratio & $q = \nicefrac{M_\mathrm{B}}{M_\mathrm{Be}}$ & \\
-- evolutionary & $q_\mathrm{evol}= \nicefrac{M_\mathrm{B, evol}}{M_\mathrm{Be, dyn}}$ & $\leq 0.14 $ \\ %= M_\mathrm{Be}/M_\mathrm{B, evol}$
-- spectroscopic & $q_\mathrm{spec}=\nicefrac{M_\mathrm{B, spec}}{M_\mathrm{Be, dyn}}$ & $\leq 0.10 $ \\
-- dynamic & $q_\mathrm{dyn}=\nicefrac{K_\mathrm{Be}}{K_\mathrm{B}}$ & $0.066 \pm 0.053$ \\

Barycentric velocity & $v_0$ [km s$^{-1}$] & $62.3 \pm 1.0$ \\
Distance & $d$ [pc] & $1730\pm260$ \\
&&\\
\multicolumn{3}{l}{\textbf{Parameters of B star}}\\
Effective temperature & $T_{\text{eff,B}}$ [kK] & $13.5 \pm 0.5$ \\
Surface gravity & $\log(g_{\mathrm{B}}/\text{cm s}^{-2})$ & $2.25 \pm 0.25$ \\
Rotation velocity & $v_\mathrm{rot,B}\sin i$ [km s$^{-1}$] & $\leq30$ \\
Macroturbulent velocity & $v_{\text{mac,B}}$ [km s$^{-1}$] & $\leq50$ \\
Microturbulent velocity & $v_{\text{mic,B}}$ [km s$^{-1}$] & $10.0$ (fixed)\\
Continuum flux ratio & $f_{\mathrm{B}}/f_{\text{tot}}$(4300 Å) & $0.60\pm0.08$ \\
Radius & $R_{\mathrm{B}}$ [$\mathrm{R}_{\odot}$] & $9.0^{+2.1}_{-1.7}$ \\
Bolometric luminosity & $L_{\mathrm{B}}\,[\mathrm{L}_{\odot}]$& $2300^{+1300}_{-800}$ \\
Spectroscopic mass & $M_\mathrm{B,spec}$ [$\mathrm{M}_{\odot}$] & $0.69^{+0.63}_{-0.31}$ \\
Evolutionary mass & $M_\mathrm{B,evol}$ [$\mathrm{M}_{\odot}$] & $0.99\pm0.11$ \\
RV semi-amplitude & $K_{\mathrm{B}}$ [km s$^{-1}$] & \\
-- orbit model  &  & $74.1 \pm 2.2$ \\ 
-- disentangling  &  & $76.0 \pm 0.9$ \\ 
&&\\
\multicolumn{3}{l}{\textbf{Parameters of the Be star}} \\
Effective temperature & $T_{\text{eff,Be}}$ [kK] & $17^{+2}_{-1}$ \\
Surface gravity & $\log(g_{\mathrm{Be}}/\text{cm s}^{-2})$ & $4.0 \pm 0.5$ \\
Rotation velocity & $v_\mathrm{rot,Be}\sin i$ [km s$^{-1}$] & $270\pm70$ \\
Macroturbulent velocity & $v_{\text{mac,Be}}$ [km s$^{-1}$] & $50$ (fixed) \\
Microturbulent velocity & $v_{\text{mic,Be}}$ [km s$^{-1}$] & $2$ (fixed)\\
Continuum flux ratio & $f_{\mathrm{Be}}/f_{\text{tot}}$(4300 Å) & $0.40\pm0.08$ \\
Radius & $R_{\mathrm{Be}}$ [$\mathrm{R}_{\odot}$] & $6.0^{+1.6}_{-1.3}$ \\
Bolometric luminosity & $L_{\mathrm{Be}}\,[\mathrm{L}_{\odot}]$ & $2700^{+1600}_{-1100}$  \\
Spectroscopic mass & $M_\mathrm{Be,spec}$ [$\mathrm{M}_{\odot}$] & $10.5^{+21.5}_{-7.1}$ \\
Evolutionary mass & $M_\mathrm{Be,evol}$ [$\mathrm{M}_{\odot}$] & $6.5\pm0.9$\\
Min. dynamic mass & $M_\mathrm{Be,dyn}$ [$\mathrm{M}_{\odot}$] & $\geq 7.0$ \\
RV semi-amplitude & $K_{\mathrm{Be}}$ [km s$^{-1}$] & $5 \pm 4$ \\

\hline
\end{tabular}
\label{tab:binary_parameters}
\setlength{\extrarowheight}{0pt}
\end{table}

\section{Spectral disentangling}
\label{sec:disentangling}

\begin{figure*}[t!]
    \centering
    \includegraphics[width=\textwidth]{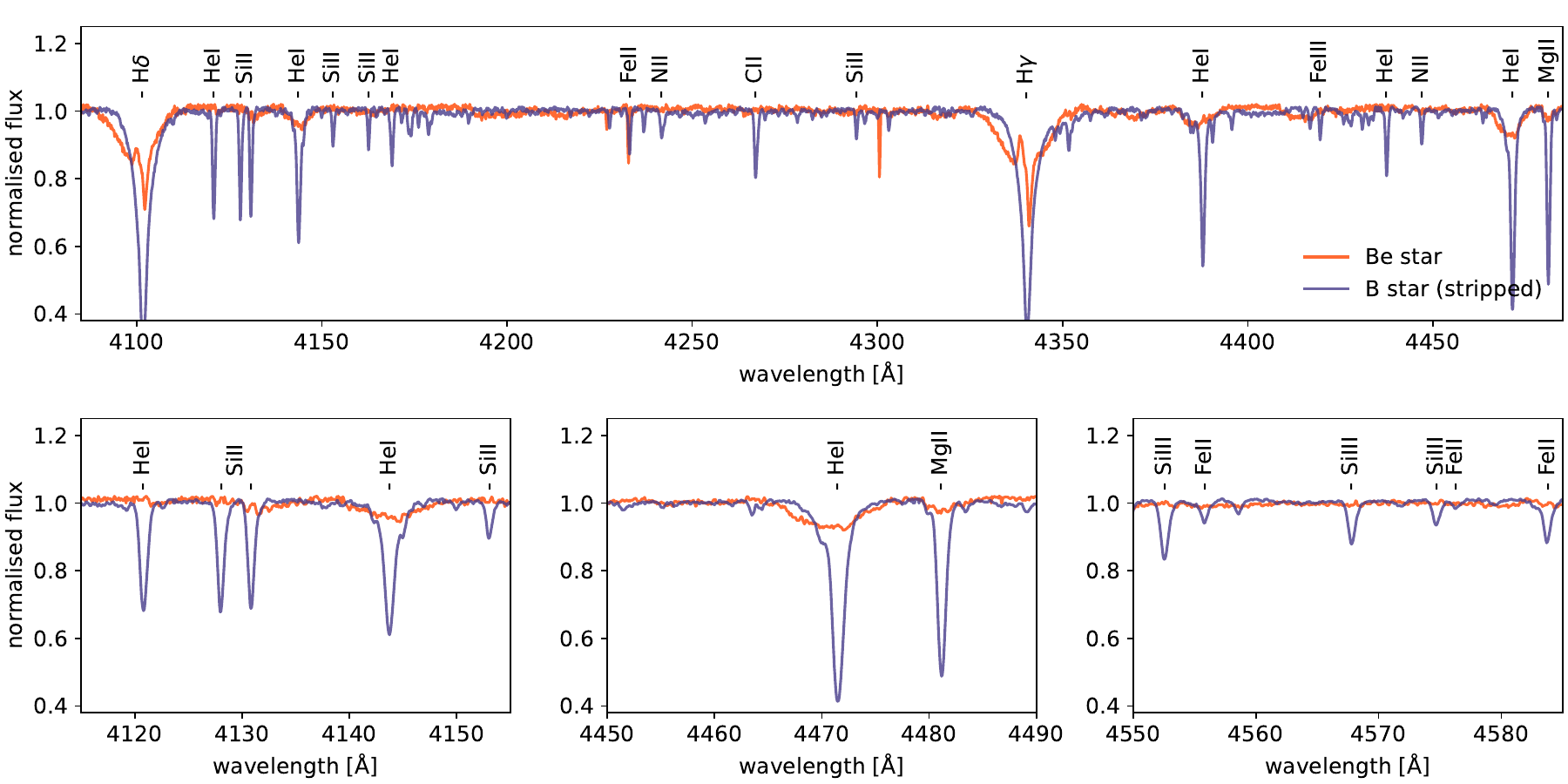}
    \caption{{Disentangled single-star spectra shifted to the presumed rest frame.} The disentangled spectrum of the narrow-lined stripped star (dark blue) and the fast-rotating Be companion star (orange) are shown in selected wavelength ranges. Both component spectra have been rescaled assuming light ratios of 0.6 and 0.4 for the narrow-lined star and the broad-lined star, respectively (see Sect.~\ref{sec:light_ratio}). The rapid rotation of the Be companion is apparent, for example, in the He I line at 4471.5\AA~(centre panel at bottom). The narrow lines in the spectrum of the Be star, including the one at 4300\,Å, are spurious and arise from disentangling artefacts and telluric or interstellar absorption.} 
    \label{fig:disentangled_spectra}
\end{figure*}

We performed spectral disentangling in an attempt to separate the observed composite spectra into the intrinsic spectra of the two stellar components. To increase the robustness of our results, we used two independent methods, namely the shift-and-add technique \citep{Marchenko+1998,Gonzalez_Levato, Shenar+2020} and a method based on singular value decomposition following \cite{simon1994disentangling}.

The shift-and-add technique has previously been used to disentangle spectra in several bloated stripped star systems \citep[e.g.][]{Bodensteiner+2020,Villasenor+2023}. This method starts with an initial guess, typically a flat spectrum for the companion and Doppler-shifted, coadded spectra for the primary star, and iteratively refines both components. The algorithm is described in \cite{Gonzalez_Levato} and we used the \href{https://github.com/TomerShenar/Disentangling_Shift_And_Add}{implementation by T. Shenar} \citep{Shenar+2020,Shenar+2022}, which requires the binary orbital parameters as input.

Using the inferred orbital parameters from Sect.~\ref{sec:orbit_analysis}, we fixed the values of orbital angles, barycentric system velocity, and period and performed the disentangling for a 2D grid in $K_{\mathrm{B}}$ and the unknown Be star RV semi-amplitude $K_{\mathrm{Be}}$. For $K_{\mathrm{B}}$, we explored values within $\pm 2\sigma$ of the inferred value $74.1 \pm 2.2\, \mathrm{km\,s}^{-1}$ with a step size of $\approx0.5\,\mathrm{km \,s}^{-1}$. 
For $K_{\mathrm{Be}}$, we initially tested a coarser grid with semi-amplitudes between $0$ and $74 \, \mathrm{km \, s^{-1}}$ in steps of \( 2 \, \mathrm{km \, s^{-1}} \). The results indicated values of $K_{\mathrm{Be}}$ close to zero $\, \mathrm{km \, s^{-1}}$, motivating a finer grid search. We refined the range to \( 0 \)–\( 36 \, \mathrm{km \, s^{-1}} \), with a step size of \( 0.5 \, \mathrm{km \, s^{-1}} \). The best-fit values $K_{\mathrm{B}}, \, K_{\mathrm{Be}}$ were determined by minimising the combined $\chi^2$ across all epochs. To avoid spurious features, we forced the disentangled spectra to remain below the continuum in regions without emission lines.
We initially performed the disentangling on photospheric helium absorption lines between 4000\,Å and 4500\,Å and the Balmer lines H$\gamma$ and H$\delta$ to derive suitable values for $K_{\mathrm{B}}$ and $K_{\mathrm{Be}}$. Subsequently, we disentangled the entire spectrum using these fixed RV semi-amplitudes.
By focussing on the blue part of the spectrum, we emphasised the region where the presumably hotter Be star companion's contribution is larger and avoided the H$_\alpha$ and H$_\beta$ Balmer lines, which show the strongest emission and variability.

For the RV semi-amplitude of the narrow-lined star, there is a minimum in the $\chi^2$ distribution for $K_{\mathrm{B}} = 76.0 \pm 0.9\,\mathrm{km \, s}^{-1}$, consistent with the results from orbital analysis. For the broad-lined companion star, the $\chi^2$ distribution is minimised for $K_{\mathrm{Be}} = 5 \pm 4 \,\mathrm{km \, s}^{-1}$, corresponding to a dynamic mass ratio of $q_\mathrm{dyn} = \frac{K_{\mathrm{Be}}}{K_{\mathrm{B}}} = 0.066 \pm 0.053$. However, as a consequence of the broad and flat spectral lines of the Be star, the disentangled spectra are nearly identical for values $K_{\mathrm{Be}} \lesssim 20$ km s$^{-1}$, as illustrated in Appendix \ref{appendix:disentangling}.

The component spectra obtained with the shift-and-add disentangling are shown in Fig.~\ref{fig:disentangled_spectra}. The spectra have been rescaled, that is, divided by their estimated continuum flux contributions (0.6 and 0.4 for the narrow-lined and broad-lined stars, respectively; see Sect.~\ref{sec:light_ratio}), to restore the equivalent widths to the values they would have if the stars were observed individually. The spectra are set to a continuum flux of one. Although the flux ratio affects the overall scaling of the spectral line depths, it cannot be directly derived from the disentangling method without additional assumptions. Instead, we estimated the light ratio as part of the spectral analysis and refined the estimate with a fit of the spectral energy distribution (SED; see Sect.~\ref{sec:sed}), taking into account the wavelength dependence of the flux contributions. 

\begin{figure*}[t]
    \centering
    \includegraphics[width=\textwidth]{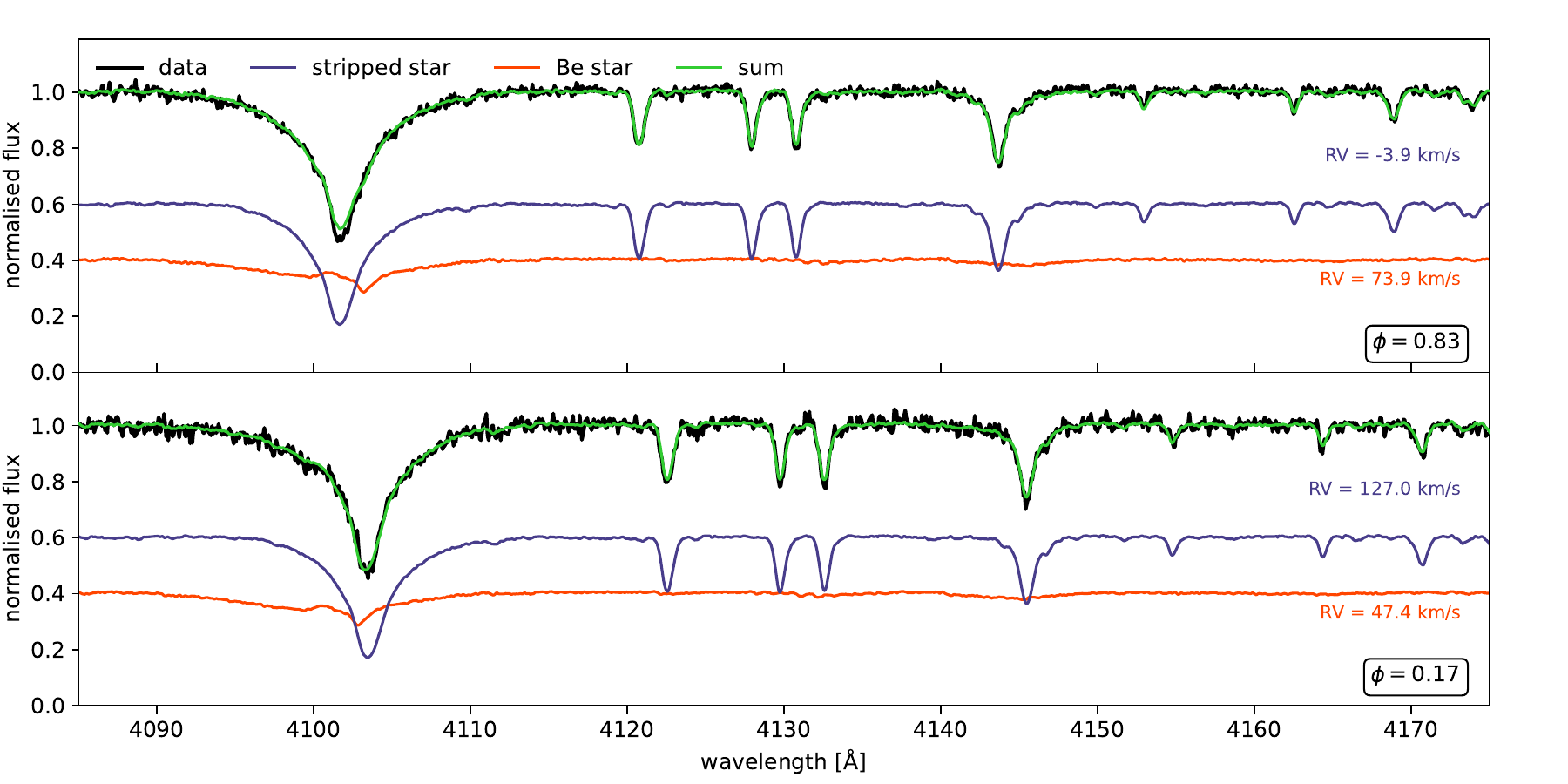}
    \caption{{Observed and disentangled spectra close to quadrature}. Observed TRES spectra with orbital phases $\phi = 0.83$ (top) and $\phi=0.17$ (bottom) are shown as black solid lines. The disentangled spectral components of the narrow-lined stripped star (dark blue) and the Be star (orange) are shown, Doppler-shifted by the inferred RVs (noted on the right). The observed spectra are well reproduced by the sum of both components (green) but are dominated by the contribution of the narrow-lined stripped star. With a flux contribution of $\approx 40\%$ (see Sect.~\ref{sec:light_ratio}) and rotationally broadened absorption lines, the Be star's contribution to the combined spectrum is barely visible, even though it is the considerably more massive star.}
    \label{fig:disentangling_quadrature}
\end{figure*}

The secondary spectrum reveals the presence of a luminous companion star with shallow H and He absorption lines and emission features in the Balmer line cores. This indicates that the companion of the narrow-lined star is indeed a luminous star and not a dark remnant like the previously hypothesised BH. 
The sum of the individual disentangled spectra superimposed on the observed composite spectra at quadrature is shown in Fig.~\ref{fig:disentangling_quadrature}, for a region around the H$\delta$ line.  A plot showing the full wavelength range of the observed, disentangled, and modelled spectra is provided in Fig.~\ref{fig:full_range_disentangled_spectra_with_models_1} in the appendix.

The second disentangling method is described in detail in \citet{seeburger2024autonomous}. This approach combines a linear algebra solver step to compute the component spectra from the input data (epoch spectra and initial RV guesses) with a non-linear optimisation step to determine the best-fit input parameters. The method yields both disentangled component spectra and RV estimates for each component. We find that the \citet{seeburger2024autonomous} method yields comparable results, both for the inferred mass ratio ($\approx 0.061$) and for the two component spectra (see Fig.~\ref{fig:disentangled_Be_star_K2_comparison} and Appendix~\ref{appendix:disentangling} for details). Slight differences are seen in some of the spectral lines, but the differences do not affect our analysis results or main conclusions. 

Both techniques assume that the spectral components remain time-invariant, except for Doppler shifts in wavelength. Although we avoided the H$\alpha$ and H$\beta$ lines during the disentangling process, it is important to note that even the lesser variability in the H$\gamma$ and H$\delta$ lines can impact the results \citep[as was the case for VFTS 291; for example, see][]{Villasenor+2023}. Consequently, the exact profiles of the Balmer lines in the component spectra may be influenced by emission and should be interpreted with caution, in particular for the broad-lined companion. 

%--------------------------------------------------------------------
\section{Spectral analysis}
\label{sec:spectral_analysis}

Through spectral disentangling of the high-resolution, multi-epoch spectra, we have revealed the presence of a second luminous component in the system. Given the lower light contribution of the companion and the broad spectral lines, the combined spectra are dominated by the narrow-lined star, resembling the composite spectra of recently discovered bloated stripped stars with Be companions.

We aim to determine stellar parameters for both stars. To this end, the disentangled spectra were compared with synthetic spectra from model atmospheres. 
For the broad-lined Be star, we used the BSTAR2006 grid based on the model atmosphere code \textsc{Tlusty} \citep{Lanz_Hubeny2007}. These plane-parallel hydrostatic model atmospheres relax the assumption of local thermodynamic equilibrium (non-LTE) and are appropriate for a wide range of B-type stars. The grid covers effective temperatures in the range 15-30\,kK in steps of 1\,kK and surface gravities between 1.75 and 4.75\,dex in steps of 0.25\,dex. We assumed a solar composition and a microturbulent velocity of $2\,\mathrm{km \, s}^{-1}$, a typical value for B-type dwarfs \citep{Nieva_Simon-Diaz2011}. 

When applying BSTAR2006 models to the narrow-lined B star, the fits approached the lower temperature limit of the grid, indicating the need for cooler model atmospheres. We therefore calculated new models in the range 10--15\,kK using the non-LTE Potsdam Wolf-Rayet (\textsc{PoWR}) code \citep{Grafener+2002,Hamann_Grafener2003,Sander+2015}. The models assumed a microturbulent velocity of 10\,km/s, typical for B supergiants \citep{Crowther+2006} and synthetic spectra were generated for stars of effective temperatures 10, 12.5, 13, 14, and 15\,kK, and with surface gravity $\log g$ between 2.0 and 3.0 with step sizes of 0.25\,dex. Further details on the assumptions and setup of the model are provided in Appendix~\ref{appendix:model_abundances}.

We computed models for solar-like metallicity with two sets of abundances: one adopting solar values from \cite{Asplund+2021} and another representing `stripped star-like' abundances, characterised by N enrichment and C and O depletion. The mass fractions for both compositions are listed in Table~\ref{tab:stripped_model_abundances}, and Fig.~\ref{fig:stripped_star_models} compares their corresponding spectra. The most noticeable spectral differences are stronger N lines and weaker O lines, with slight changes in the Si and Fe lines. Since the stripped star-like models better reproduced the strong N lines in the narrow-lined star, we adopted these for the subsequent analysis.

\subsection{The narrow-lined stripped B star}
\label{sec:stripped_star_analysis}

Based on observations with the Low Dispersion Survey Spectrograph at the William Herschel Telescope on La Palma, \cite{Navarro+2012} classify HIP~15429 as a B5Ib star. Another spectrum was obtained by \cite{Gaia+2023} with the HERMES spectrograph to confirm the Gaia DR3 SB1 orbital solution. The authors confirm the spectral classification and interpret the star as a $4.9\pm0.2\,\mathrm{M}_\odot$ blue supergiant with a radius of $20.16\,\mathrm{R}_\odot$, suggesting that it has recently left the main sequence. The mass estimate is based on a comparison of the dereddened position of the binary in the colour-magnitude diagram with single star \textsc{PARSEC} \citep{Bressan+2012} evolutionary tracks. Here, we performed a spectral analysis of the disentangled spectrum and reevaluated this classification of the narrow-lined star.

\subsubsection{Temperature and surface gravity}
\label{sec:stripped_star_teff}

The temperature of the narrow-lined star was determined via the ionisation balance of Si and Fe. Specifically, we measured the EWs of Si~II (4128, 4131, 6347\,Å), Si~III (4552, 4568, 4575\,Å), Fe~II (4584 Å), and Fe~III (5074, 5127, 5156\,Å) lines. For both observed and model spectra, we calculated all possible combinations of EW(Si~III)/EW(Si~II) and EW(Fe~III)/EW(Fe~II) using these lines.
By comparing these ratios, we identified the model temperature that best matched the observed values for a given surface gravity. We avoided the commonly used EW ratios He~I 4471/Mg~II 4481 and He~I 4121/Si~II 4128-32 as temperature indicators, as the He lines may be influenced by a past mass transfer episode \citep[e.g.][Chapter 4.2]{Gray_Corbally2009}.

\begin{figure}[t]%[15]{r}{0.55\columnwidth}%[!htb]
    \centering
    \begin{minipage}[t]{\columnwidth}
        \centering
        \includegraphics[width=\columnwidth]{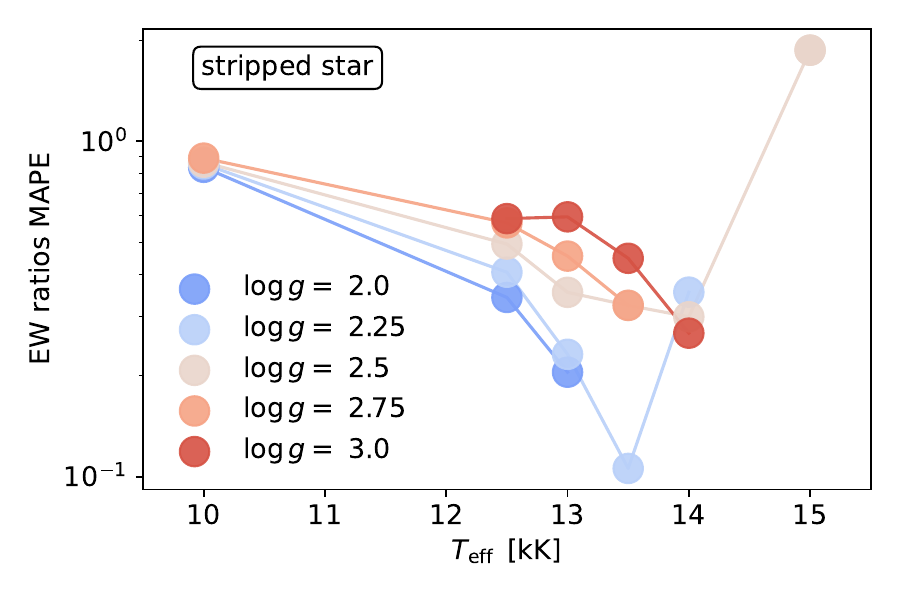}
    \end{minipage}
    \begin{minipage}[t]{\columnwidth}
        \centering
        \includegraphics[width=\columnwidth]{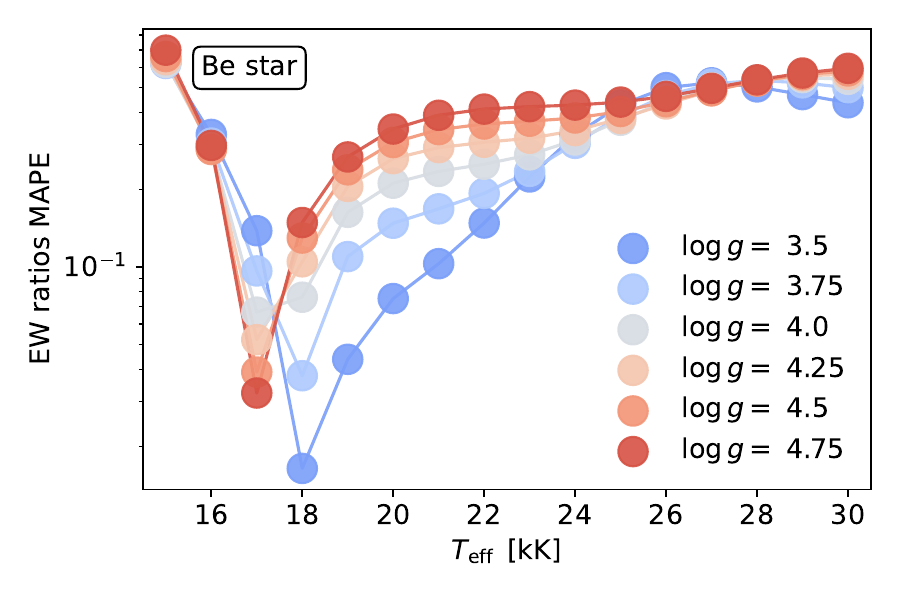}
    \end{minipage}
    \caption{{Temperature determination of the narrow-lined stripped B star (top) and broad-lined Be star (bottom) from ionisation equilibrium}. Markers indicate the MAPE values averaged over all measured EW ratios. For the narrow-lined star, this comprises EW ratios of Si~III over Si~II lines and Fe~III over Fe~II lines. For the Be star, results are shown for the EW ratio of Mg~II over He~I lines. Values for model spectra of varying surface gravities are indicated with different colours (see legends). With an estimated value of $\log g$ in the range 2.0 to 2.5 for the narrow-lined star, the best fit in terms of minimal MAPE is achieved for effective temperatures between 13 and 14\,kK. For the Be star with estimated surface gravity in the range $3.5$ to $4.5$, the best fit is found for effective temperatures between 16 and 19\,kK.} 
    \label{fig:stripped_star_temperature_fit}
\end{figure}

Since the disentangled spectra show no contribution from the Be star in these lines, we assumed that they originate solely from the narrow-lined star. Furthermore, as the flux ratio is not expected to vary substantially across the wavelength range (see Sect.~\ref{sec:light_ratio}), the EW ratios should remain unaffected by the adopted flux ratio and can be directly measured from the observed spectra without rescaling. A detailed overview plot of all measured observed and model EW ratios can be found in Appendix~\ref{appendix:temperature_determination}, see Fig.~\ref{fig:stripped_star_temperature}. In Fig.~\ref{fig:stripped_star_temperature_fit}, the mean absolute percentage error (MAPE) between the EW ratios measured from the observed and model spectra is plotted as a function of the model's effective temperature and surface gravity. For a $\log g$ value of 2.25, MAPE is minimised at an effective temperature of 13.5\,kK. Higher surface gravity would imply higher temperature and vice versa. 

The primary diagnostic for the surface gravity of the narrow-lined B star is the profile of the Balmer line wings. For a fixed temperature, we determined the value of $\log g$ for which the model best matches the observed line wings. We focused on the Balmer lines $H_{\gamma}$ and $H_{\delta}$, where the Be star emission is weakest. 

We employed an iterative approach to determine the optimal combinations of temperature and surface gravity. This involved measurement of the EW ratios for the Si and Fe lines in the observed spectra, comparison of these ratios to those in the model spectra to estimate temperature, adjustment of surface gravity to best match the wings of the Balmer lines and
repeating the process to refine the temperature and gravity estimates.\footnote{We note that we later adopt more helium-enriched models (Sect.~\ref{sec:abundances}). We expect the derived $T_\mathrm{eff}$ and $\log g$ values to remain valid because (a) the Balmer line wings used to constrain $\log g$ are largely insensitive to He abundance (Fig.~\ref{fig:helium_content}), and (b) the ionisation balance is not substantially affected by moderate He enrichment at fixed $T_\mathrm{eff}$ and $\log g$, which we tested by comparing the Si II/III EW ratios in the more helium-rich ($X=0.3$) model to the values quoted here.}
We find that model spectra with a temperature of $T_\mathrm{eff} = 13.5 \pm 0.5 \,$kK
and surface gravity $\log g = 2.25 \pm 0.25$ best reproduce the observations. These results are consistent with the previously determined spectral type B5Ib \citep{Navarro+2012}. Fig.~\ref{fig:stripped_star_models} shows the disentangled spectrum of the narrow-lined star overplotted with the best-fitting \textsc{PoWR} model. The disentangled spectrum has been scaled for its flux contribution, and the model spectra are rotationally broadened (see Sect.~\ref{sec:stripped_star_rotation}).

\subsubsection{Continuum light ratio}
\label{sec:light_ratio}
Spectral disentangling can separate the normalised spectra into the contributions of individual stars but does not constrain their continuum flux ratio. To estimate the flux contribution of the narrow-lined star, we scaled the disentangled spectrum to match the Balmer line depths of the model spectra. Specifically, we used the H$\gamma$ and H$\delta$ absorption lines and determined the flux ratio by averaging the scaling factors required to match their equivalent widths in the model and disentangled spectra. This yields a light contribution of 60\% for the narrow-lined star. The inferred flux ratio varies by approximately 8\% when varying the temperature and surface gravity within their uncertainties, which we adopt as the uncertainty in the flux ratio.

The inferred flux ratio is remarkably insensitive to the assumed hydrogen abundance, as the depth of the Balmer lines remains nearly unchanged across a broad range of plausible values (see Appendix~\ref{appendix:helium_content}).
While the flux ratio varies slightly with wavelength, our SED fit (Sect.~\ref{sec:sed_fit}) shows that this variation is less than 3\% within the 4000 to 7000\,Å range. For the spectral analysis, we adopt a flux ratio of 60\% for the narrow-lined B star and 40\% for the Be companion. Changes in this ratio impact the inferred stellar parameters -- for example, a lower flux contribution for the narrow-lined star would result in smaller radius, luminosity, and mass estimates, and vice versa. However, moderate variations ($\lesssim 10\%$) have no significant impact on the results.\footnote{For instance, increasing the B star’s contribution by 10\% alters the final mass estimate by only $+0.1\,\text{M}_\odot$ ($\sim10\%$), well within the reported uncertainties.}

\subsubsection{Rotational velocity and macroturbulence}
\label{sec:stripped_star_rotation}

We determined the rotational velocities of both stars by fitting rotationally broadened models to individual absorption lines in
the disentangled spectra. For the narrow-lined B-type star, we used the Mg~II $4481\,\AA$, Si~II 4128, $4131\,\AA$ and Si~III 4552, 4567, $4574\,\AA$ absorption lines. The \textsc{PoWR} model spectra were convolved with a rotational kernel, varying the value of $v_\mathrm{rot} \sin i$, and a radial-tangential kernel, with varying values of the macroturbulent velocity $v_\mathrm{mac}$ \citep{Gray1977,Gray2005}. The numerical implementation used for both line-broadening mechanisms is an adaption of the Fortran code \href{http://tlusty.oca.eu/Synspec49/synspec.html}{\texttt{rotin3}} to Python, and the best-fitting parameters for each absorption line were determined from a least-squares fit.

For a fixed macroturbulent velocity of $v_\mathrm{mac} = 0\,\mathrm{km\,s}^{-1}$, the resulting weighted mean and standard deviation over all lines of the best-fit rotational velocities are $v_\mathrm{rot} \sin i = 29 \pm 1\,\mathrm{km\,s}^{-1}$. In contrast, for vanishing rotation ($v_\mathrm{rot} \sin i = 0\,\mathrm{km\,s}^{-1}$), we find $v_\mathrm{mac} = 33\pm 1\,\mathrm{km\,s}^{-1}$. Leaving both parameters to vary freely produces degenerate results, with $v_\mathrm{rot} \sin i$ in the range of 5 to $19\,\mathrm{km\,s}^{-1}$ and values for $v_\mathrm{mac}$ between 26 and $49\,\mathrm{km\,s}^{-1}$. The results are illustrated for the Si~III $4552\,\AA$ line in Fig.~\ref{fig:stripped_star_rotation} in the appendix. 

Given the degeneracy between macroturbulent and rotational broadening, we only placed an upper limit on the rotation of the B star at $v_\mathrm{rot} \sin i \lesssim 30\,\mathrm{km\,s}^{-1}$ and proceeded with median values of $v_\mathrm{rot} \sin i = 10\,\mathrm{km\,s}^{-1}$ and $v_\mathrm{mac} = 30\,\mathrm{km\,s}^{-1}$ for the rest of the analysis. If the B star were tidally synchronised from a recent mass transfer phase, it would be expected to rotate with $v_\mathrm{rot} \equiv 2\pi R_* / P \approx 2\,\mathrm{km\,s}^{-1}$\,, with $R_*$ the stellar radius.

\begin{figure*}
    \centering
    \includegraphics[width=\textwidth]{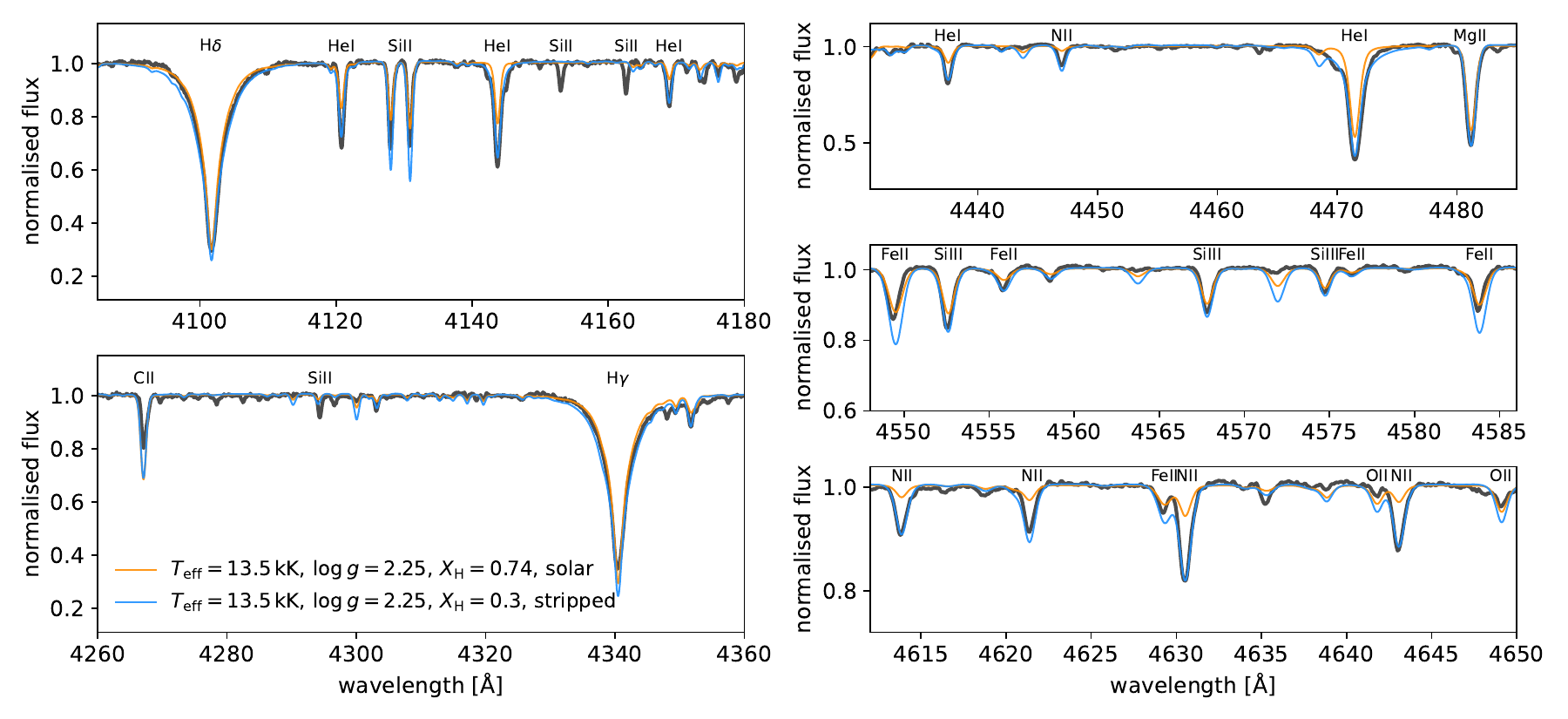}
    \caption{Spectral constraints on the stellar parameters of the narrow-lined stripped star. The panels show the disentangled spectrum (black) compared to PoWR models with the best-fit parameters ($T_\mathrm{eff} = 13.5,\text{kK}$, $\log g = 2.25$). The models differ in composition: one assumes solar hydrogen, helium, and metal abundances \citep{Asplund+2021}, while our fiducial model (blue) is He- and N-enriched, characteristic of a stripped star. The narrow H-Balmer absorption lines support a low surface gravity, while the stripped-star model provides a better fit to the He and N absorption features. Metal lines not present in the model spectra (e.g. Si~II 4153, 4163\,\AA) were not included in the atomic line lists.}
    \label{fig:stripped_star_models}
\end{figure*}

\subsubsection{Stellar abundances}
\label{sec:abundances}

The initial comparison with a reference star of approximately solar metallicity and composition (Sect.~\ref{sec:variability}) suggested a peculiar abundance pattern in the narrow-lined star. In particular, it indicated significant carbon and oxygen depletion in the star's photosphere. Spectral analysis confirms this and further shows that nitrogen is enhanced in the narrow-lined star, evident from the stronger nitrogen lines in the disentangled spectrum compared to the solar abundance model (Fig.~\ref{fig:stripped_star_models}). The enhanced nitrogen was not clearly visible in the composite spectrum (Fig.~\ref{fig:CNO_comparison_reference_star}) due to the flux contribution of the Be companion, but becomes apparent in the disentangled spectrum. The observed abundance pattern -- nitrogen enhancement along with carbon and oxygen depletion -- is consistent with that seen in other bloated stripped stars \citep{Bodensteiner+2020,ElBadry_Quataert2021,Villasenor+2023,Ramachandran+2024}.

The Balmer lines are well reproduced in the model spectra by construction, since they were used to determine the flux ratio. However, the helium lines in the disentangled spectrum are significantly stronger than in the solar abundance model, indicating an elevated helium mass fraction and corresponding hydrogen depletion.
To estimate the degree of hydrogen depletion, we computed additional model spectra for the best-fit $T_\mathrm{eff}$ and $\log g$, varying the hydrogen mass fraction between $X = 0.01$ and $X = 0.7$ in steps of 0.1 (see Appendix~\ref{appendix:helium_content} for details). We adopted the $X = 0.3$ model as our fiducial model, as it best reproduces the observed equivalent widths of helium lines (see Figures \ref{fig:stripped_star_models} and \ref{fig:helium_content}). This corresponds to a helium enhancement factor of 2.8 relative to solar values, higher than what has been reported for other bloated stripped stars \citep{Bodensteiner+2020,ElBadry_Quataert2021,Villasenor+2023,Ramachandran+2024}. The He abundance estimate is sensitive to the flux ratio of the two components. Our reported flux ratio uncertainty of approximately 0.08 translates into an uncertainty in the estimated He mass fraction of about $\pm0.1$.

Regarding individual metal abundances, our stripped-star models assume a carbon depletion factor of 0.29 and a slightly subsolar oxygen abundance (factor of 0.86). The carbon and oxygen lines in the observed spectrum are weaker than in the models, indicating even lower actual abundances for these elements. In contrast, the nitrogen lines are well matched by the models, with an assumed enhancement factor of 10, implying a carbon-to-nitrogen ratio at least 34 times lower than the solar value. Other elements, such as silicon and magnesium, appear to be at approximately solar levels, with their absorption lines well reproduced in the models.

Overall, the observed abundance pattern suggests the presence of material processed via the CNO cycle in the convective core during the main sequence. This is consistent with the scenario where the donor star's core was exposed in the photosphere when the star lost its outer layers during a mass-transfer phase.

\subsection{The broad-lined Be star}
\label{sec:be_star_analysis}
\begin{figure}[t]
    \centering
    \includegraphics[width=\columnwidth]{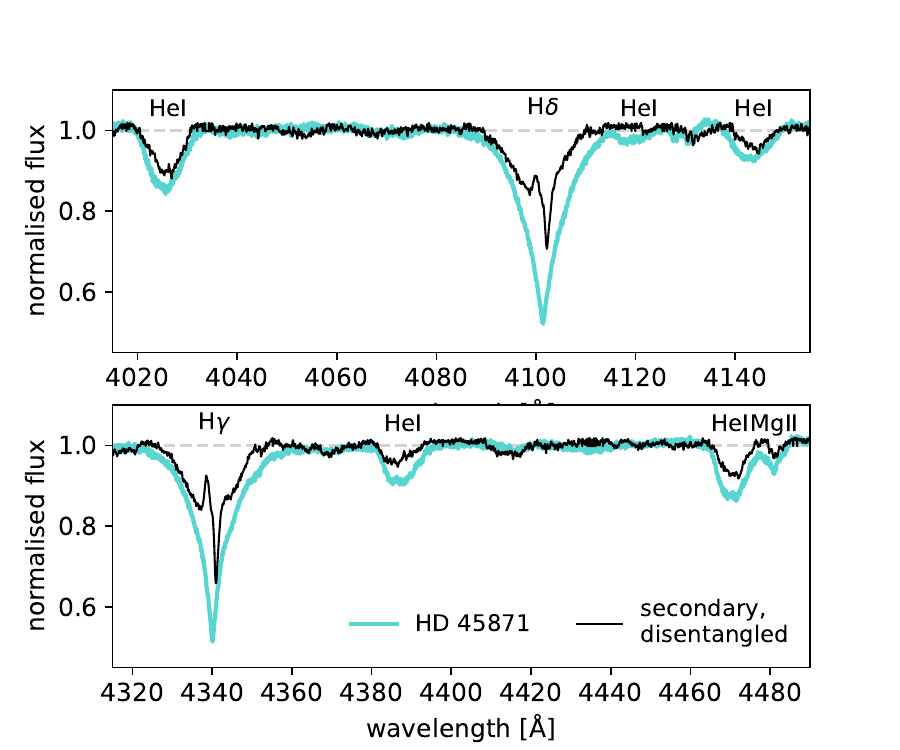}
    \caption{Comparison of the disentangled spectrum of the Be companion star (black) with that of the Be star HD~45871, which has a spectral type of B3Ve \citep{Levenhagen+2006}. The reference spectrum was obtained from the ESO archive. It was observed by M. Borges Ferndandes with the FEROS Échelle spectrograph at the
MPG/2.2m telescope and can be found under the archive ID ADP.2016-09-28T06:54:50.234.}
    \label{fig:be_star_reference_spectrum}
\end{figure}

The Be nature of the secondary star is apparent from the broad-lined emission in the Balmer lines. However, spectral analysis is complicated by the otherwise shallow absorption lines. The only clear absorption lines in the spectrum come from the H Balmer lines, He~I absorption and a Mg~II line at 4481\,\AA. 

We started by qualitatively comparing the disentangled spectrum to the archival spectra of classical Be stars with known spectral types. A comparison of the HIP~15429 Be star spectrum to HD~45871 is shown in Fig.~\ref{fig:be_star_reference_spectrum}. HD~45871 is a B3Ve star with measured temperature of $T_\mathrm{eff} = (20\pm0.5)$\,kK, surface gravity of $\log g = 3.72 \pm 0.10$ and $v_\mathrm{rot} \sin i = (275\pm15)\,{\rm km\,s^{-1}}$ \citep{Levenhagen+2006}. The similarities between the absorption lines and the line profiles suggest a similar spectral type and properties for the HIP~15429 Be companion. 

For a more quantitative fit of the stellar parameters, we compared the disentangled spectrum to the BSTAR2006 grid of \textsc{Tlusty} model atmospheres \citep{Lanz_Hubeny2007}. We continuum-normalised the model spectra in the same way as the observed spectra and resampled them to the same wavelength grid as the disentangled spectrum. 

\subsubsection{Temperature and surface gravity}
\label{sec:be_star_teff}

Similarly to what was done for the analysis of the narrow-lined star, we estimated the effective temperature based on the equivalent width ratios of different absorption lines. In the absence of suitable Si~II, Si~III or Fe lines, we opted for the ratio of Mg~II to He~I at 4481 and 4472\,\AA, respectively. For the narrow-lined B star, we could estimate the surface gravity on the basis of the width of the Balmer line wings. As explained in Sect.~\ref{sec:disentangling}, this is not feasible for the broad-lined Be star because the Balmer lines, including H$_\gamma$ and H$_\delta$, in the disentangled spectrum are likely compromised by variable emission from the circumstellar disc. Instead we assume that the Be star has surface gravity typical of a B-type main-sequence star on the order of $\log g \approx 4.0$. From Figures~\ref{fig:stripped_star_temperature_fit} and \ref{fig:Be_star_temperature}, we see that for a plausible range of surface gravities $3.5 \lesssim \log g \lesssim 4.5$, the best-fit temperature is in the range $16$ to $19$\,kK. We assumed these parameter ranges for the further analysis but point out the substantial uncertainties.

\subsubsection{Rotational velocity}
\label{sec:be_star_rotation}

Following the same approach used for the narrow-lined star, we determined the rotational velocity of the Be star by fitting individual absorption lines in the spectrum. We used the \textsc{Tlusty} model spectrum with $T_\mathrm{eff} = 17\,$kK, $\log g = 4.0$ and convolved it with a rotational kernel \citep{Gray2005} with varying values of $v_\mathrm{rot} \sin i$. A best fit is determined by minimising least squares. Because of the low S/N and strong rotation, the macroturbulent velocity of the Be star cannot be meaningfully constrained. We therefore set it at a fixed value of $50\,\mathrm{km}$/s (following, e.g. \citealt{Ramachandran+2024,Bodensteiner+2020}). In the absence of other suitable absorption lines in the spectrum, we used He~I $4026\,\AA$, $4144\,\AA$, $4388\,\AA$, and $4472\,\AA$ lines of the Be star to estimate the rotational velocity. The method yields $v_\mathrm{rot} \sin i$ values in the range $\sim250$ to $425$\,km/s for the individual lines and $v_\mathrm{rot} \sin i = 270 \pm 70\,\mathrm{km \, s}^{-1}$ as the weighted mean and standard deviation of the weighted mean. The fit results are illustrated for the He~I 4026\,Å line in Appendix \ref{appendix:rotation}.
Helium lines, though affected by pressure broadening, are commonly used for rotational velocity estimates in massive stars \citep{Dufton+2013, Ramirez-Agudelo+2013}. However, given the noise level and the limited number of available lines, the uncertainties derived for $v_\mathrm{rot} \sin i$ should be interpreted with caution.

\begin{figure*}[t]
    \centering
    \includegraphics[width=\textwidth]{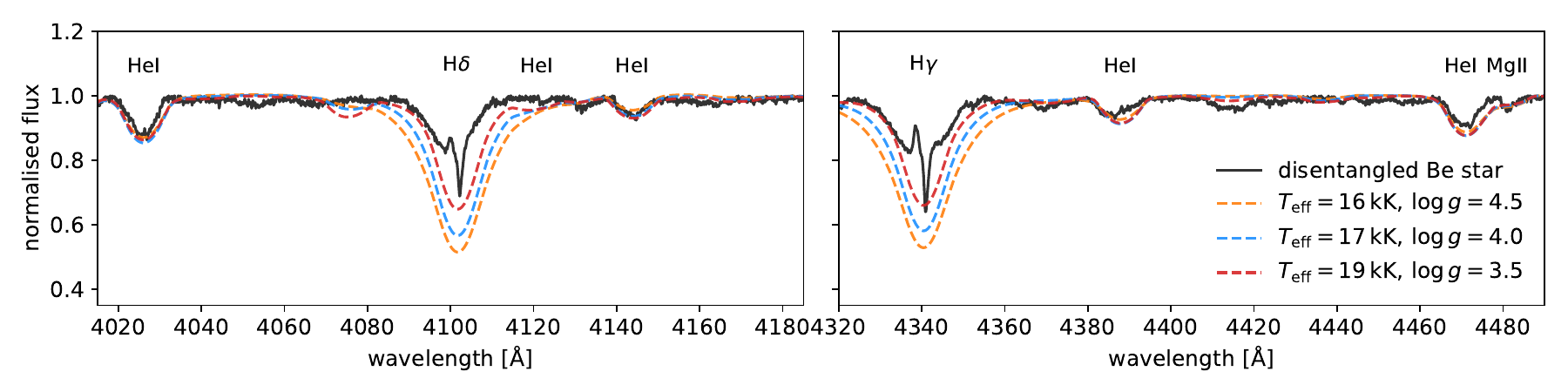}
    \caption{Spectral constraints on the temperature and surface gravity of the Be companion star. The panels show a comparison between the disentangled Be star spectrum (black) with models of different $T_\mathrm{eff}$ - $\log g$ combinations (see the legend for the respective $T_\mathrm{eff}$ and $\log g$ values). The blue dashed line represents our fiducial model. The shape of the Balmer line wings favours a surface gravity of $\log g$ of 3.5 - 4.0, as would be expected for a star that is slightly inflated due to recent mass transfer.} 
    \label{fig:Be_star_spectrum}
\end{figure*}

In Fig.~\ref{fig:Be_star_spectrum}, the disentangled spectrum of the Be star is compared to the model spectra with temperatures and surface gravities within the derived parameter ranges and a rotational velocity $v_\mathrm{rot} \sin i = 270\,\mathrm{km \, s}^{-1}$. The models match the observations well, except for the Balmer absorption line cores, which are strongly affected by the variable emission, and support the initial classification as a B3V star. 

As discussed in Sect.~\ref{sec:luminosities_masses}, the system is possibly observed close to edge on, that is, it is unlikely that the value of $\sin i$ is significantly less than one. With that in mind, the Be star's rotational velocity of $v_\mathrm{rot} \approx 270\,\mathrm{km \, s}^{-1}$ is low, corresponding to about 60\% of the critical rotation velocity, $v_\mathrm{crit} = \sqrt{\nicefrac{2 G M_*}{3 R_*}}$, assuming typical masses and radii for B-type main-sequence stars (i.e. $ 5\,\mathrm{M}_\odot, \, 3.5\,\mathrm{R}_\odot $). For Be stars, one would typically expect $v_\mathrm{rot}/v_\mathrm{crit} \geq 0.8$ \citep{Townsend+2004}. This is a first indicator that the Be star is somewhat inflated - as also suggested by its narrower Balmer lines - because this would reduce its critical rotation velocity. However, \cite{Townsend+2004} note that $v_\mathrm{rot} \sin i $ determined in this way may underestimate the true projected equatorial rotation velocity due to gravity darkening. On the other hand, \cite{Zorec+2016} shed doubt on the fact that Be stars are near-critical rotators and instead found a wide range of velocity ratios $0.3 \lesssim v_\mathrm{rot}/v_\mathrm{crit} \lesssim 0.95$ among Be stars.

\section{Photometric analysis}
\label{sec:sed}
\subsection{Spectral energy distribution fit}
\label{sec:sed_fit}
To estimate the stellar radii, we fitted the SED of the system.  We queried archival photometric data for HIP~15429 using the VO Sed Analyzer \citep[VOSA;][]{Bayo+2008}.  We found near-infrared photometry that includes J-, H-, and Ks-band data from the 2MASS all-sky catalogue \citep{Cutri+2003}, and far-infrared photometry that covers the W1, W2, W3, and W4 bands from the Wide-field Infrared Survey Explorer \citep[WISE;][]{Wright+2010}. Optical photometry was obtained from Gaia DR3 in the G, G\_RP, and G\_BP filters, together with synthetic photometry in the Sloan u, g, r, and i filters derived from Gaia BP/RP mean spectra \citep{Gaia_synthphot}. Additionally, we acquired near-UV photometry with the Swift Ultra-Violet/Optical Telescope \citep[UVOT;][]{Roming+2005}. On 2 March 2023, we obtained a single exposure in the UVM2 filter band with an exposure time of 505\,s and 2x2 binning \footnote{Obs. ID: 00015861001, Target ID: 15861}. The raw data were processed and calibrated with the standard pipeline at the Swift Data Center (version 3.19.01). We measured flux densities and uncertainties via manually placed apertures using the \texttt{uvotsource} routine as part of the \textsc{HEASOFT} software package \citep{HEASOFT}. 
Table~\ref{tab:phot_overview} lists all flux measurements and filter bands. 

For the narrow-lined star, we used the grid of PoWR stellar atmosphere models, while for the companion Be star, we used the \textsc{Tlusty} spectra (see the beginning of Sect.~\ref{sec:spectral_analysis} and Appendix~\ref{appendix:model_abundances} for model specifications). We adopted the extinction law of \cite{Fitzpatrick_Edward1999} and performed two-dimensional linear interpolation between the model SEDs, with synthetic photometry computed using \texttt{pyphot} \citep{Fouesneau2024}.

We used a nested sampling framework to infer the posterior PDFs for the stellar radii. The radius of the narrow-lined star, $R_B$, was assigned a uniform prior of $[1,15]\,\mathrm{R}_\odot$. Since extinction affects the shape of the SED, we also inferred the extinction parameters, assuming uniform priors for $E(B-V) \in [0.5, 1.0]\,\textrm{mag}$ and $R_V \in [2.0, 4.0]$. The broad-lined Be star's radius was constrained through the flux ratio and the inferred $R_B$. The stellar parameters ($T_\mathrm{eff}$, $\log g$, and the flux ratio) were determined from spectral analysis; rather than independently redetermining them, we adopted normal priors based on their measured values and uncertainties (see Table~\ref{tab:binary_parameters}), allowing them to vary only to propagate their uncertainties to the radius estimates. For distance, we assumed a normal prior centred on the Gaia DR3 estimate of $(1693 \pm 89)$\,pc, with the standard deviation set to three times the Gaia uncertainty. This accounts for the typical underestimation of parallax (and hence distance) uncertainties by up to a factor of 2 for bright sources and those with $\mathrm{RUWE} > 1.4$ \citep{ElBadry+2021,Nagarajan_ElBadry2024}. Our analysis indicates an infrared excess, likely originating from the companion star's circumstellar disc. Therefore, we restricted the SED fit to the optical and UV wavelength bands, excluding WISE and 2MASS photometry.

\begin{figure}
    \centering
    \includegraphics[width=\columnwidth]{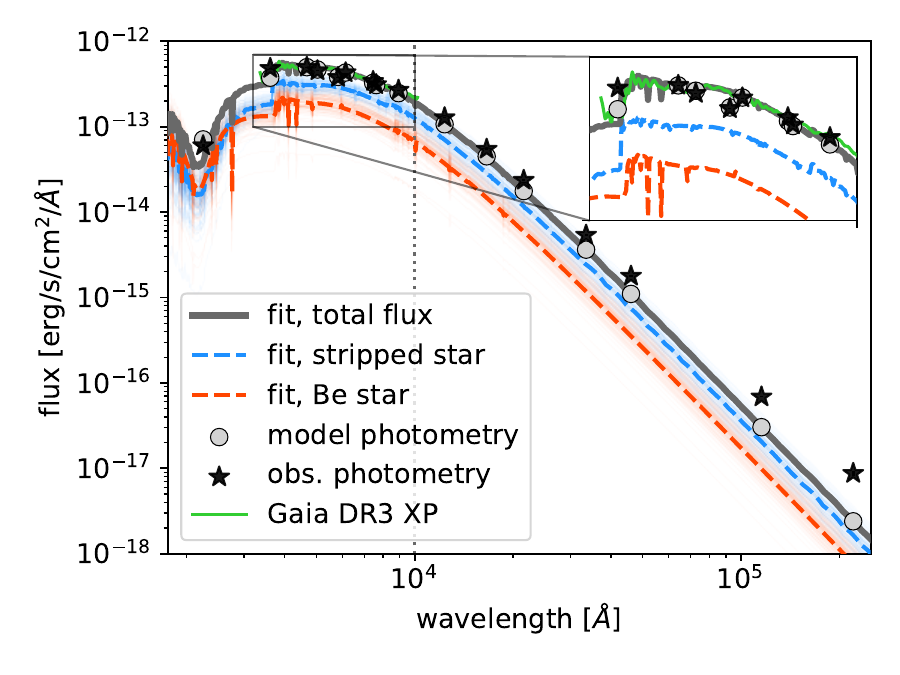}
    \caption{{Comparison of model SEDs with observed photometry of HIP~15429}. Black stars indicate photometric measurements (see Table~\ref{tab:phot_overview}), grey circles show synthetic photometry generated from the model SED in the same filters. Only data points leftwards of the grey dotted line were used in the fitting process. As a grey line, the model fit of the SED is plotted which is the sum of contributions from the narrow-lined stripped star (blue) and Be star companion (red). For the stellar components, example draws from the posterior are shown as thin lines, the median values as the thicker, dashed curves. The inset axis shows a zoom-in on the optical wavelength regime with the Gaia XP spectrum overplotted in green. An infrared excess is visible on the right-hand side of the plot.}
    \label{fig:sed_fit}
\end{figure}

\begin{figure*}[t]
    \centering
    \begin{minipage}[t]{\columnwidth}
        \centering
        \includegraphics[width=\columnwidth]{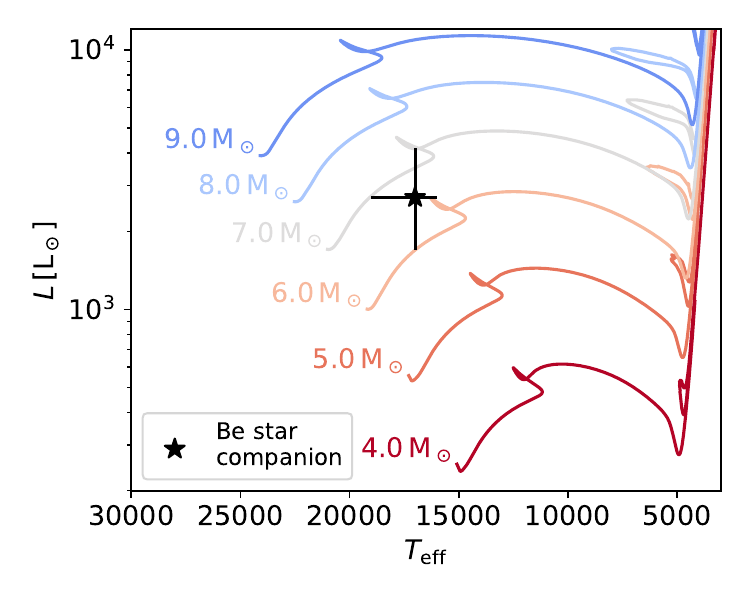}
    \end{minipage}%
    \begin{minipage}[t]{\columnwidth}
        \centering
        \includegraphics[width=\columnwidth]{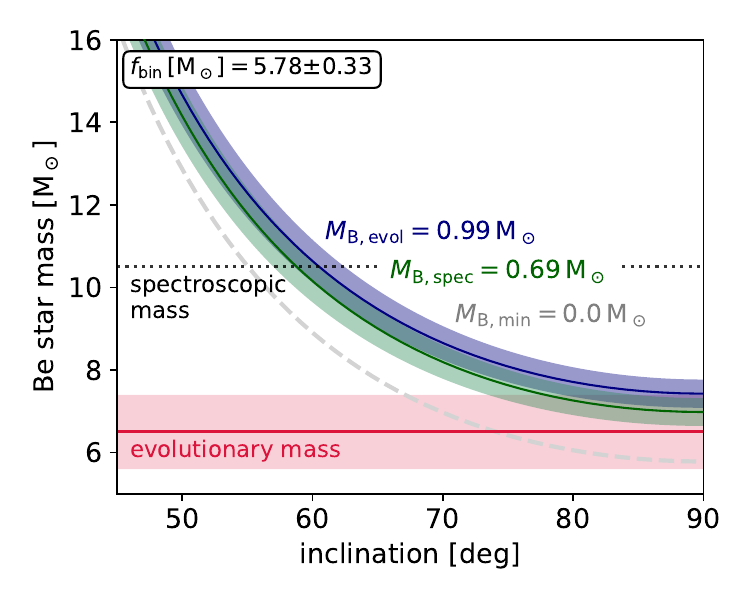}
    \end{minipage}
    \caption{\textit{Left:} {Comparison of the inferred parameters of the Be star companion to MIST single star evolutionary tracks.} The position of the Be star in the HRD is consistent with a $M_{\mathrm{Be,evol}} = 6.5\pm0.9\,\mathrm{M}_\odot$ B-type dwarf. \\
    \textit{Right:} {Summary of constraints on stellar mass for the Be star companion.} The temperature and luminosity-based mass estimate is shown as a red horizontal line, with the shaded region indicating the uncertainty estimate. The spectroscopic mass is shown in black with the uncertainties exceeding the plotted mass range. Blue, green, and grey curves indicate Be star masses inferred from the binary mass function ($f_\mathrm{bin}$) as a function of inclination angle. The different colours correspond to different estimates of the mass of the stripped star: $M_{\mathrm{B,spec}} = 0.69\,\mathrm{M}_\odot$, as inferred from the spectroscopic analysis, and $M_\mathrm{B,evol} = 0.99\,\mathrm{M}_\odot$, the stripped star mass of the best-fit MESA evolutionary model. We note that for the Be star, the minimum allowed dynamical mass, $M_\mathrm{Be,dyn} \geq 7.0\,\mathrm{M}_\odot$ assuming $M_\mathrm{B} = 0.69\,\mathrm{M}_\odot$, is somewhat larger than the inferred evolutionary mass.}
    \label{fig:be_star_stellar_tracks}
\end{figure*}

Figure~\ref{fig:sed_fit} compares the fitted model SED to observed photometry. A corner plot of the posterior PDFs for all parameters is provided in Fig.~\ref{fig:sed_corner_plot} in the appendix. We find good agreement between models and data for a reddening of $E(B-V) = 0.69\pm0.06$\,mag with $R_V = 2.7\pm 0.5$, although the latter is less constrained. The inferred extinction $A_V = E(B-V) R_V \approx 1.88^{+0.36}_{-0.32}$\,mag is consistent with values inferred in the Milky Way dust extinction map by \cite{Zhang_Green2024}. 
We infer a radius of $R_\mathrm{B} = 9.0^{+2.1}_{-1.7}\, \mathrm{R}_\odot$ for the narrow-lined B star. With a light ratio of $f_\mathrm{B}/(f_\mathrm{B}+f_\mathrm{Be}) = 0.6\pm0.08$ at $4300\,\AA$, this yields a radius of the broad-lined Be star of $R_\mathrm{Be} = 6.0^{+1.6}_{-1.3}\, \mathrm{R}_\odot$. All values quoted here are the medians of the posterior PDFs, with the uncertainties corresponding to the 68\% quantiles.

\subsection{Stellar masses and luminosities}
\label{sec:luminosities_masses}

We applied the Stefan-Boltzmann law to infer the stellar luminosities of both stars from the samples in the equally weighted posterior distribution of the SED fit. We report median values with 1-$\sigma$ uncertainties, yielding ${L_\mathrm{B}}/{\mathrm{L}_\odot} = 2300^{+1300}_{-800}$ for the narrow-lined B star and ${L_\mathrm{Be}}/{\mathrm{L}_\odot} = 2700^{+1600}_{-1100}$ for the broad-lined Be companion. Similarly, we combined the posterior samples for radii and surface gravities to estimate stellar masses using Newton's law of gravitation;  $g = \nicefrac{GM}{R^2}$, yielding ${M_{\mathrm{B,spec}}}/{\mathrm{M}_\odot} = 0.69^{+0.63}_{-0.31}$ and ${M_{\mathrm{Be,spec}}}/{\mathrm{M}_\odot} = 10.5^{+21.5}_{-7.1}$.

The inferred spectroscopic mass of the narrow-lined B-type star is much lower than the value of $4.9 \pm 0.2\,\mathrm{M}_\odot$ inferred by \cite{Gaia+2023} based on single-star evolutionary tracks, and it is unreasonably low for a regular B5 supergiant \citep[e.g.][]{Haucke+2018}. Together with the low inferred surface gravity and large radius, this suggests that, similar to LB-1 \citep{Shenar+2020} and HR~6819 \citep{Bodensteiner+2020, ElBadry_Quataert2021}, the narrow-lined star in HIP~15429 is a bloated stripped star. In this scenario, the initially more massive star in the binary was stripped of its envelope, transferring mass to its companion and leaving it in a short evolutionary stage with an inflated size, hence the term `bloated', and temperatures in the mid- to late-B regime. For a more detailed discussion of the evolutionary history of the system (see Sect.~\ref{sec:evolution}). 

The inferred spectroscopic mass of the Be star is higher than those of typical B stars of similar temperatures \citep{Pecaut_Mamajek2013}, but we note the substantial uncertainties due to poorly constrained surface gravity. However, we can get independent evolutionary and dynamic mass estimates of the companion. The evolutionary mass estimate is illustrated in Fig.~\ref{fig:be_star_stellar_tracks}, where the inferred spectroscopic temperature and luminosity of the Be star are compared with a selection of stellar evolution tracks of single B-type stars from the MESA Isochrones \& Stellar Tracks (MIST) library \citep{Dotter2016,Choi+2016,Paxton+2011,Paxton+2013,Paxton+2015,Paxton+2018,Paxton+2019,Jermyn+2023}. 
We used a grid of stellar evolution tracks of solar metallicity with masses between 4 and 9\,M$_\odot$ in steps of 0.1\,M$_\odot$. Each track was sampled at 1000 evenly spaced time steps, resulting in a regular grid of predicted stellar effective temperatures, luminosities, and masses. The best-fit mass was determined by minimising the $\chi^2$ statistic, which quantifies the deviation between observed and predicted values, weighted by their respective uncertainties. To estimate uncertainties, we fitted a parabola to the $\chi^2$ distribution around its minimum and identified the parameter values corresponding to $\chi_\mathrm{min}^2+1$. This yielded an evolutionary mass of $M_{\mathrm{Be,evol}} = 6.5\pm0.9\,\mathrm{M}_\odot$ for the Be star.

We can also estimate the minimum mass of the companion dynamically, by combining the inferred stripped star mass of ${M_{\mathrm{B,spec}}}/{\mathrm{M}_\odot} = 0.69^{+0.63}_{-0.31}$ with the binary mass function 
\begin{equation}
    f_M \equiv \frac{M_\mathrm{Be}^3 \sin^3 i}{(M_\mathrm{B}+M_\mathrm{Be})^2}
    \equiv \frac{P K^3}{2\pi G}\left(1-e^2\right)^{3/2} = (5.78 \pm 0.33)\,\mathrm{M}_\odot \,.
\end{equation}
This yields a minimum companion mass of ${M_{\mathrm{Be,dyn}}} \geq 7.0\,\mathrm{M}_\odot$, assuming a stripped star mass of ${M_{\mathrm{B,spec}}} = 0.69\,{\mathrm{M}_\odot}$ and would imply an RV semi-amplitude of $K_\mathrm{Be} \leq 7.5\,\mathrm{km/s}$. 

All three mass constraints (spectroscopic, evolutionary, dynamical) and the inclination dependence of the inferred dynamical companion mass are illustrated in Fig.~\ref{fig:be_star_stellar_tracks}.
We note that the minimum implied dynamical mass is slightly higher than the evolutionary luminosity-based mass estimate. This may indicate that the system is viewed at high inclination and that the Be star is less luminous in the optical than expected for its mass. This could be due to recent accretion or self-absorption from a circumstellar disc, or the star could be slightly inflated, with a lower temperature and larger inferred radius compared to typical main-sequence stars, possibly as a result of recent mass transfer \citep{Kippenhahn_Meyer-Hofmeister1977,Lau+2024}.

\section{Evolutionary history}
\label{sec:evolution}

\begin{figure}
    \centering
    \includegraphics[width=\columnwidth]{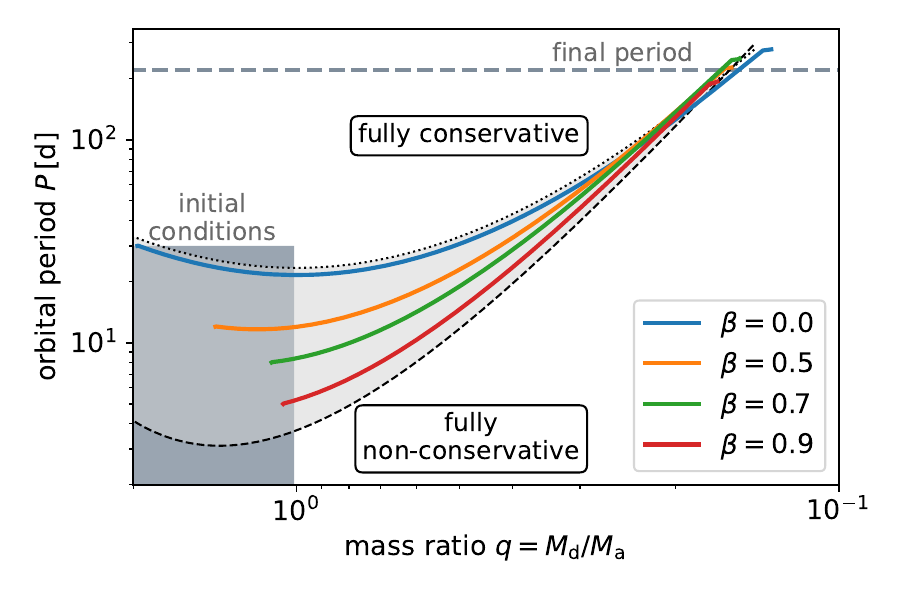}
    \caption{{Period versus mass ratio evolution models for HIP~15429 assuming isotropic re-emission.} The estimated present-day period is shown as the grey dashed line, while the range of tested initial parameters of the pre-mass transfer system are indicated as the grey-shaded region. Theoretical tracks for the case of fully (non-) conservative mass transfer are shown as (dashed) dotted black lines. They are chosen to match the estimated present-day binary properties. Coloured lines indicate binary evolution models computed with varying initial periods, mass ratios and mass accretion efficiencies, $\beta$ (see models M3a to M3d in Table~\ref{tab:mesa_models}). Time evolves from left to right, with the left-most point representing the initial values. The evolutionary tracks assume circularised orbits during mass transfer. For discussion of eccentric mass transfer scenarios (see Sect.~\ref{sec:discussion}).}
    \label{fig:period_evolution}
\end{figure}

Our aim was to constrain both the properties of the progenitor system and the future evolution of the HIP~15429 binary. To do so, we created binary evolution models for the system using the 1D stellar evolution code Modules for Experiments in Stellar Astrophysics \citep[MESA;][version 24.03.1]{Paxton+2011,Paxton+2013,Paxton+2015,Paxton+2018,Paxton+2019,Jermyn+2023}. 
In a small grid of models, we varied initial masses, mass ratios, and initial orbital periods trying to match the current observed values. 
A detailed description of the simulation setup can be found in Appendix~\ref{appendix:mesa_set-up}.

We denote the initial parameters of the simulated binary stars with the subscript `initial' and refer to the post-mass transfer properties of the stripped and Be stars with the subscripts `stripped' and `Be'. The donor and accretor stars are indicated with the suffixes `d' and `a', respectively. The present-day properties of the stripped and Be stars were extracted at the model step after mass transfer, where the difference between the simulated and observed effective temperatures of the stripped star was minimised.
At any point during evolution, we define the binary mass ratio as $q \equiv \nicefrac{M_\mathrm{d}}{M_\mathrm{a}}$. Since the donor star is initially the more massive one, we have $q_\mathrm{initial} > 1$. 

In response to the changing mass ratio, the system's orbital period evolves. We computed mass transfer during Roche-lobe overflow using the implicit mass transfer approach proposed by \cite{Kolb_Ritter1990} and adopted the $\alpha,\, \beta,\, \gamma,\,\delta$ formalism as described in \cite{Tauris_vandenHeuvel2006}. We chose to set $\alpha,\, \gamma,\,\delta$ to zero and tested different values of the parameter $\beta$, that is, the fraction of mass lost from the vicinity of the accretor as a fast wind. This scenario, known as isotropic re-emission \citep{Tauris_vandenHeuvel2006,VandenHeuvel+2017}, is a common assumption in binary stable mass transfer modelling and has been proposed as the evolutionary channel also for other stripped star + Be star binaries \citep[e.g. HR6819,][]{Bodensteiner+2020}. However, the actual physics underlying binary mass transfer and angular momentum loss remains highly uncertain. Our models do not include tides, rotation, or its related effects (e.g. rotational mixing, rotation-limited accretion), allowing the accretion efficiency to be treated as a free parameter\footnote{The evolution of the stripped star, which is the primary focus of our models, is not expected to be significantly affected by rotation \citep{Gotberg+2018}. However, we note that this simplification introduces substantial uncertainties in the envelope structure of the Be star \citep[e.g.][]{Ramachandran+2023}.}.
 Under the assumption of circular orbits and for a constant mass transfer efficiency, the period evolution can be analytically derived by solving the equation of angular momentum balance \citep{Soberman+1997}.
In the limit of fully conservative mass transfer, that is, $\beta = 0.0$, the period as a function of the mass ratio is given by
\begin{equation}
\frac{P}{P_\mathrm{initial}} = \left(\frac{q_\mathrm{initial}}{q}\right)^{3} \left(\frac{1+q}{1+q_\mathrm{initial}}\right)^{6}\,.    
\end{equation}
Using the estimated current values of $P \approx 221\,$d and $q \lesssim 0.14$, this implies an initial period of less than $30$\,d, assuming an initial mass ratio of $q_\mathrm{initial} \leq 2.0$. 
For fully non-conservative mass transfer ($\beta = 1.0$), the period evolution is described by
\begin{equation}
\frac{P}{P_\mathrm{initial}} = \left(\frac{q_\mathrm{initial}}{q}\right)^{3} \left(\frac{1+q}{1+q_\mathrm{initial}}\right)^{-2}  \exp\left(3\left(q-q_\mathrm{initial}\right)\right)\,.    
\end{equation}
In this case, a lower initial period is required to match the observed values. 
The evolution of the period for both mass transfer regimes is illustrated in Fig.~\ref{fig:period_evolution}, along with four MESA models for intermediate values of $\beta$ and initial periods ranging from 5 to 30\,d. To avoid unstable mass transfer and common envelope evolution that result in short-period systems, the initial mass ratio must be comparable to or smaller than a critical value, on the order of $q_{\mathrm{crit}} \approx 4.0$ \citep{Temmink+2023}.

We first iterated through a grid of initial masses for the donor star from $4$ to $7\,\mathrm{M}_\odot$. We found that an initial mass of approximately $M_\mathrm{d, initial} = 5.5\,\mathrm{M}_\odot$ reproduces the estimated luminosity, temperature, and surface gravity of the stripped star. Fixing the initial mass of the primary star to $5.5\,\mathrm{M}_\odot$, we then computed a second set of models and chose values for $\beta$ and $M_{\mathrm{a,initial}}$ such that the final mass of the Be star matched our estimate of $M_\mathrm{a, Be} \approx 6-8\,\mathrm{M}_\odot$. Given that the current mass of the companion star is $M_\mathrm{a,Be} = M_\mathrm{a,initial} + (1-\beta)(M_\mathrm{d,initial} - M_\mathrm{d, stripped})$, this requires less conservative mass transfer for higher initial masses of the companion. The parameter values for all models are summarised in Table~\ref{tab:mesa_models} in the appendix.

\begin{figure}[t]
    \centering
    \includegraphics[width=\columnwidth]{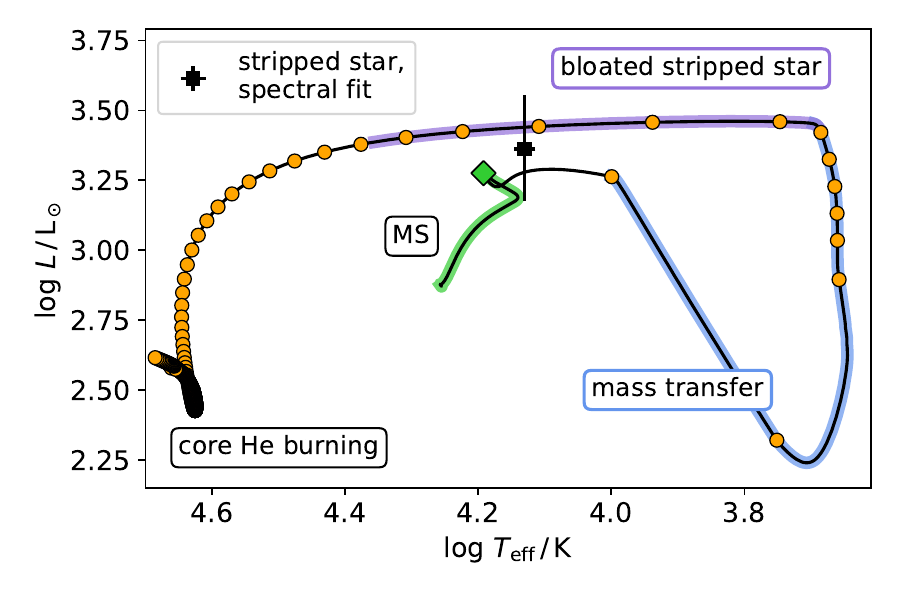}
    \caption{{Evolution of the stripped donor star of HIP~15429 in the HRD.} The model track (black line) was computed for a binary system with initial masses of $5.5\,\mathrm{M}_\odot$ and $4.95\,\mathrm{M}_\odot$ and an orbital period of 8 days. The black square marker indicates the present-day values of the stripped star, estimated from spectroscopy (effective temperature) and SED fitting (luminosity). Coloured sections mark critical evolutionary episodes. Orange markers are spaced in regular time intervals of $5\times10^4$ years starting at the time of first Roche-lobe overflow.}
    \label{fig:mesa_model_HRD}
\end{figure}

Figure~\ref{fig:mesa_model_HRD} illustrates the evolution of the stripped star for a model with initial donor mass of $5.5\,\mathrm{M}_\odot$ (model M3 in Table~\ref{tab:mesa_models}). The main-sequence evolution of the donor star closely follows the evolution of a single star. After core-hydrogen depletion, hydrogen shell burning begins, the core contracts, the outer layers expand, and the star moves to cooler effective temperatures in the HRD. The expansion continues until the star fills its Roche lobe, initiating a phase of stable Case B mass transfer to its companion. This mass transfer episode lasts approximately 0.37\,Myr. The donor star loses $>4\,\mathrm{M}_\odot$ of its hydrogen-rich envelope, reducing its mass to $0.99\,\mathrm{M}_\odot$ and leaving an envelope mass of approximately $0.14\,\mathrm{M}_\odot$. Here, envelope refers to the layers above the core-envelope boundary, which we define as the mass coordinate where the hydrogen abundance drops below $X_H = 0.1$, following \citet{Kruckow+2016}.

After mass transfer ends, the donor star contracts and evolves towards higher effective temperatures to become a compact helium star. During this transition phase, referred to as the bloated stripped star phase in the recent literature, the donor resembles a regular supergiant in the HRD but is actually much less massive. Following \cite{Dutta_Klencki2023} and defining the bloated stripped star phase as the time between the end of the Roche-lobe overflow and the time when the stripped donor becomes hotter than its ZAMS position plus 0.1 dex in $\log T_\mathrm{eff}$, the bloated stripped star phase of the donor lasts for $\approx 0.27$\,Myr; corresponding to less than 1\% of the main-sequence lifetime. The star continues to contract and heat up before settling as a core helium burning hot subdwarf star with $T_\mathrm{eff} \approx 44\,\mathrm{kK},\, \log g \approx 5.2\,$. We modelled its evolution up to the point of core-helium depletion, after which the stripped star will evolve into a white dwarf. This transition is expected to occur before the Be star completes its main-sequence evolution. However, for our models with initial mass ratios close to unity ($q_\mathrm{initial}\lesssim1.1$) and highly nonconservative mass transfer ($\beta\gtrsim0.7$), for example model M3d, the Be star completed its main-sequence evolution first. Although the subsequent evolution of the system has not been computed, it might then undergo a phase of inverse (and likely unstable) mass transfer. A common envelope phase could ultimately lead to the formation of a close double white dwarf binary or a merger event.

\begin{figure}[t]
    \centering
    \begin{minipage}[t]{\columnwidth}
        \centering
        \includegraphics[width=\columnwidth]{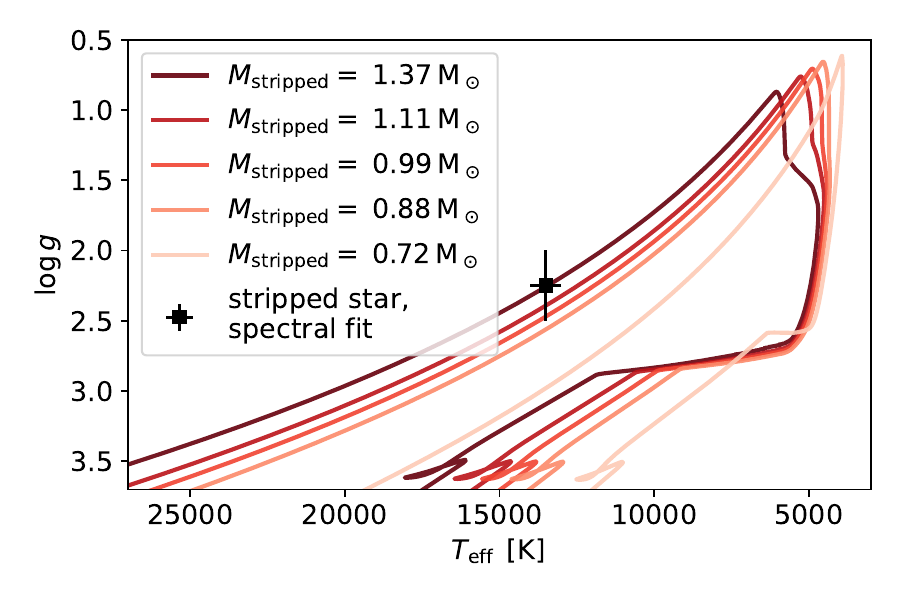}
    \end{minipage}
    \begin{minipage}[t]{\columnwidth}
        \centering
        \includegraphics[width=\columnwidth]{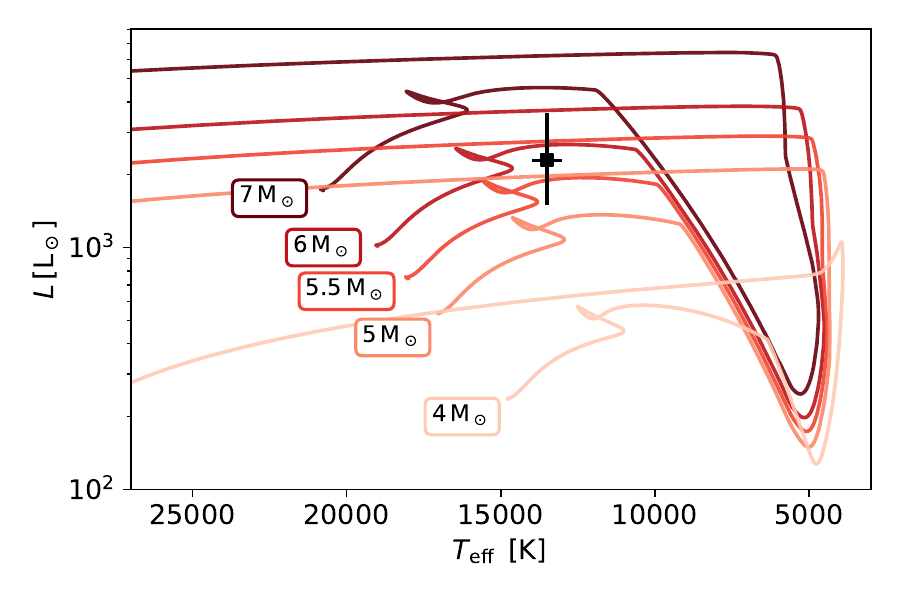}
    \end{minipage}
    
    \caption{{Evolutionary models for the stripped donor star with varying initial masses.} The panels show the model tracks on a Kiel diagram (\textit{top}) and in the HRD (\textit{bottom}). The model tracks are plotted for a range of initial masses between 4 and $7\,\mathrm{M}_\odot$ (models M1 - M5 in Table~\ref{tab:mesa_models}), which yields stripped star masses of $0.7 \leq M_\mathrm{d,stripped} \,/\, \mathrm{M}_\odot \leq 1.4$, as indicated in the legend. The present-day values of HIP~15429, inferred from spectral analysis and SED fitting, are plotted as black squares.}
    \label{fig:mesa_stripped_teff_logg}
\end{figure}

In Fig.~\ref{fig:mesa_stripped_teff_logg}, we show a Kiel diagram of stripped star models with masses between 0.71 and $1.39\,\mathrm{M}_\odot$ alongside the observationally determined parameters of the HIP~15429 stripped star. The models were produced from calculations with initial donor masses of 4.0, 5.0, 5.5, 6.0, and $7.0\,\mathrm{M}_\odot$. The stripped star mass depends on the core mass, which increases with initial mass and is sensitive to internal mixing processes (e.g. overshooting, semi-convection) during the main sequence. Based on surface gravity and effective temperature (Fig.~\ref{fig:mesa_stripped_teff_logg}, upper panel), no tight constraints on the stripped star mass are possible, and the measured values for the stripped star in HIP~15429 are consistent with a broad range of (initial) masses. We find, however, that the luminosity of the stripped star is best reproduced with a model of initial mass $5.5\,\mathrm{M}_\odot$ and stripped star mass $0.99\,\mathrm{M}_\odot$ (see Fig.~\ref{fig:mesa_stripped_teff_logg} in the lower panel). This value is higher than the inferred spectroscopic mass ${M_{\mathrm{B,spec}}} = 0.69^{+0.63}_{-0.31}\,{\mathrm{M}_\odot}$ but is consistent within the uncertainties. To match the spectroscopic mass more closely, models with an initial donor mass around $\sim4\,\mathrm{M}_\odot$ would be required. However, such models would underpredict the observed luminosity.

Assuming this initial mass for the stripped star, we varied initial masses and mass transfer efficiencies for the accreting companion star. Overall we find a good match between the estimated parameters of the binary stars in HIP~15429 and those of the MESA simulated binary during the bloated stripped star phase. When the simulated stripped star (again model M3; see Fig.~\ref{fig:mesa_model_HRD}) reaches an effective temperature of $\approx 13.5\,$kK, it has $\log g = 2.47$, radius $R = 9.5\,\mathrm{R}_\odot$ and $L = 2750\,\mathrm{L}_\odot$, all consistent within the uncertainties with the observed values. At this point the now more massive companion star has a temperature of $\approx 17.9\,$kK, it has $\log g = 3.9$, radius $R = 4.4\,\mathrm{R}_\odot$ and $L = 1790\,\mathrm{L}_\odot$. The mass ratio $q = 0.16$ and the period $P = 247\,$d are comparable to observational estimates as well. 

Without having conducted an extensive parameter search, we cannot put tight constrains on the properties of the progenitor system, and there may well be other regions of parameter space that can reproduce observed parameter values. However, the fact that we can construct a model in good agreement with observations also with a small grid speaks to the feasibility and plausibility of the evolutionary scenario.

\section{Discussion on orbital eccentricity}
\label{sec:discussion}

\begin{figure}[t]
    \centering
    \includegraphics[width=\columnwidth]{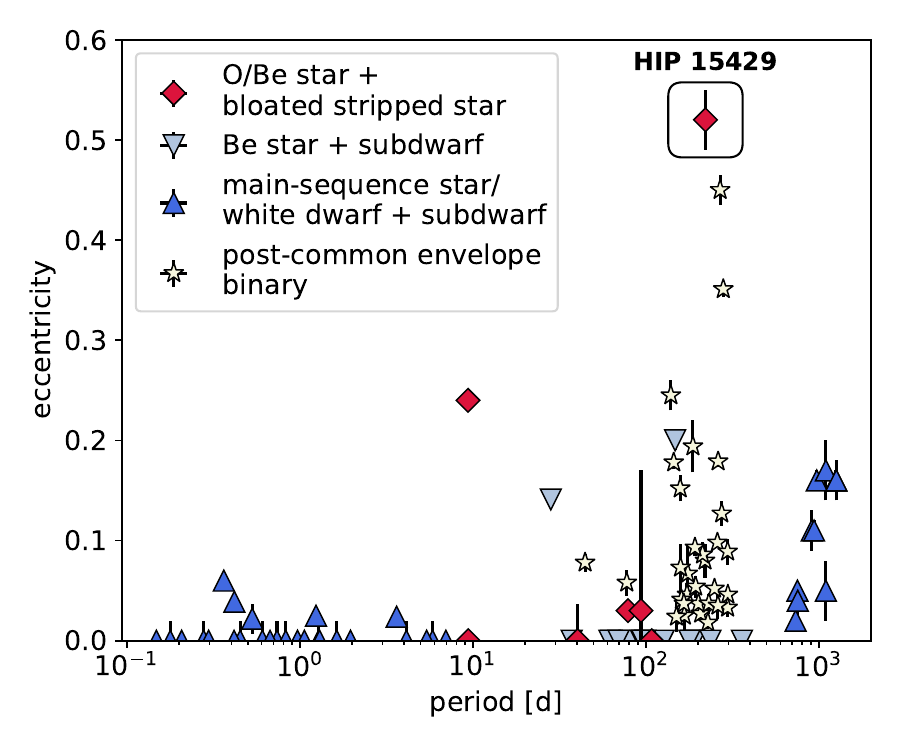}
    \caption{{Period-eccentricity diagram for known post-mass transfer binaries.} Recently discovered bloated stripped star binaries with O/Be companions \citep{Shenar+2020,ElBadry_Quataert2021,Villasenor+2023,Ramachandran+2024,Pauli+2022} are indicated with red diamond markers. These include the parameter values for HIP~15429 determined in this work (Müller-Horn et al. 2025). Be + subdwarf binaries with known orbital parameters are shown with light blue triangles \citep{Mourard+2015,Chojnowski+2018,Peters+2008,Peters+2013,Klement+2022,Klement+2024,Wang+2023}. We plot the short-period subdwarf binaries compiled by \cite{Edelmann+2005} and \cite{Kupfer+2015} and long-period ($>500\,$d orbits) subdwarf binaries discovered by \cite{Deca+2012,Barlow+2013,Vos+2012,Vos+2013,Vos+2017} with dark blue triangles. Likely post-common envelope binaries with white dwarf or main-sequence stars analysed by \cite{Yamaguchi+2024} are shown with yellow stars.} 
    \label{fig:period_eccentricity_comparison}
\end{figure}

Fig.~\ref{fig:period_eccentricity_comparison} places the orbital eccentricity of HIP~15429 in the context of physically related systems. We find it to be empirically exceptional, warranting a detailed discussion. There are a number of conceivable mechanisms to explain the high eccentricity, which we discuss in turn without identifying an obvious favourite.

We compare HIP~15429 to other bloated stripped stars with constrained periods and eccentricities and O/Be companions (red diamond markers), namely LB-1 \citep{Shenar+2020} and HR~6819 \citep{ElBadry_Quataert2021} in the lower mass regime and VFTS~291 \citep{Villasenor+2023}, 2dFS 2553, Sk -71$^\circ$35 \citep{Ramachandran+2024}
and AzV 476 \citep{Pauli+2022} in the intermediate mass range. HIP~15429 stands out as the bloated stripped star with the longest period and by a factor of more than two the highest eccentricity. The as yet only other identified bloated stripped star binary with substantial eccentricity is AzV 476 \citep[$e = 0.24\pm 0.002$;][]{Pauli+2022}.

As a likely progenitor of a subdwarf + Be binary system itself, we also compare HIP~15429 to other such binaries that have constrained orbital solutions \citep{Mourard+2015,Chojnowski+2018,Peters+2008,Peters+2013,Klement+2022,Klement+2024,Wang+2023}. For most of these systems, the orbits are consistent with being circular (see the light blue triangles in Fig.~\ref{fig:period_eccentricity_comparison}). The two notable exceptions are the binaries 59~Cyg and 60~Cyg with eccentricities of $0.141\pm0.008$ and $0.2\pm0.01$, respectively \citep{Peters+2013,Klement+2024}. For 59~Cyg an outer third component has been detected that may have tidally affected the inner Be + subdwarf binary. For 60~Cyg the origin of the nonzero eccentricity remains unknown \citep{Klement+2024}.

Figure~\ref{fig:period_eccentricity_comparison} further shows a selection of subdwarf binaries with lower-mass main-sequence or white dwarf companions. The orbital period distribution of this population is bimodal, with a short-period subset ($\lesssim10$\,d) compiled by \cite{Edelmann+2005} and \cite{Kupfer+2015}, and a group of long-period ($\gtrsim 500\,$d) subdwarf binaries \citep{Deca+2012,Barlow+2013,Vos+2013,Vos+2017}. The short-period binaries are thought to result from common envelope evolution, leading to near-circular orbits \citep{Paczynski1976}, while the long-period systems likely formed via stable mass transfer, with low but non-zero eccentricity. 
For subdwarfs formed via stable Roche-lobe overflow, final orbital periods depend on the onset of mass transfer and the mechanisms of angular momentum loss, producing a wide range of predicted periods \citep{Han+2002,Han+2003}. The intermediate period of HIP~15429 is therefore not unexpected, but its unusually high eccentricity (the highest by more than a factor of two) is.

We describe several mechanisms that have been proposed to explain eccentric orbits in post-interaction binary systems and discuss their applicability to HIP~15429.

\subsection{Phase-dependent Roche-lobe overflow}

One possible mechanism to excite eccentricity in mass-transferring binaries is phase-dependent mass loss during Roche-lobe overflow. If the binary is in a somewhat eccentric orbit to begin with, a varying mass loss rate that is greater during the periastron than the apastron can increase the orbital eccentricity \citep{Vos+2015}. The strength of eccentricity pumping will depend on the mass loss rate and the fraction of mass accreted by the companion. The mechanism has been invoked, for example, by \cite{Bonacic+2008} to explain the observed eccentricity of post-AGB binaries. Simulations of \cite{Vos+2015} show the potential to increase eccentricities to $e \approx 0.15$ at periods of close to a thousand days. However, it is unclear whether the mechanism is sufficient to excite eccentricities as high as that of HIP~15429 at shorter orbital periods. Furthermore, there would need to be a specific reason why HIP~15429 is more eccentric than other similar systems.

\subsection{Circumbinary disc evolution}
An alternative mechanism for eccentricity pumping is the temporary formation of a circumbinary disc (CBD) during the mass transfer phase. Interactions between binary and CBD, including gas and momentum exchange and gravitational torques, can significantly alter the binary orbital parameters \citep[e.g.][]{Goldreich_Tremaine1980}.

The CBDs have been observed around post-AGB binaries with periods of hundreds to thousands of days \citep{Kluska+2022}. Many post-AGB binaries exhibit substantial eccentricities despite the expected orbital circularisation during the AGB phase \citep{VanWinckel2018,Oomen+2018}, potentially arising from gravitational interactions with a CBD \citep{Dermine+2013}.

The outcome of disc-binary interactions depends on the binary parameters, in particular the mass ratio and eccentricity \citep[e.g.][]{Goldreich_Tremaine1980,Artymowicz+1991}. Hydrodynamical simulations indicate that disc-binary interactions drive the system towards an equilibrium eccentricity, determined by a balance between Lindblad and corotation resonances \citep{Valli+2024}. For binaries with mass ratios near unity, the equilibrium eccentricity is predicted to be roughly 0.5 but it may be lower for systems with unequal mass ratios \citep{Siwek+2023}. The timescale to reach the equilibrium state is estimated to be on the order of a few megayears.

For HIP~15429, a CBD may have formed temporarily from the mass lost through the donor's L2 Lagrange point \citep[or possibly also the accretor star in case of high mass transfer rates;][]{Lu+2023}. Dynamical interactions could have excited the orbital eccentricity, although the observed value of 0.52 exceeds the predicted equilibrium eccentricity for the low mass ratio of the system ($q \lesssim 0.1$), which is closer to 0.2 \citep{Siwek+2023}. Furthermore, the short duration of the mass transfer phase ($\lesssim 0.4\,$Myr) may have been insufficient to significantly increase eccentricity. As in the phase-dependent mass transfer scenario, it remains unclear why HIP~15429 exhibits higher eccentricity compared to other binaries with similar periods and CBDs.

\subsection{Inefficient tidal dissipation}
\label{sec:tidal_dissipation}
An alternative explanation for the current eccentricity of HIP~15429 is inefficient tidal dissipation, where rapid mass transfer left insufficient time for full orbital circularisation. 
Evidence from post-common envelope binaries, composed of main-sequence stars with white dwarf companions \citep[][see also Fig.~\ref{fig:period_eccentricity_comparison}]{Yamaguchi+2024}, and millisecond pulsars with massive white dwarf companions \citep[e.g.][]{Lorimer+2015}, indicates that systems with high mass loss rates often retain moderate to high eccentricities, suggesting incomplete tidal circularisation \citep{Glanz+2021}. If the binary was eccentric during mass transfer, the mass transfer rates and orbital evolution could differ from our explored regime (Sect.~\ref{sec:evolution}). Although eccentric mass transfer has not yet been implemented for normal stars in MESA, \cite{Rocha+2025} recently incorporated such physics for compact object binaries, demonstrating that eccentricity can persist through mass transfer. A more detailed treatment of eccentric mass transfer for systems like HIP~15429 remains a goal for future work.

\subsection{Kozai-Lidov cycles}
\label{sec:kozai-lidov}

The presence of a third star in a hierarchical triple system can significantly affect the evolution of a binary. Secular three-body dynamics, specifically Lidov-Kozai cycles, can induce periodic variations in the eccentricity and inclination of the inner and outer orbits through angular momentum exchange \citep{vonZeipel1910,Kozai1962, Lidov1962,Harrington1968}. While the semi-major axes remain constant due to the conserved orbital energy, these cycles depend on the mutual inclination between the orbits \citep{Toonen+2016}. Initially derived for triples with outer companions in circular orbits, the Kozai-Lidov effect was later extended to eccentric outer orbits, where more chaotic dynamics may arise \citep[see][for a review on the eccentric Kozai-Lidov effect]{Naoz2016}.

Given that approximately 50\% of B-type stars belong to triple- or higher-order systems \citep{Moe_DiStefano2017, Offner+2023}, we investigated whether a triple system could explain the orbital eccentricity of HIP~15429. We used the Python package \texttt{rebound} \citep{Rein_Liu2012} to simulate in a simple model the dynamical evolution of the B+Be binary, assuming a hierarchical triple configuration. We used the IAS15 integrator, which is well suited for high-centricity orbits and ensures energy conservation to machine precision \citep{Rein_Spiegel2015}. The simulations varied the mass (0.1–5$,\mathrm{M}_\odot$, log-uniformly spaced in 15 steps), period (10, 20, and 50 times the inner period), and inclination (45–90$^\circ$ in 5$^\circ$ steps) of the outer companion. The inner binary was initialised with parameters consistent with observations and circular orbits, assuming tidal circularisation during mass transfer. The outer orbit was also assumed to be circular. The simulations covered $10^5$ years, which approximately corresponds to the estimated time since the end of the mass transfer phase. 
The results indicate that a triple system could excite the observed eccentricity, albeit with a low likelihood. However, the limited parameter space explored cautions against strong conclusions.

\begin{figure}[t]
    \centering
    \includegraphics[width=\columnwidth]{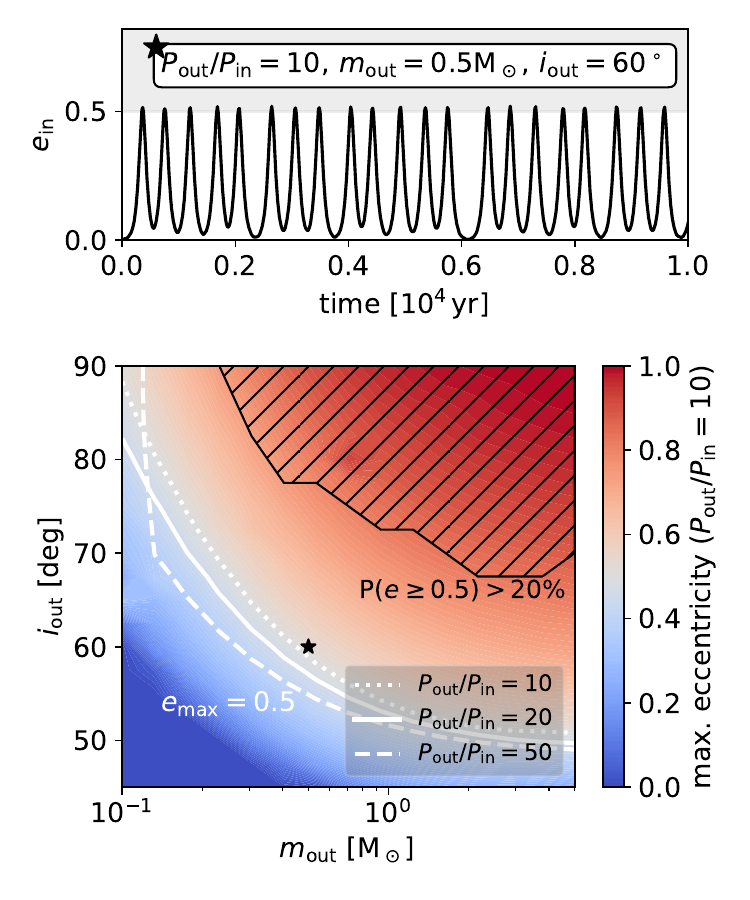}
    \caption{\textit{Upper panel}: Simulated eccentricity evolution of the binary HIP~15429 under the influence of a tertiary outer companion with a mass of 0.5\,$M_\odot$, inclination of $i_\mathrm{out} = 60^\circ$, and period ratio of ten during the first $10^4$\,yr since the end of the mass transfer phase. \textit{Lower panel:} Heat map visualising the maximum eccentricity excited in the inner binary as a function of outer companion mass and inclination. The heat map has been computed for a fixed outer to inner period ratio of ten. The white dotted line marks the region of parameter space where the inner binary's maximum eccentricity exceeds 0.5. An increase of the outer companion's period shifts this line to slightly lower inclinations, shown for $P_\mathrm{out}/P_\mathrm{in} = 20$ and 50 with solid and dashed white lines, respectively. The black hashed region indicates the parameter space for which the probability of observing $e>0.5$ for the inner binary at any given point exceeds 20\%.} 
    \label{fig:ecc_grid_triple}
\end{figure}

In Fig.~\ref{fig:ecc_grid_triple}, the colour map shows the maximum eccentricity achieved by the inner binary due to interaction with an outer companion, as a function of the companion's mass and inclination, with a fixed period ratio of 10 ($P_\mathrm{out} \approx 2200\,$d). The white dotted line marks where the inner binary's eccentricity exceeds 0.5. Increasing the period ratio to 20 or 50 shifts this threshold to lower inclinations, shown by solid and dashed white lines. A broad range of parameters can excite eccentricities above 0.5, with low-mass companions ($\approx 0.3\,M_\odot$) sufficient at high inclinations ($i \gtrsim 70^\circ$) and moderate masses ($m_\mathrm{out} \gtrsim 1\,M_\odot$) requiring inclinations of $i_\mathrm{out} \gtrsim 50^\circ$. The upper panel illustrates the eccentricity evolution for a specific configuration ($m_\mathrm{out} = 0.5\,M_\odot$, $i_\mathrm{out} = 60^\circ$, $P_\mathrm{out}/P_\mathrm{in} = 10$), where eccentricities greater than 0.5 are achieved but only briefly. Requiring a probability of at least 20\% for observing $e > 0.5$, estimated as the fraction of time spent at such high eccentricities, imposes stricter constraints. For $P_\mathrm{out} = 10 \times P_\mathrm{in}$, inclinations >65$^\circ$ are needed even at the highest masses tested (black-hatched region in Fig.~\ref{fig:ecc_grid_triple}), while for $P_\mathrm{out} = 50 \times P_\mathrm{in}$, inclinations >80$^\circ$ are required. Thus, if an outer companion induced the eccentricity, we are likely to observe the system near an eccentricity peak or in a less probable high-inclination configuration.
The period of eccentricity oscillations scales inversely with the mass of the outer companion and quadratically with its period \citep{Antognini2015}; ranging from $\mathcal{O}(10^2\,\mathrm{yr})$ to $\mathcal{O}(10^4\,\mathrm{yr})$ for the configurations tested.

To test the triple scenario observationally, interferometric observations with the Center for High Angular Resolution Astronomy (CHARA) Array could search for a distant outer companion. Using the MIRC-X beam combiner in the H-band and the longest baseline, CHARA can resolve companions with a minimum separation of 0.5\,mas \citep{Gallenne+2015}, which corresponds to about 0.87\,AU at the distance of HIP~15429 (smaller even than the inner binary's semi-major axis $a\approx0.94\,$AU). The inner binary, with an H-band magnitude difference of 1.2\,mag, meets the resolution criteria of CHARA, and outer companions with masses greater than $\sim 0.6\,\mathrm{M}_\odot$ should also be detectable\footnote{The mass estimate is based on calculations using MIST stellar evolution tracks and ATLAS9 Model Atmospheres \citep{Castelli_Kurucz2003} in combination with the Python packages \texttt{pystellibs} \citep{Fouesneau2022} and \texttt{pyphot} \citep{Fouesneau2024}.}. However, the combined H-band magnitude of the system of $\sim8.5\,$mag is at CHARA's sensitivity limit \citep{Anugu+2020}, making it uncertain whether a sufficiently high signal-to-noise ratio can be achieved.

\section{Summary and conclusions}
\label{sec:conclusion}

We have conducted a detailed spectroscopic and evolutionary analysis of HIP~15429. The system was initially proposed as a B star + dormant BH binary because of its high binary mass function and single-lined spectroscopic nature. Our results reveal that HIP~15429 is more likely a post-interaction binary with two non-degenerate components. Similar to recently discovered systems such as HR 6819, LB-1, and VFTS 291, HIP~15429 comprises a bloated stripped star and a Be star companion recently detached from a mass transfer episode.

Disentangling the obtained multi-epoch spectra revealed a second luminous component. It was previously undetected because of its shallow rotationally broadened absorption lines and low total flux contribution. The only notable feature in the composite spectra was a broad, almost stationary emission line in H$\alpha$, a typical feature of Be star decretion discs.

Spectroscopic analysis of the disentangled spectra yielded a temperature of $T_\mathrm{eff} = 13.5 \pm 0.5\,$kK and a surface gravity of $\log g = 2.25 \pm 0.25$ for the stripped donor star. Due to the degeneracy between rotational and macroturbulent broadening, we could only set an upper limit on the rotational velocity, with $v_\mathrm{rot} \sin i \leq 30$\,km/s.

The companion star's spectrum resembles that of classical Be stars, specifically of spectral type $\approx$ B3V, matching an inferred temperature of $T_\text{eff} = 17^{+2}_{-1}$\,kK for estimated surface gravity of a main-sequence star $\log g \approx 4.0 \pm 0.5$. Assuming a fixed macroturbulent velocity of 50\,km/s, we estimate the projected rotational velocity, $v_\mathrm{rot} \sin i$, to be $270 \pm 70$\,km/s.

Fitting the spectral energy distribution yielded luminosities and radii for both stars, allowing us to derive their spectroscopic masses. For the stripped star, we found a mass of $M_{\text{B,spec}} = 0.69^{+0.63}_{-0.31} \,\mathrm{M}_\odot$ and a radius of $9.0^{+2.1}_{-1.7}\,\mathrm{R}_\odot$. The mass of the Be star was estimated at $M_{\text{Be,spec}} = 10.5^{+21.5}_{-7.1} \,\mathrm{M}_\odot$, though this value is highly uncertain because of the poorly constrained surface gravity. Comparing the Be star's luminosity and temperature to evolutionary tracks suggests a mass of $M_{\mathrm{Be,evol}} = 6.5\pm0.9\,\mathrm{M}_\odot$, while estimates based on the binary mass function indicate a somewhat higher mass of at least $M_{\mathrm{Be,dyn}} \geq 7.0\,\mathrm{M}_\odot$. We deem the dynamic mass to be the more robust measurement and conclude that the discrepancy between evolutionary and dynamical mass is likely a consequence of recent accretion or self-absorption from a circumstellar disc, rendering the Be star less luminous in the optical than expected for its mass.

Our MESA modelling showed that the observed parameters are consistent with a shorter-period progenitor system that underwent Case B mass transfer. We propose that the B star is the remnant core of a star with an initial mass of about $5.5\,\mathrm{M}_\odot$ whose envelope was stripped during a recent phase of stable mass transfer. In this scenario, the accretion of mass and angular momentum from the donor star's stripped envelope onto the companion led to its rapid rotation and the observed Be star phenomenon.
The donor star is currently in a brief contraction phase towards the extreme horizontal branch and will eventually evolve into a core helium-burning subdwarf star, while the Be star continues core-hydrogen burning on the main sequence.

Although we have not measured quantitative abundances for the stripped star, we could make qualitative statements about its surface composition. Comparison of the disentangled spectra with model spectra of solar abundances suggests helium and nitrogen enrichment along with carbon and oxygen depletion, as expected for stars stripped of their outer layers.

The high orbital eccentricity ($e = 0.52\pm0.03$) is unusual for a post-mass transfer binary, as the tidal interactions typically circularise the orbit. We explored mechanisms such as phase-dependent Roche-lobe overflow, the temporary formation of a circumbinary disc, inefficient tidal dissipation, and interaction with a tertiary outer companion through the Kozai-Lidov effect to explain the eccentric orbit. We conclude that distance- and magnitude-based detection limits imposed by interferometric telescopes such as the CHARA Array make observational tests of the triple scenario challenging.

Further modelling is required to accurately determine the masses and evolutionary states of the components in this system. Although our MESA calculations are consistent with the inferred properties of HIP~15429, there are still discrepancies between the dynamically determined and evolutionary masses, particularly for the companion Be star. From an observational perspective, future astrometric data from Gaia DR4 could help refine the system parameters, including the currently unconstrained orbital inclination.

\section*{Data availability}
The BSTAR2006 spectral models for solar metallicity and abundances are available for download from the \href{http://tlusty.oca.eu/Tlusty2002/tlusty-frames-BS06.html}{\textsc{Tlusty} webpage}. 
The newly calculated stripped-star PoWR model spectra will be provided as supplementary material on \href{https://doi.org/10.5281/zenodo.15176294}{Zenodo}. 
The \href{https://docs.mesastar.org/en/latest/index.html}{MESA} code is publicly accessible, and we will share the input files required to reproduce our MESA simulations on \href{https://doi.org/10.5281/zenodo.15176294}{Zenodo}. All spectroscopic data will be shared upon reasonable request to the authors.

\begin{acknowledgements}
The authors sincerely thank the anonymous referee for their thorough review and thoughtful and constructive comments. The authors thank S. Janssens, H. Van Winckel and A. Jorissen for their contributions of HERMES optical spectra, which were a crucial part in the analysis. J. M\"uller-Horn and H. W. Rix acknowledge support from the European Research Council for the ERC Advanced Grant [101054731]. This research was supported in part by NSF grant AST-2307232. This project has received funding from the European Research Council (ERC) under the European Union's Horizon 2020 research and innovation programme (grant agreement 101164755/METAL) and was supported by the Israel Science Foundation (ISF) under grant number 2434/24.
This work has made use of data from the European Space Agency (ESA) mission Gaia (https://www.cosmos.esa.int/gaia), processed by the Gaia Data Processing and Analysis Consortium (DPAC, https://www.cosmos.esa.int/web/gaia/dpac/consortium). Funding for the DPAC has been provided by national institutions, in particular the institutions participating in the Gaia Multilateral Agreement. The analysis is based on observations made with the Mercator Telescope, operated on the island of La Palma by the Flemish Community, at the Spanish Observatorio del Roque de los Muchachos of the Instituto de Astrofísica de Canarias. In particular, several of the spectra used were obtained with the HERMES spectrograph, which is supported by the Research Foundation - Flanders (FWO), Belgium, the Research Council of KU Leuven, Belgium, the Fonds National de la Recherche Scientifique (F.R.S.-FNRS), Belgium, the Royal Observatory of Belgium, the Observatoire de Genève, Switzerland and the Thüringer Landessternwarte Tautenburg, Germany.
This publication makes use of VOSA, developed under the Spanish Virtual Observatory (https://svo.cab.inta-csic.es) project funded by MCIN/AEI/10.13039/501100011033/ through grant PID2020-112949GB-I00.
VOSA has been partially updated by using funding from the European Union's Horizon 2020 Research and Innovation Programme, under Grant Agreement nº 776403 (EXOPLANETS-A). This work used the following software packages: \texttt{astropy } \citep{astropy:2013, astropy:2018, astropy:2022}, \texttt{numpy} \citep{numpy}, \texttt{python} \citep{python}, and \texttt{scipy
} \citep{2020SciPy-NMeth, scipy_10909890}. Software citation information has been aggregated using \texttt{\href{https://www.tomwagg.com/software-citation-station/}{The Software Citation Station}} \citep{software-citation-station-paper, software-citation-station-zenodo}.

\end{acknowledgements}

\bibliographystyle{aa}
\bibliography{references.bib}

\begin{thebibliography}{194}
\expandafter\ifx\csname natexlab\endcsname\relax\def\natexlab#1{#1}\fi

\bibitem[{{Aerts} {et~al.}(2014){Aerts}, {Sim{\'o}n-D{\'\i}az}, {Groot}, \& {Degroote}}]{Aerts+2014}
{Aerts}, C., {Sim{\'o}n-D{\'\i}az}, S., {Groot}, P.~J., \& {Degroote}, P. 2014, \aap, 569, A118

\bibitem[{{Angulo} {et~al.}(1999){Angulo}, {Arnould}, {Rayet}, {Descouvemont}, {Baye}, {Leclercq-Willain}, {Coc}, {Barhoumi}, {Aguer}, {Rolfs}, {Kunz}, {Hammer}, {Mayer}, {Paradellis}, {Kossionides}, {Chronidou}, {Spyrou}, {degl'Innocenti}, {Fiorentini}, {Ricci}, {Zavatarelli}, {Providencia}, {Wolters}, {Soares}, {Grama}, {Rahighi}, {Shotter}, \& {Lamehi Rachti}}]{Angulo1999}
{Angulo}, C., {Arnould}, M., {Rayet}, M., {et~al.} 1999, \nphysa, 656, 3

\bibitem[{{Antognini}(2015)}]{Antognini2015}
{Antognini}, J.~M.~O. 2015, \mnras, 452, 3610

\bibitem[{{Anugu} {et~al.}(2020){Anugu}, {Le Bouquin}, {Monnier}, {Kraus}, {Setterholm}, {Labdon}, {Davies}, {Lanthermann}, {Gardner}, {Ennis}, {Johnson}, {Ten Brummelaar}, {Schaefer}, \& {Sturmann}}]{Anugu+2020}
{Anugu}, N., {Le Bouquin}, J.-B., {Monnier}, J.~D., {et~al.} 2020, \aj, 160, 158

\bibitem[{{Artymowicz} {et~al.}(1991){Artymowicz}, {Clarke}, {Lubow}, \& {Pringle}}]{Artymowicz+1991}
{Artymowicz}, P., {Clarke}, C.~J., {Lubow}, S.~H., \& {Pringle}, J.~E. 1991, \apjl, 370, L35

\bibitem[{{Asplund} {et~al.}(2021){Asplund}, {Amarsi}, \& {Grevesse}}]{Asplund+2021}
{Asplund}, M., {Amarsi}, A.~M., \& {Grevesse}, N. 2021, \aap, 653, A141

\bibitem[{{Astropy Collaboration} {et~al.}(2022){Astropy Collaboration}, {Price-Whelan}, {Lim}, {Earl}, {Starkman}, {Bradley}, {Shupe}, {Patil}, {Corrales}, {Brasseur}, {N{"o}the}, {Donath}, {Tollerud}, {Morris}, {Ginsburg}, {Vaher}, {Weaver}, {Tocknell}, {Jamieson}, {van Kerkwijk}, {Robitaille}, {Merry}, {Bachetti}, {G{"u}nther}, {Aldcroft}, {Alvarado-Montes}, {Archibald}, {B{'o}di}, {Bapat}, {Barentsen}, {Baz{'a}n}, {Biswas}, {Boquien}, {Burke}, {Cara}, {Cara}, {Conroy}, {Conseil}, {Craig}, {Cross}, {Cruz}, {D'Eugenio}, {Dencheva}, {Devillepoix}, {Dietrich}, {Eigenbrot}, {Erben}, {Ferreira}, {Foreman-Mackey}, {Fox}, {Freij}, {Garg}, {Geda}, {Glattly}, {Gondhalekar}, {Gordon}, {Grant}, {Greenfield}, {Groener}, {Guest}, {Gurovich}, {Handberg}, {Hart}, {Hatfield-Dodds}, {Homeier}, {Hosseinzadeh}, {Jenness}, {Jones}, {Joseph}, {Kalmbach}, {Karamehmetoglu}, {Ka{l}uszy{'n}ski}, {Kelley}, {Kern}, {Kerzendorf}, {Koch}, {Kulumani}, {Lee}, {Ly}, {Ma}, {MacBride}, {Maljaars}, {Muna}, {Murphy}, {Norman}, {O'Steen},
  {Oman}, {Pacifici}, {Pascual}, {Pascual-Granado}, {Patil}, {Perren}, {Pickering}, {Rastogi}, {Roulston}, {Ryan}, {Rykoff}, {Sabater}, {Sakurikar}, {Salgado}, {Sanghi}, {Saunders}, {Savchenko}, {Schwardt}, {Seifert-Eckert}, {Shih}, {Jain}, {Shukla}, {Sick}, {Simpson}, {Singanamalla}, {Singer}, {Singhal}, {Sinha}, {Sip{H{o}}cz}, {Spitler}, {Stansby}, {Streicher}, {{{S}}umak}, {Swinbank}, {Taranu}, {Tewary}, {Tremblay}, {Val-Borro}, {Van Kooten}, {Vasovi{'c}}, {Verma}, {de Miranda Cardoso}, {Williams}, {Wilson}, {Winkel}, {Wood-Vasey}, {Xue}, {Yoachim}, {Zhang}, {Zonca}, \& {Astropy Project Contributors}}]{astropy:2022}
{Astropy Collaboration}, {Price-Whelan}, A.~M., {Lim}, P.~L., {et~al.} 2022, \apj, 935, 167

\bibitem[{{Astropy Collaboration} {et~al.}(2018){Astropy Collaboration}, {Price-Whelan}, {Sip{\H{o}}cz}, {G{\"u}nther}, {Lim}, {Crawford}, {Conseil}, {Shupe}, {Craig}, {Dencheva}, {Ginsburg}, {Vand erPlas}, {Bradley}, {P{\'e}rez-Su{\'a}rez}, {de Val-Borro}, {Aldcroft}, {Cruz}, {Robitaille}, {Tollerud}, {Ardelean}, {Babej}, {Bach}, {Bachetti}, {Bakanov}, {Bamford}, {Barentsen}, {Barmby}, {Baumbach}, {Berry}, {Biscani}, {Boquien}, {Bostroem}, {Bouma}, {Brammer}, {Bray}, {Breytenbach}, {Buddelmeijer}, {Burke}, {Calderone}, {Cano Rodr{\'\i}guez}, {Cara}, {Cardoso}, {Cheedella}, {Copin}, {Corrales}, {Crichton}, {D'Avella}, {Deil}, {Depagne}, {Dietrich}, {Donath}, {Droettboom}, {Earl}, {Erben}, {Fabbro}, {Ferreira}, {Finethy}, {Fox}, {Garrison}, {Gibbons}, {Goldstein}, {Gommers}, {Greco}, {Greenfield}, {Groener}, {Grollier}, {Hagen}, {Hirst}, {Homeier}, {Horton}, {Hosseinzadeh}, {Hu}, {Hunkeler}, {Ivezi{\'c}}, {Jain}, {Jenness}, {Kanarek}, {Kendrew}, {Kern}, {Kerzendorf}, {Khvalko}, {King}, {Kirkby}, {Kulkarni},
  {Kumar}, {Lee}, {Lenz}, {Littlefair}, {Ma}, {Macleod}, {Mastropietro}, {McCully}, {Montagnac}, {Morris}, {Mueller}, {Mumford}, {Muna}, {Murphy}, {Nelson}, {Nguyen}, {Ninan}, {N{\"o}the}, {Ogaz}, {Oh}, {Parejko}, {Parley}, {Pascual}, {Patil}, {Patil}, {Plunkett}, {Prochaska}, {Rastogi}, {Reddy Janga}, {Sabater}, {Sakurikar}, {Seifert}, {Sherbert}, {Sherwood-Taylor}, {Shih}, {Sick}, {Silbiger}, {Singanamalla}, {Singer}, {Sladen}, {Sooley}, {Sornarajah}, {Streicher}, {Teuben}, {Thomas}, {Tremblay}, {Turner}, {Terr{\'o}n}, {van Kerkwijk}, {de la Vega}, {Watkins}, {Weaver}, {Whitmore}, {Woillez}, {Zabalza}, \& {Astropy Contributors}}]{astropy:2018}
{Astropy Collaboration}, {Price-Whelan}, A.~M., {Sip{\H{o}}cz}, B.~M., {et~al.} 2018, \aj, 156, 123

\bibitem[{{Astropy Collaboration} {et~al.}(2013){Astropy Collaboration}, {Robitaille}, {Tollerud}, {Greenfield}, {Droettboom}, {Bray}, {Aldcroft}, {Davis}, {Ginsburg}, {Price-Whelan}, {Kerzendorf}, {Conley}, {Crighton}, {Barbary}, {Muna}, {Ferguson}, {Grollier}, {Parikh}, {Nair}, {Unther}, {Deil}, {Woillez}, {Conseil}, {Kramer}, {Turner}, {Singer}, {Fox}, {Weaver}, {Zabalza}, {Edwards}, {Azalee Bostroem}, {Burke}, {Casey}, {Crawford}, {Dencheva}, {Ely}, {Jenness}, {Labrie}, {Lim}, {Pierfederici}, {Pontzen}, {Ptak}, {Refsdal}, {Servillat}, \& {Streicher}}]{astropy:2013}
{Astropy Collaboration}, {Robitaille}, T.~P., {Tollerud}, E.~J., {et~al.} 2013, \aap, 558, A33

\bibitem[{{Baklanov} {et~al.}(2005){Baklanov}, {Blinnikov}, \& {Pavlyuk}}]{Baklanov2005}
{Baklanov}, P.~V., {Blinnikov}, S.~I., \& {Pavlyuk}, N.~N. 2005, Astronomy Letters, 31, 429

\bibitem[{{Barlow} {et~al.}(2013){Barlow}, {Liss}, {Wade}, \& {Green}}]{Barlow+2013}
{Barlow}, B.~N., {Liss}, S.~E., {Wade}, R.~A., \& {Green}, E.~M. 2013, \apj, 771, 23

\bibitem[{{Bayo} {et~al.}(2008){Bayo}, {Rodrigo}, {Barrado Y Navascu{\'e}s}, {Solano}, {Guti{\'e}rrez}, {Morales-Calder{\'o}n}, \& {Allard}}]{Bayo+2008}
{Bayo}, A., {Rodrigo}, C., {Barrado Y Navascu{\'e}s}, D., {et~al.} 2008, \aap, 492, 277

\bibitem[{{Blinnikov} \& {Sorokina}(2004)}]{Blinnikov2004}
{Blinnikov}, S. \& {Sorokina}, E. 2004, \apss, 290, 13

\bibitem[{{Blinnikov} {et~al.}(2006){Blinnikov}, {R{\"o}pke}, {Sorokina}, {Gieseler}, {Reinecke}, {Travaglio}, {Hillebrandt}, \& {Stritzinger}}]{Blinnikov2006}
{Blinnikov}, S.~I., {R{\"o}pke}, F.~K., {Sorokina}, E.~I., {et~al.} 2006, \aap, 453, 229

\bibitem[{{Blouin} {et~al.}(2020){Blouin}, {Shaffer}, {Saumon}, \& {Starrett}}]{Blouin2020}
{Blouin}, S., {Shaffer}, N.~R., {Saumon}, D., \& {Starrett}, C.~E. 2020, \apj, 899, 46

\bibitem[{{Bodensteiner} {et~al.}(2020{\natexlab{a}}){Bodensteiner}, {Shenar}, {Mahy}, {Fabry}, {Marchant}, {Abdul-Masih}, {Banyard}, {Bowman}, {Dsilva}, {Frost}, {Hawcroft}, {Reggiani}, \& {Sana}}]{Bodensteiner+2020}
{Bodensteiner}, J., {Shenar}, T., {Mahy}, L., {et~al.} 2020{\natexlab{a}}, \aap, 641, A43

\bibitem[{{Bodensteiner} {et~al.}(2020{\natexlab{b}}){Bodensteiner}, {Shenar}, \& {Sana}}]{Bodensteiner+2020b}
{Bodensteiner}, J., {Shenar}, T., \& {Sana}, H. 2020{\natexlab{b}}, \aap, 641, A42

\bibitem[{{B{\"o}hm-Vitense}(1958)}]{Bohm-Vitense1958}
{B{\"o}hm-Vitense}, E. 1958, \zap, 46, 108

\bibitem[{{Bona{\v{c}}i{\'c} Marinovi{\'c}} {et~al.}(2008){Bona{\v{c}}i{\'c} Marinovi{\'c}}, {Glebbeek}, \& {Pols}}]{Bonacic+2008}
{Bona{\v{c}}i{\'c} Marinovi{\'c}}, A.~A., {Glebbeek}, E., \& {Pols}, O.~R. 2008, \aap, 480, 797

\bibitem[{{Bressan} {et~al.}(2012){Bressan}, {Marigo}, {Girardi}, {Salasnich}, {Dal Cero}, {Rubele}, \& {Nanni}}]{Bressan+2012}
{Bressan}, A., {Marigo}, P., {Girardi}, L., {et~al.} 2012, \mnras, 427, 127

\bibitem[{{Buchhave} {et~al.}(2010){Buchhave}, {Bakos}, {Hartman}, {Torres}, {Kov{\'a}cs}, {Latham}, {Noyes}, {Esquerdo}, {Everett}, {Howard}, {Marcy}, {Fischer}, {Johnson}, {Andersen}, {F{\H{u}}r{\'e}sz}, {Perumpilly}, {Sasselov}, {Stefanik}, {B{\'e}ky}, {L{\'a}z{\'a}r}, {Papp}, \& {S{\'a}ri}}]{Buchhave+2010}
{Buchhave}, L.~A., {Bakos}, G.~{\'A}., {Hartman}, J.~D., {et~al.} 2010, \apj, 720, 1118

\bibitem[{{Buchner}(2016)}]{Buchner2016}
{Buchner}, J. 2016, Statistics and Computing, 26, 383

\bibitem[{{Buchner}(2019)}]{Buchner2019}
{Buchner}, J. 2019, \pasp, 131, 108005

\bibitem[{{Buchner}(2021)}]{Buchner2021}
{Buchner}, J. 2021, The Journal of Open Source Software, 6, 3001

\bibitem[{{Cassisi} {et~al.}(2007){Cassisi}, {Potekhin}, {Pietrinferni}, {Catelan}, \& {Salaris}}]{Cassisi2007}
{Cassisi}, S., {Potekhin}, A.~Y., {Pietrinferni}, A., {Catelan}, M., \& {Salaris}, M. 2007, \apj, 661, 1094

\bibitem[{{Castelli} \& {Kurucz}(2003)}]{Castelli_Kurucz2003}
{Castelli}, F. \& {Kurucz}, R.~L. 2003, in Modelling of Stellar Atmospheres, ed. N.~{Piskunov}, W.~W. {Weiss}, \& D.~F. {Gray}, Vol. 210, A20

\bibitem[{{Choi} {et~al.}(2016){Choi}, {Dotter}, {Conroy}, {Cantiello}, {Paxton}, \& {Johnson}}]{Choi+2016}
{Choi}, J., {Dotter}, A., {Conroy}, C., {et~al.} 2016, \apj, 823, 102

\bibitem[{{Chojnowski} {et~al.}(2018){Chojnowski}, {Labadie-Bartz}, {Rivinius}, {Gies}, {Panoglou}, {Borges Fernandes}, {Wisniewski}, {Whelan}, {Mennickent}, {McMillan}, {Dembicky}, {Gray}, {Rudyk}, {Stringfellow}, {Lester}, {Hasselquist}, {Zharikov}, {Levenhagen}, {Souza}, {Leister}, {Stassun}, {Siverd}, \& {Majewski}}]{Chojnowski+2018}
{Chojnowski}, S.~D., {Labadie-Bartz}, J., {Rivinius}, T., {et~al.} 2018, \apj, 865, 76

\bibitem[{{Christensen-Dalsgaard}(2008)}]{ChristensenDalsgaard2008}
{Christensen-Dalsgaard}, J. 2008, \apss, 316, 113

\bibitem[{{Chugunov} {et~al.}(2007){Chugunov}, {Dewitt}, \& {Yakovlev}}]{Chugunov2007}
{Chugunov}, A.~I., {Dewitt}, H.~E., \& {Yakovlev}, D.~G. 2007, \prd, 76, 025028

\bibitem[{{Crowther} {et~al.}(2006){Crowther}, {Lennon}, \& {Walborn}}]{Crowther+2006}
{Crowther}, P.~A., {Lennon}, D.~J., \& {Walborn}, N.~R. 2006, \aap, 446, 279

\bibitem[{{Cutri} {et~al.}(2003){Cutri}, {Skrutskie}, {van Dyk}, {Beichman}, {Carpenter}, {Chester}, {Cambresy}, {Evans}, {Fowler}, {Gizis}, {Howard}, {Huchra}, {Jarrett}, {Kopan}, {Kirkpatrick}, {Light}, {Marsh}, {McCallon}, {Schneider}, {Stiening}, {Sykes}, {Weinberg}, {Wheaton}, {Wheelock}, \& {Zacarias}}]{Cutri+2003}
{Cutri}, R.~M., {Skrutskie}, M.~F., {van Dyk}, S., {et~al.} 2003, {VizieR Online Data Catalog: 2MASS All-Sky Catalog of Point Sources (Cutri+ 2003)}, VizieR On-line Data Catalog: II/246. Originally published in: 2003yCat.2246....0C

\bibitem[{{Cyburt} {et~al.}(2010){Cyburt}, {Amthor}, {Ferguson}, {Meisel}, {Smith}, {Warren}, {Heger}, {Hoffman}, {Rauscher}, {Sakharuk}, {Schatz}, {Thielemann}, \& {Wiescher}}]{Cyburt2010}
{Cyburt}, R.~H., {Amthor}, A.~M., {Ferguson}, R., {et~al.} 2010, \apjs, 189, 240

\bibitem[{{de Kool} \& {Ritter}(1993)}]{Kolb_Ritter1993}
{de Kool}, M. \& {Ritter}, H. 1993, \aap, 267, 397

\bibitem[{{de Mink} {et~al.}(2013){de Mink}, {Langer}, {Izzard}, {Sana}, \& {de Koter}}]{deMink+2013}
{de Mink}, S.~E., {Langer}, N., {Izzard}, R.~G., {Sana}, H., \& {de Koter}, A. 2013, \apj, 764, 166

\bibitem[{{Deca} {et~al.}(2012){Deca}, {Marsh}, {{\O}stensen}, {Morales-Rueda}, {Copperwheat}, {Wade}, {Stark}, {Maxted}, {Nelemans}, \& {Heber}}]{Deca+2012}
{Deca}, J., {Marsh}, T.~R., {{\O}stensen}, R.~H., {et~al.} 2012, \mnras, 421, 2798

\bibitem[{{Dermine} {et~al.}(2013){Dermine}, {Izzard}, {Jorissen}, \& {Van Winckel}}]{Dermine+2013}
{Dermine}, T., {Izzard}, R.~G., {Jorissen}, A., \& {Van Winckel}, H. 2013, \aap, 551, A50

\bibitem[{{Dotter}(2016)}]{Dotter2016}
{Dotter}, A. 2016, \apjs, 222, 8

\bibitem[{{Drout} {et~al.}(2023){Drout}, {G{\"o}tberg}, {Ludwig}, {Groh}, {de Mink}, {O'Grady}, \& {Smith}}]{Drout+2023}
{Drout}, M.~R., {G{\"o}tberg}, Y., {Ludwig}, B.~A., {et~al.} 2023, Science, 382, 1287

\bibitem[{{Dufton} {et~al.}(2013){Dufton}, {Langer}, {Dunstall}, {Evans}, {Brott}, {de Mink}, {Howarth}, {Kennedy}, {McEvoy}, {Potter}, {Ram{\'\i}rez-Agudelo}, {Sana}, {Sim{\'o}n-D{\'\i}az}, {Taylor}, \& {Vink}}]{Dufton+2013}
{Dufton}, P.~L., {Langer}, N., {Dunstall}, P.~R., {et~al.} 2013, \aap, 550, A109

\bibitem[{{Dutta} \& {Klencki}(2024)}]{Dutta_Klencki2023}
{Dutta}, D. \& {Klencki}, J. 2024, \aap, 687, A215

\bibitem[{{Edelmann} {et~al.}(2005){Edelmann}, {Heber}, {Altmann}, {Karl}, \& {Lisker}}]{Edelmann+2005}
{Edelmann}, H., {Heber}, U., {Altmann}, M., {Karl}, C., \& {Lisker}, T. 2005, \aap, 442, 1023

\bibitem[{{Eggleton}(1983)}]{Eggleton1983}
{Eggleton}, P.~P. 1983, \apj, 268, 368

\bibitem[{{El-Badry} \& {Quataert}(2020)}]{ElBadry_Quataert2020}
{El-Badry}, K. \& {Quataert}, E. 2020, \mnras, 493, L22

\bibitem[{{El-Badry} \& {Quataert}(2021)}]{ElBadry_Quataert2021}
{El-Badry}, K. \& {Quataert}, E. 2021, \mnras, 502, 3436

\bibitem[{{El-Badry} {et~al.}(2021){El-Badry}, {Rix}, \& {Heintz}}]{ElBadry+2021}
{El-Badry}, K., {Rix}, H.-W., \& {Heintz}, T.~M. 2021, \mnras, 506, 2269

\bibitem[{{Ferguson} {et~al.}(2005){Ferguson}, {Alexander}, {Allard}, {Barman}, {Bodnarik}, {Hauschildt}, {Heffner-Wong}, \& {Tamanai}}]{Ferguson2005}
{Ferguson}, J.~W., {Alexander}, D.~R., {Allard}, F., {et~al.} 2005, \apj, 623, 585

\bibitem[{F\H{u}r\'esz(2008)}]{Furesz2008}
F\H{u}r\'esz, G. 2008, PhD thesis, Univ. Szeged, Hungary

\bibitem[{{Fitzpatrick}(1999)}]{Fitzpatrick_Edward1999}
{Fitzpatrick}, E.~L. 1999, \pasp, 111, 63

\bibitem[{Fouesneau(2022)}]{Fouesneau2022}
Fouesneau, M. 2022, pystellibs

\bibitem[{Fouesneau(2024)}]{Fouesneau2024}
Fouesneau, M. 2024, pyphot

\bibitem[{{Fuller} {et~al.}(1985){Fuller}, {Fowler}, \& {Newman}}]{Fuller1985}
{Fuller}, G.~M., {Fowler}, W.~A., \& {Newman}, M.~J. 1985, \apj, 293, 1

\bibitem[{{Gaia Collaboration} {et~al.}(2023{\natexlab{a}}){Gaia Collaboration}, {Arenou}, {Babusiaux}, {Barstow}, {Faigler}, {Jorissen}, {Kervella}, {Mazeh}, {Mowlavi}, {Panuzzo}, {Sahlmann}, {Shahaf}, {Sozzetti}, {Bauchet}, {Damerdji}, {Gavras}, {Giacobbe}, {Gosset}, {Halbwachs}, {Holl}, {Lattanzi}, {Leclerc}, {Morel}, {Pourbaix}, {Re Fiorentin}, {Sadowski}, {S{\'e}gransan}, {Siopis}, {Teyssier}, {Zwitter}, {Planquart}, {Brown}, {Vallenari}, {Prusti}, {de Bruijne}, {Biermann}, {Creevey}, {Ducourant}, {Evans}, {Eyer}, {Guerra}, {Hutton}, {Jordi}, {Klioner}, {Lammers}, {Lindegren}, {Luri}, {Mignard}, {Panem}, {Randich}, {Sartoretti}, {Soubiran}, {Tanga}, {Walton}, {Bailer-Jones}, {Bastian}, {Drimmel}, \& {Jansen}}]{Gaia+2023}
{Gaia Collaboration}, {Arenou}, F., {Babusiaux}, C., {et~al.} 2023{\natexlab{a}}, \aap, 674, A34

\bibitem[{{Gaia Collaboration} {et~al.}(2023{\natexlab{b}}){Gaia Collaboration}, {Montegriffo}, {Bellazzini}, {De Angeli}, {Andrae}, {Barstow}, {Bossini}, {Bragaglia}, {Burgess}, {Cacciari}, {Carrasco}, {Chornay}, {Delchambre}, {Evans}, \& {Fouesneau}}]{Gaia_synthphot}
{Gaia Collaboration}, {Montegriffo}, P., {Bellazzini}, M., {et~al.} 2023{\natexlab{b}}, \aap, 674, A33

\bibitem[{{Gallenne} {et~al.}(2015){Gallenne}, {M{\'e}rand}, {Kervella}, {Monnier}, {Schaefer}, {Baron}, {Breitfelder}, {Le Bouquin}, {Roettenbacher}, {Gieren}, {Pietrzy{\'n}ski}, {McAlister}, {ten Brummelaar}, {Sturmann}, {Sturmann}, {Turner}, {Ridgway}, \& {Kraus}}]{Gallenne+2015}
{Gallenne}, A., {M{\'e}rand}, A., {Kervella}, P., {et~al.} 2015, \aap, 579, A68

\bibitem[{{Gies} {et~al.}(1998){Gies}, {Bagnuolo}, {Ferrara}, {Kaye}, {Thaller}, {Penny}, \& {Peters}}]{Gies+1998}
{Gies}, D.~R., {Bagnuolo}, William~G., J., {Ferrara}, E.~C., {et~al.} 1998, \apj, 493, 440

\bibitem[{{Glanz} \& {Perets}(2021)}]{Glanz+2021}
{Glanz}, H. \& {Perets}, H.~B. 2021, \mnras, 507, 2659

\bibitem[{{Glebbeek} {et~al.}(2009){Glebbeek}, {Gaburov}, {de Mink}, {Pols}, \& {Portegies Zwart}}]{Glebbeek+2009}
{Glebbeek}, E., {Gaburov}, E., {de Mink}, S.~E., {Pols}, O.~R., \& {Portegies Zwart}, S.~F. 2009, \aap, 497, 255

\bibitem[{{Goldreich} \& {Tremaine}(1980)}]{Goldreich_Tremaine1980}
{Goldreich}, P. \& {Tremaine}, S. 1980, \apj, 241, 425

\bibitem[{Gommers {et~al.}(2024)Gommers, Virtanen, Haberland, Burovski, Weckesser, Reddy, Oliphant, Cournapeau, Nelson, alexbrc, Roy, Peterson, Polat, Wilson, endolith, Mayorov, van~der Walt, Brett, Laxalde, Larson, Sakai, Millman, Lars, peterbell10, Carey, van Mulbregt, eric jones, McKibben, Kai, \& Kern}]{scipy_10909890}
Gommers, R., Virtanen, P., Haberland, M., {et~al.} 2024, scipy/scipy: SciPy 1.13.0

\bibitem[{{Gonz{\'a}lez} \& {Levato}(2006)}]{Gonzalez_Levato}
{Gonz{\'a}lez}, J.~F. \& {Levato}, H. 2006, \aap, 448, 283

\bibitem[{{G{\"o}tberg} {et~al.}(2018){G{\"o}tberg}, {de Mink}, {Groh}, {Kupfer}, {Crowther}, {Zapartas}, \& {Renzo}}]{Gotberg+2018}
{G{\"o}tberg}, Y., {de Mink}, S.~E., {Groh}, J.~H., {et~al.} 2018, \aap, 615, A78

\bibitem[{{G{\"o}tberg} {et~al.}(2019){G{\"o}tberg}, {de Mink}, {Groh}, {Leitherer}, \& {Norman}}]{Gotberg+2019}
{G{\"o}tberg}, Y., {de Mink}, S.~E., {Groh}, J.~H., {Leitherer}, C., \& {Norman}, C. 2019, \aap, 629, A134

\bibitem[{{G{\"o}tberg} {et~al.}(2023){G{\"o}tberg}, {Drout}, {Ji}, {Groh}, {Ludwig}, {Crowther}, {Smith}, {de Koter}, \& {de Mink}}]{Gotberg+2023}
{G{\"o}tberg}, Y., {Drout}, M.~R., {Ji}, A.~P., {et~al.} 2023, \apj, 959, 125

\bibitem[{{Gr{\"a}fener} {et~al.}(2002){Gr{\"a}fener}, {Koesterke}, \& {Hamann}}]{Grafener+2002}
{Gr{\"a}fener}, G., {Koesterke}, L., \& {Hamann}, W.~R. 2002, \aap, 387, 244

\bibitem[{{Gray}(1977)}]{Gray1977}
{Gray}, D.~F. 1977, \apj, 218, 530

\bibitem[{{Gray}(2005)}]{Gray2005}
{Gray}, D.~F. 2005, {The Observation and Analysis of Stellar Photospheres}

\bibitem[{{Gray} \& {Corbally}(2009)}]{Gray_Corbally2009}
{Gray}, R.~O. \& {Corbally}, Christopher, J. 2009, {Stellar Spectral Classification}

\bibitem[{{Hainich} {et~al.}(2019){Hainich}, {Ramachandran}, {Shenar}, {Sander}, {Todt}, {Gruner}, {Oskinova}, \& {Hamann}}]{Hainich+2019}
{Hainich}, R., {Ramachandran}, V., {Shenar}, T., {et~al.} 2019, \aap, 621, A85

\bibitem[{{Hamann} \& {Gr{\"a}fener}(2003)}]{Hamann_Grafener2003}
{Hamann}, W.~R. \& {Gr{\"a}fener}, G. 2003, \aap, 410, 993

\bibitem[{{Han} {et~al.}(2003){Han}, {Podsiadlowski}, {Maxted}, \& {Marsh}}]{Han+2003}
{Han}, Z., {Podsiadlowski}, P., {Maxted}, P.~F.~L., \& {Marsh}, T.~R. 2003, \mnras, 341, 669

\bibitem[{{Han} {et~al.}(2002){Han}, {Podsiadlowski}, {Maxted}, {Marsh}, \& {Ivanova}}]{Han+2002}
{Han}, Z., {Podsiadlowski}, P., {Maxted}, P.~F.~L., {Marsh}, T.~R., \& {Ivanova}, N. 2002, \mnras, 336, 449

\bibitem[{{Harrington}(1968)}]{Harrington1968}
{Harrington}, R.~S. 1968, \aj, 73, 190

\bibitem[{Harris {et~al.}(2020)Harris, Millman, van~der Walt, Gommers, Virtanen, Cournapeau, Wieser, Taylor, Berg, Smith, Kern, Picus, Hoyer, van Kerkwijk, Brett, Haldane, del R{\'{i}}o, Wiebe, Peterson, G{\'{e}}rard-Marchant, Sheppard, Reddy, Weckesser, Abbasi, Gohlke, \& Oliphant}]{numpy}
Harris, C.~R., Millman, K.~J., van~der Walt, S.~J., {et~al.} 2020, Nature, 585, 357

\bibitem[{{Haucke} {et~al.}(2018){Haucke}, {Cidale}, {Venero}, {Cur{\'e}}, {Kraus}, {Kanaan}, \& {Arcos}}]{Haucke+2018}
{Haucke}, M., {Cidale}, L.~S., {Venero}, R.~O.~J., {et~al.} 2018, \aap, 614, A91

\bibitem[{{Heber}(2009)}]{Heber2009}
{Heber}, U. 2009, \araa, 47, 211

\bibitem[{{Herwig}(2000)}]{Herwig2000}
{Herwig}, F. 2000, \aap, 360, 952

\bibitem[{{Hiltner}(1956)}]{Hiltner1956}
{Hiltner}, W.~A. 1956, \apjs, 2, 389

\bibitem[{{Howard} {et~al.}(2010){Howard}, {Johnson}, {Marcy}, {Fischer}, {Wright}, {Bernat}, {Henry}, {Peek}, {Isaacson}, {Apps}, {Endl}, {Cochran}, {Valenti}, {Anderson}, \& {Piskunov}}]{Howard+2010}
{Howard}, A.~W., {Johnson}, J.~A., {Marcy}, G.~W., {et~al.} 2010, \apj, 721, 1467

\bibitem[{{Iglesias} \& {Rogers}(1993)}]{Iglesias1993}
{Iglesias}, C.~A. \& {Rogers}, F.~J. 1993, \apj, 412, 752

\bibitem[{{Iglesias} \& {Rogers}(1996)}]{Iglesias1996}
{Iglesias}, C.~A. \& {Rogers}, F.~J. 1996, \apj, 464, 943

\bibitem[{{Irrgang} {et~al.}(2020){Irrgang}, {Geier}, {Kreuzer}, {Pelisoli}, \& U.}]{Irrgang+2020}
{Irrgang}, A., {Geier}, S., {Kreuzer}, S., {Pelisoli}, I., \& U., H. 2020, \aap, 633, L5

\bibitem[{{Irwin}(2004)}]{Irwin2004}
{Irwin}, A.~W. 2004, The FreeEOS Code for Calculating the Equation of State for Stellar Interiors

\bibitem[{{Itoh} {et~al.}(1996){Itoh}, {Hayashi}, {Nishikawa}, \& {Kohyama}}]{Itoh1996}
{Itoh}, N., {Hayashi}, H., {Nishikawa}, A., \& {Kohyama}, Y. 1996, \apjs, 102, 411

\bibitem[{{Jacoby} {et~al.}(1984){Jacoby}, {Hunter}, \& {Christian}}]{Jacoby_Hunter1984}
{Jacoby}, G.~H., {Hunter}, D.~A., \& {Christian}, C.~A. 1984, \apjs, 56, 257

\bibitem[{{Jermyn} {et~al.}(2023){Jermyn}, {Bauer}, {Schwab}, {Farmer}, {Ball}, {Bellinger}, {Dotter}, {Joyce}, {Marchant}, {Mombarg}, {Wolf}, {Sunny Wong}, {Cinquegrana}, {Farrell}, {Smolec}, {Thoul}, {Cantiello}, {Herwig}, {Toloza}, {Bildsten}, {Townsend}, \& {Timmes}}]{Jermyn+2023}
{Jermyn}, A.~S., {Bauer}, E.~B., {Schwab}, J., {et~al.} 2023, \apjs, 265, 15

\bibitem[{{Jermyn} {et~al.}(2021){Jermyn}, {Schwab}, {Bauer}, {Timmes}, \& {Potekhin}}]{Jermyn2021}
{Jermyn}, A.~S., {Schwab}, J., {Bauer}, E., {Timmes}, F.~X., \& {Potekhin}, A.~Y. 2021, \apj, 913, 72

\bibitem[{{Kippenhahn} \& {Meyer-Hofmeister}(1977)}]{Kippenhahn_Meyer-Hofmeister1977}
{Kippenhahn}, R. \& {Meyer-Hofmeister}, E. 1977, \aap, 54, 539

\bibitem[{{Kippenhahn} {et~al.}(1980){Kippenhahn}, {Ruschenplatt}, \& {Thomas}}]{Kippenhahn+1980}
{Kippenhahn}, R., {Ruschenplatt}, G., \& {Thomas}, H.~C. 1980, \aap, 91, 175

\bibitem[{{Kippenhahn} \& {Weigert}(1967)}]{Kippenhahn_Weigert1967}
{Kippenhahn}, R. \& {Weigert}, A. 1967, \zap, 65, 251

\bibitem[{{Klement} {et~al.}(2022){Klement}, {Baade}, {Rivinius}, {Gies}, {Wang}, {Labadie-Bartz}, {Ticiani dos Santos}, {Monnier}, {Carciofi}, {M{\'e}rand}, {Anugu}, {Schaefer}, {Le Bouquin}, {Davies}, {Ennis}, {Gardner}, {Kraus}, {Setterholm}, \& {Labdon}}]{Klement+2022}
{Klement}, R., {Baade}, D., {Rivinius}, T., {et~al.} 2022, \apj, 940, 86

\bibitem[{{Klement} {et~al.}(2024){Klement}, {Rivinius}, {Gies}, {Baade}, {M{\'e}rand}, {Monnier}, {Schaefer}, {Lanthermann}, {Anugu}, {Kraus}, \& {Gardner}}]{Klement+2024}
{Klement}, R., {Rivinius}, T., {Gies}, D.~R., {et~al.} 2024, \apj, 962, 70

\bibitem[{{Klencki} {et~al.}(2020){Klencki}, {Nelemans}, {Istrate}, \& {Pols}}]{Klencki+2020}
{Klencki}, J., {Nelemans}, G., {Istrate}, A.~G., \& {Pols}, O. 2020, \aap, 638, A55

\bibitem[{{Kluska} {et~al.}(2022){Kluska}, {Van Winckel}, {Copp{\'e}e}, {Oomen}, {Dsilva}, {Kamath}, {Bujarrabal}, \& {Min}}]{Kluska+2022}
{Kluska}, J., {Van Winckel}, H., {Copp{\'e}e}, Q., {et~al.} 2022, \aap, 658, A36

\bibitem[{{Kobulnicky} \& {Fryer}(2007)}]{Kobulnicky+2007}
{Kobulnicky}, H.~A. \& {Fryer}, C.~L. 2007, \apj, 670, 747

\bibitem[{{Kolb} \& {Ritter}(1990)}]{Kolb_Ritter1990}
{Kolb}, U. \& {Ritter}, H. 1990, \aap, 236, 385

\bibitem[{{Kozai}(1962)}]{Kozai1962}
{Kozai}, Y. 1962, \aj, 67, 591

\bibitem[{{Kruckow} {et~al.}(2016){Kruckow}, {Tauris}, {Langer}, {Sz{\'e}csi}, {Marchant}, \& {Podsiadlowski}}]{Kruckow+2016}
{Kruckow}, M.~U., {Tauris}, T.~M., {Langer}, N., {et~al.} 2016, \aap, 596, A58

\bibitem[{{Kupfer} {et~al.}(2015){Kupfer}, {Geier}, {Heber}, {{\O}stensen}, {Barlow}, {Maxted}, {Heuser}, {Schaffenroth}, \& {G{\"a}nsicke}}]{Kupfer+2015}
{Kupfer}, T., {Geier}, S., {Heber}, U., {et~al.} 2015, \aap, 576, A44

\bibitem[{{Langanke} \& {Mart{\'{\i}}nez-Pinedo}(2000)}]{Langanke2000}
{Langanke}, K. \& {Mart{\'{\i}}nez-Pinedo}, G. 2000, Nuclear Physics A, 673, 481

\bibitem[{{Lanz} \& {Hubeny}(2007)}]{Lanz_Hubeny2007}
{Lanz}, T. \& {Hubeny}, I. 2007, \apjs, 169, 83

\bibitem[{{Laplace} {et~al.}(2021){Laplace}, {Justham}, {Renzo}, {G{\"o}tberg}, {Farmer}, {Vartanyan}, \& {de Mink}}]{Laplace+2021}
{Laplace}, E., {Justham}, S., {Renzo}, M., {et~al.} 2021, \aap, 656, A58

\bibitem[{{Lau} {et~al.}(2024){Lau}, {Hirai}, {Mandel}, \& {Tout}}]{Lau+2024}
{Lau}, M. Y.~M., {Hirai}, R., {Mandel}, I., \& {Tout}, C.~A. 2024, \apjl, 966, L7

\bibitem[{{Levenhagen} \& {Leister}(2006)}]{Levenhagen+2006}
{Levenhagen}, R.~S. \& {Leister}, N.~V. 2006, \mnras, 371, 252

\bibitem[{{Lidov}(1962)}]{Lidov1962}
{Lidov}, M.~L. 1962, \planss, 9, 719

\bibitem[{{Liu} {et~al.}(2019){Liu}, {Zhang}, {Howard}, {Bai}, {Lu}, {Soria}, {Justham}, {Li}, {Zheng}, {Wang}, {Belczynski}, {Casares}, {Zhang}, {Yuan}, {Dong}, {Lei}, {Isaacson}, {Wang}, {Bai}, {Shao}, {Gao}, {Wang}, {Niu}, {Cui}, {Zheng}, {Mu}, {Zhang}, {Wang}, {Heger}, {Qi}, {Liao}, {Lattanzi}, {Gu}, {Wang}, {Wu}, {Shao}, {Shen}, {Wang}, {Bregman}, {Di Stefano}, {Liu}, {Han}, {Zhang}, {Wang}, {Ren}, {Zhang}, {Zhang}, {Wang}, {Cabrera-Lavers}, {Corradi}, {Rebolo}, {Zhao}, {Zhao}, {Chu}, \& {Cui}}]{Liu+2019}
{Liu}, J., {Zhang}, H., {Howard}, A.~W., {et~al.} 2019, \nat, 575, 618

\bibitem[{{Lorimer} {et~al.}(2015){Lorimer}, {Esposito}, {Manchester}, {Possenti}, {Lyne}, {McLaughlin}, {Kramer}, {Hobbs}, {Stairs}, {Burgay}, {Eatough}, {Keith}, {Faulkner}, {D'Amico}, {Camilo}, {Corongiu}, \& {Crawford}}]{Lorimer+2015}
{Lorimer}, D.~R., {Esposito}, P., {Manchester}, R.~N., {et~al.} 2015, \mnras, 450, 2185

\bibitem[{{Lu} {et~al.}(2023){Lu}, {Fuller}, {Quataert}, \& {Bonnerot}}]{Lu+2023}
{Lu}, W., {Fuller}, J., {Quataert}, E., \& {Bonnerot}, C. 2023, \mnras, 519, 1409

\bibitem[{{Marchant} \& {Bodensteiner}(2024)}]{Marchant_Bodensteiner2023}
{Marchant}, P. \& {Bodensteiner}, J. 2024, \araa, 62, 21

\bibitem[{{Marchenko} {et~al.}(1998){Marchenko}, {Moffat}, \& {Eenens}}]{Marchenko+1998}
{Marchenko}, S.~V., {Moffat}, A. F.~J., \& {Eenens}, P. R.~J. 1998, \pasp, 110, 1416

\bibitem[{{Mason} {et~al.}(2009){Mason}, {Hartkopf}, {Gies}, {Henry}, \& {Helsel}}]{Mason+2009}
{Mason}, B.~D., {Hartkopf}, W.~I., {Gies}, D.~R., {Henry}, T.~J., \& {Helsel}, J.~W. 2009, \aj, 137, 3358

\bibitem[{{Moe} \& {Di Stefano}(2017)}]{Moe_DiStefano2017}
{Moe}, M. \& {Di Stefano}, R. 2017, \apjs, 230, 15

\bibitem[{{Mourard} {et~al.}(2015){Mourard}, {Monnier}, {Meilland}, {Gies}, {Millour}, {Benisty}, {Che}, {Grundstrom}, {Ligi}, {Schaefer}, {Baron}, {Kraus}, {Zhao}, {Pedretti}, {Berio}, {Clausse}, {Nardetto}, {Perraut}, {Spang}, {Stee}, {Tallon-Bosc}, {McAlister}, {ten Brummelaar}, {Ridgway}, {Sturmann}, {Sturmann}, {Turner}, \& {Farrington}}]{Mourard+2015}
{Mourard}, D., {Monnier}, J.~D., {Meilland}, A., {et~al.} 2015, \aap, 577, A51

\bibitem[{{Nagarajan} \& {El-Badry}(2024)}]{Nagarajan_ElBadry2024}
{Nagarajan}, P. \& {El-Badry}, K. 2024, \pasp, 136, 094203

\bibitem[{{Naoz}(2016)}]{Naoz2016}
{Naoz}, S. 2016, \araa, 54, 441

\bibitem[{{Nasa High Energy Astrophysics Science Archive Research Center (Heasarc)}(2014)}]{HEASOFT}
{Nasa High Energy Astrophysics Science Archive Research Center (Heasarc)}. 2014, {HEAsoft: Unified Release of FTOOLS and XANADU}, Astrophysics Source Code Library, record ascl:1408.004

\bibitem[{{Navarro} {et~al.}(2012){Navarro}, {Corradi}, \& {Mampaso}}]{Navarro+2012}
{Navarro}, S.~G., {Corradi}, R.~L.~M., \& {Mampaso}, A. 2012, \aap, 538, A76

\bibitem[{{Nieva} \& {Sim{\'o}n-D{\'\i}az}(2011)}]{Nieva_Simon-Diaz2011}
{Nieva}, M.~F. \& {Sim{\'o}n-D{\'\i}az}, S. 2011, \aap, 532, A2

\bibitem[{{Oda} {et~al.}(1994){Oda}, {Hino}, {Muto}, {Takahara}, \& {Sato}}]{Oda1994}
{Oda}, T., {Hino}, M., {Muto}, K., {Takahara}, M., \& {Sato}, K. 1994, Atomic Data and Nuclear Data Tables, 56, 231

\bibitem[{{Offner} {et~al.}(2023){Offner}, {Moe}, {Kratter}, {Sadavoy}, {Jensen}, \& {Tobin}}]{Offner+2023}
{Offner}, S.~S.~R., {Moe}, M., {Kratter}, K.~M., {et~al.} 2023, in Astronomical Society of the Pacific Conference Series, Vol. 534, Protostars and Planets VII, ed. S.~{Inutsuka}, Y.~{Aikawa}, T.~{Muto}, K.~{Tomida}, \& M.~{Tamura}, 275

\bibitem[{{Oomen} {et~al.}(2018){Oomen}, {Van Winckel}, {Pols}, {Nelemans}, {Escorza}, {Manick}, {Kamath}, \& {Waelkens}}]{Oomen+2018}
{Oomen}, G.-M., {Van Winckel}, H., {Pols}, O., {et~al.} 2018, \aap, 620, A85

\bibitem[{{Paczy{\'n}ski}(1967)}]{Paczynski1967}
{Paczy{\'n}ski}, B. 1967, \actaa, 17, 355

\bibitem[{{Paczy{\'n}ski}(1971)}]{Paczynski1971}
{Paczy{\'n}ski}, B. 1971, \araa, 9, 183

\bibitem[{{Paczynski}(1976)}]{Paczynski1976}
{Paczynski}, B. 1976, in Structure and Evolution of Close Binary Systems, ed. P.~{Eggleton}, S.~{Mitton}, \& J.~{Whelan}, Vol.~73, 75

\bibitem[{{Pauli} {et~al.}(2022){Pauli}, {Oskinova}, {Hamann}, {Ramachandran}, {Todt}, {Sander}, {Shenar}, {Rickard}, {Ma{\'\i}z Apell{\'a}niz}, \& {Prinja}}]{Pauli+2022}
{Pauli}, D., {Oskinova}, L.~M., {Hamann}, W.~R., {et~al.} 2022, \aap, 659, A9

\bibitem[{{Paxton} {et~al.}(2011){Paxton}, {Bildsten}, {Dotter}, {Herwig}, {Lesaffre}, \& {Timmes}}]{Paxton+2011}
{Paxton}, B., {Bildsten}, L., {Dotter}, A., {et~al.} 2011, \apjs, 192, 3

\bibitem[{{Paxton} {et~al.}(2013){Paxton}, {Cantiello}, {Arras}, {Bildsten}, {Brown}, {Dotter}, {Mankovich}, {Montgomery}, {Stello}, {Timmes}, \& {Townsend}}]{Paxton+2013}
{Paxton}, B., {Cantiello}, M., {Arras}, P., {et~al.} 2013, \apjs, 208, 4

\bibitem[{{Paxton} {et~al.}(2015){Paxton}, {Marchant}, {Schwab}, {Bauer}, {Bildsten}, {Cantiello}, {Dessart}, {Farmer}, {Hu}, {Langer}, {Townsend}, {Townsley}, \& {Timmes}}]{Paxton+2015}
{Paxton}, B., {Marchant}, P., {Schwab}, J., {et~al.} 2015, \apjs, 220, 15

\bibitem[{{Paxton} {et~al.}(2018){Paxton}, {Schwab}, {Bauer}, {Bildsten}, {Blinnikov}, {Duffell}, {Farmer}, {Goldberg}, {Marchant}, {Sorokina}, {Thoul}, {Townsend}, \& {Timmes}}]{Paxton+2018}
{Paxton}, B., {Schwab}, J., {Bauer}, E.~B., {et~al.} 2018, \apjs, 234, 34

\bibitem[{{Paxton} {et~al.}(2019){Paxton}, {Smolec}, {Schwab}, {Gautschy}, {Bildsten}, {Cantiello}, {Dotter}, {Farmer}, {Goldberg}, {Jermyn}, {Kanbur}, {Marchant}, {Thoul}, {Townsend}, {Wolf}, {Zhang}, \& {Timmes}}]{Paxton+2019}
{Paxton}, B., {Smolec}, R., {Schwab}, J., {et~al.} 2019, \apjs, 243, 10

\bibitem[{{Pecaut} \& {Mamajek}(2013)}]{Pecaut_Mamajek2013}
{Pecaut}, M.~J. \& {Mamajek}, E.~E. 2013, \apjs, 208, 9

\bibitem[{{Peters} {et~al.}(2008){Peters}, {Gies}, {Grundstrom}, \& {McSwain}}]{Peters+2008}
{Peters}, G.~J., {Gies}, D.~R., {Grundstrom}, E.~D., \& {McSwain}, M.~V. 2008, \apj, 686, 1280

\bibitem[{{Peters} {et~al.}(2013){Peters}, {Pewett}, {Gies}, {Touhami}, \& {Grundstrom}}]{Peters+2013}
{Peters}, G.~J., {Pewett}, T.~D., {Gies}, D.~R., {Touhami}, Y.~N., \& {Grundstrom}, E.~D. 2013, \apj, 765, 2

\bibitem[{{Pols} {et~al.}(1991){Pols}, {Cote}, {Waters}, \& {Heise}}]{Pols+1991}
{Pols}, O.~R., {Cote}, J., {Waters}, L.~B.~F.~M., \& {Heise}, J. 1991, \aap, 241, 419

\bibitem[{{Porter} \& {Rivinius}(2003)}]{Porter_Rivinius2003}
{Porter}, J.~M. \& {Rivinius}, T. 2003, \pasp, 115, 1153

\bibitem[{{Potekhin} \& {Chabrier}(2010)}]{Potekhin2010}
{Potekhin}, A.~Y. \& {Chabrier}, G. 2010, Contributions to Plasma Physics, 50, 82

\bibitem[{{Poutanen}(2017)}]{Poutanen2017}
{Poutanen}, J. 2017, \apj, 835, 119

\bibitem[{{Ramachandran} {et~al.}(2023){Ramachandran}, {Klencki}, {Sander}, {Pauli}, {Shenar}, {Oskinova}, \& {Hamann}}]{Ramachandran+2023}
{Ramachandran}, V., {Klencki}, J., {Sander}, A.~A.~C., {et~al.} 2023, \aap, 674, L12

\bibitem[{{Ramachandran} {et~al.}(2024){Ramachandran}, {Sander}, {Pauli}, {Klencki}, {Backs}, {Tramper}, {Bernini-Peron}, {Crowther}, {Hamann}, {Ignace}, {Kuiper}, {Oey}, {Oskinova}, {Shenar}, {Todt}, {Vink}, {Wang}, {Wofford}, \& {the XShootU Collaboration}}]{Ramachandran+2024}
{Ramachandran}, V., {Sander}, A.~A.~C., {Pauli}, D., {et~al.} 2024, \aap, 692, A90

\bibitem[{{Ram{\'\i}rez-Agudelo} {et~al.}(2013){Ram{\'\i}rez-Agudelo}, {Sim{\'o}n-D{\'\i}az}, {Sana}, {de Koter}, {Sab{\'\i}n-Sanjul{\'\i}an}, {de Mink}, {Dufton}, {Gr{\"a}fener}, {Evans}, {Herrero}, {Langer}, {Lennon}, {Ma{\'\i}z Apell{\'a}niz}, {Markova}, {Najarro}, {Puls}, {Taylor}, \& {Vink}}]{Ramirez-Agudelo+2013}
{Ram{\'\i}rez-Agudelo}, O.~H., {Sim{\'o}n-D{\'\i}az}, S., {Sana}, H., {et~al.} 2013, \aap, 560, A29

\bibitem[{{Raskin} {et~al.}(2011){Raskin}, {van Winckel}, {Hensberge}, {Jorissen}, {Lehmann}, {Waelkens}, {Avila}, {de Cuyper}, {Degroote}, {Dubosson}, {Dumortier}, {Fr{\'e}mat}, {Laux}, {Michaud}, {Morren}, {Perez Padilla}, {Pessemier}, {Prins}, {Smolders}, {van Eck}, \& {Winkler}}]{Raskin+2011}
{Raskin}, G., {van Winckel}, H., {Hensberge}, H., {et~al.} 2011, \aap, 526, A69

\bibitem[{{Reimers} {et~al.}(1975){Reimers}, {Baschek}, {Kegel}, \& {Traving (eds)}}]{Reimers+1975}
{Reimers}, D., {Baschek}, B., {Kegel}, W.~H., \& {Traving (eds)}, G. 1975, {Problems in stellar atmospheres and envelopes}, 229

\bibitem[{{Rein} \& {Liu}(2012)}]{Rein_Liu2012}
{Rein}, H. \& {Liu}, S.~F. 2012, \aap, 537, A128

\bibitem[{{Rein} \& {Spiegel}(2015)}]{Rein_Spiegel2015}
{Rein}, H. \& {Spiegel}, D.~S. 2015, \mnras, 446, 1424

\bibitem[{{Ritter}(1988)}]{Ritter1988}
{Ritter}, H. 1988, \aap, 202, 93

\bibitem[{{Rivinius} {et~al.}(2020){Rivinius}, {Baade}, {Hadrava}, {Heida}, \& {Klement}}]{Rivinius+2020}
{Rivinius}, T., {Baade}, D., {Hadrava}, P., {Heida}, M., \& {Klement}, R. 2020, \aap, 637, L3

\bibitem[{{Rivinius} {et~al.}(2013){Rivinius}, {Carciofi}, \& {Martayan}}]{Rivinius+2013}
{Rivinius}, T., {Carciofi}, A.~C., \& {Martayan}, C. 2013, \aapr, 21, 69

\bibitem[{{Rivinius} \& {Klement}(2024)}]{Rivinius_Klement2024}
{Rivinius}, T. \& {Klement}, R. 2024, to be published by Elsevier as a Reference Module, arXiv:2411.06882

\bibitem[{{Rocha} {et~al.}(2025){Rocha}, {Hur}, {Kalogera}, {Gossage}, {Sun}, {Doctor}, {Andrews}, {Bavera}, {Briel}, {Fragos}, {Kovlakas}, {Kruckow}, {Misra}, {Xing}, \& {Zapartas}}]{Rocha+2025}
{Rocha}, K.~A., {Hur}, R., {Kalogera}, V., {et~al.} 2025, \apj, 983, 39

\bibitem[{{Rogers} \& {Nayfonov}(2002)}]{Rogers2002}
{Rogers}, F.~J. \& {Nayfonov}, A. 2002, \apj, 576, 1064

\bibitem[{{Roming} {et~al.}(2005){Roming}, {Kennedy}, {Mason}, {Nousek}, {Ahr}, {Bingham}, {Broos}, {Carter}, {Hancock}, {Huckle}, {Hunsberger}, {Kawakami}, {Killough}, {Koch}, {McLelland}, {Smith}, {Smith}, {Soto}, {Boyd}, {Breeveld}, {Holland}, {Ivanushkina}, {Pryzby}, {Still}, \& {Stock}}]{Roming+2005}
{Roming}, P. W.~A., {Kennedy}, T.~E., {Mason}, K.~O., {et~al.} 2005, \ssr, 120, 95

\bibitem[{{Sana} {et~al.}(2012){Sana}, {de Mink}, {de Koter}, {Langer}, {Evans}, {Gieles}, {Gosset}, {Izzard}, {Le Bouquin}, \& {Schneider}}]{Sana+2012}
{Sana}, H., {de Mink}, S.~E., {de Koter}, A., {et~al.} 2012, Science, 337, 444

\bibitem[{{Sana} {et~al.}(2008){Sana}, {Gosset}, {Naz{\'e}}, {Rauw}, \& {Linder}}]{Sana+2008}
{Sana}, H., {Gosset}, E., {Naz{\'e}}, Y., {Rauw}, G., \& {Linder}, N. 2008, \mnras, 386, 447

\bibitem[{{Sander} {et~al.}(2015){Sander}, {Shenar}, {Hainich}, {G{\'\i}menez-Garc{\'\i}a}, {Todt}, \& {Hamann}}]{Sander+2015}
{Sander}, A., {Shenar}, T., {Hainich}, R., {et~al.} 2015, \aap, 577, A13

\bibitem[{{Saumon} {et~al.}(1995){Saumon}, {Chabrier}, \& {van Horn}}]{Saumon1995}
{Saumon}, D., {Chabrier}, G., \& {van Horn}, H.~M. 1995, \apjs, 99, 713

\bibitem[{{Schootemeijer} {et~al.}(2019){Schootemeijer}, {Langer}, {Grin}, \& {Wang}}]{Schootemeijer+2019}
{Schootemeijer}, A., {Langer}, N., {Grin}, N.~J., \& {Wang}, C. 2019, \aap, 625, A132

\bibitem[{Seeburger {et~al.}(2024)Seeburger, Rix, El-Badry, Xiang, \& Fouesneau}]{seeburger2024autonomous}
Seeburger, R., Rix, H.-W., El-Badry, K., Xiang, M., \& Fouesneau, M. 2024, Monthly Notices of the Royal Astronomical Society, 530, 1935

\bibitem[{{Shao} \& {Li}(2021)}]{Shao_Li2021}
{Shao}, Y. \& {Li}, X.-D. 2021, \apj, 908, 67

\bibitem[{{Shenar} {et~al.}(2020){Shenar}, {Bodensteiner}, {Abdul-Masih}, {Fabry}, {Mahy}, {Marchant}, {Banyard}, {Bowman}, {Dsilva}, {Hawcroft}, {Reggiani}, \& {Sana}}]{Shenar+2020}
{Shenar}, T., {Bodensteiner}, J., {Abdul-Masih}, M., {et~al.} 2020, \aap, 639, L6

\bibitem[{{Shenar} {et~al.}(2022){Shenar}, {Sana}, {Mahy}, {Ma{\'\i}z Apell{\'a}niz}, {Crowther}, {Gromadzki}, {Herrero}, {Langer}, {Marchant}, {Schneider}, {Sen}, {Soszy{\'n}ski}, \& {Toonen}}]{Shenar+2022}
{Shenar}, T., {Sana}, H., {Mahy}, L., {et~al.} 2022, \aap, 665, A148

\bibitem[{Simon \& Sturm(1994)}]{simon1994disentangling}
Simon, K. \& Sturm, E. 1994, Astronomy and Astrophysics (ISSN 0004-6361), vol. 281, no. 1, p. 286-291, 281, 286

\bibitem[{{Siwek} {et~al.}(2023){Siwek}, {Weinberger}, \& {Hernquist}}]{Siwek+2023}
{Siwek}, M., {Weinberger}, R., \& {Hernquist}, L. 2023, \mnras, 522, 2707

\bibitem[{{Smith} {et~al.}(2011){Smith}, {Li}, {Filippenko}, \& {Chornock}}]{Smith+2011}
{Smith}, N., {Li}, W., {Filippenko}, A.~V., \& {Chornock}, R. 2011, \mnras, 412, 1522

\bibitem[{{Smolec} \& {Moskalik}(2008)}]{Smolec2008}
{Smolec}, R. \& {Moskalik}, P. 2008, \actaa, 58, 193

\bibitem[{{Soberman} {et~al.}(1997){Soberman}, {Phinney}, \& {van den Heuvel}}]{Soberman+1997}
{Soberman}, G.~E., {Phinney}, E.~S., \& {van den Heuvel}, E.~P.~J. 1997, \aap, 327, 620

\bibitem[{{Tauris} \& {van den Heuvel}(2006)}]{Tauris_vandenHeuvel2006}
{Tauris}, T.~M. \& {van den Heuvel}, E.~P.~J. 2006, in Compact stellar X-ray sources, Vol.~39, 623--665

\bibitem[{{Temmink} {et~al.}(2023){Temmink}, {Pols}, {Justham}, {Istrate}, \& {Toonen}}]{Temmink+2023}
{Temmink}, K.~D., {Pols}, O.~R., {Justham}, S., {Istrate}, A.~G., \& {Toonen}, S. 2023, \aap, 669, A45

\bibitem[{{Timmes} \& {Swesty}(2000)}]{Timmes2000}
{Timmes}, F.~X. \& {Swesty}, F.~D. 2000, \apjs, 126, 501

\bibitem[{{Toonen} {et~al.}(2016){Toonen}, {Hamers}, \& {Portegies Zwart}}]{Toonen+2016}
{Toonen}, S., {Hamers}, A., \& {Portegies Zwart}, S. 2016, Computational Astrophysics and Cosmology, 3, 6

\bibitem[{Townsend(2019)}]{Townsend2019}
Townsend, R. 2019, MESA SDK for Mac OS

\bibitem[{{Townsend} {et~al.}(2018){Townsend}, {Goldstein}, \& {Zweibel}}]{Townsend2018}
{Townsend}, R.~H.~D., {Goldstein}, J., \& {Zweibel}, E.~G. 2018, \mnras, 475, 879

\bibitem[{{Townsend} {et~al.}(2004){Townsend}, {Owocki}, \& {Howarth}}]{Townsend+2004}
{Townsend}, R.~H.~D., {Owocki}, S.~P., \& {Howarth}, I.~D. 2004, \mnras, 350, 189

\bibitem[{{Townsend} \& {Teitler}(2013)}]{Townsend2013}
{Townsend}, R.~H.~D. \& {Teitler}, S.~A. 2013, \mnras, 435, 3406

\bibitem[{{Valli} {et~al.}(2024){Valli}, {Tiede}, {Vigna-G{\'o}mez}, {Cuadra}, {Siwek}, {Ma}, {D'Orazio}, {Zrake}, \& {de Mink}}]{Valli+2024}
{Valli}, R., {Tiede}, C., {Vigna-G{\'o}mez}, A., {et~al.} 2024, \aap, 688, A128

\bibitem[{{van den Heuvel} {et~al.}(2017){van den Heuvel}, {Portegies Zwart}, \& {de Mink}}]{VandenHeuvel+2017}
{van den Heuvel}, E.~P.~J., {Portegies Zwart}, S.~F., \& {de Mink}, S.~E. 2017, \mnras, 471, 4256

\bibitem[{Van~Rossum \& Drake(2009)}]{python}
Van~Rossum, G. \& Drake, F.~L. 2009, Python 3 Reference Manual (Scotts Valley, CA: CreateSpace)

\bibitem[{{Van Winckel}(2018)}]{VanWinckel2018}
{Van Winckel}, H. 2018, Submitted Manuscript Under Review: To appear in \textit{The Impact of Binaries on Stellar Evolution}, Beccari G. \& Boffin H.M.J. (Eds.), arXiv:1809.00871

\bibitem[{{Villase{\~n}or} {et~al.}(2023){Villase{\~n}or}, {Lennon}, {Picco}, {Shenar}, {Marchant}, {Langer}, {Dufton}, {Nardini}, {Evans}, {Bodensteiner}, {de Mink}, {G{\"o}tberg}, {Soszy{\'n}ski}, {Taylor}, \& {Sana}}]{Villasenor+2023}
{Villase{\~n}or}, J.~I., {Lennon}, D.~J., {Picco}, A., {et~al.} 2023, \mnras, 525, 5121

\bibitem[{Virtanen {et~al.}(2020)Virtanen, Gommers, Oliphant, Haberland, Reddy, Cournapeau, Burovski, Peterson, Weckesser, Bright, {van der Walt}, Brett, Wilson, Millman, Mayorov, Nelson, Jones, Kern, Larson, Carey, Polat, Feng, Moore, {VanderPlas}, Laxalde, Perktold, Cimrman, Henriksen, Quintero, Harris, Archibald, Ribeiro, Pedregosa, {van Mulbregt}, \& {SciPy 1.0 Contributors}}]{2020SciPy-NMeth}
Virtanen, P., Gommers, R., Oliphant, T.~E., {et~al.} 2020, Nature Methods, 17, 261

\bibitem[{Vogt {et~al.}(1994)Vogt, Allen, Bigelow, Bresee, Brown, Cantrall, Conrad, Couture, Delaney, Epps, Hilyard, Hilyard, Horn, Jern, Kanto, Keane, Kibrick, Lewis, Osborne, Pardeilhan, Pfister, Ricketts, Robinson, Stover, Tucker, Ward, \& Wei}]{Vogt+1994}
Vogt, S.~S., Allen, S.~L., Bigelow, B.~C., {et~al.} 1994, in Instrumentation in Astronomy VIII, ed. D.~L. Crawford \& E.~R. Craine, Vol. 2198, International Society for Optics and Photonics (SPIE), 362 -- 375

\bibitem[{{von Zeipel}(1910)}]{vonZeipel1910}
{von Zeipel}, H. 1910, Astronomische Nachrichten, 183, 345

\bibitem[{{Vos} {et~al.}(2012){Vos}, {{\O}stensen}, {Degroote}, {De Smedt}, {Green}, {Heber}, {Van Winckel}, {Acke}, {Bloemen}, {De Cat}, {Exter}, {Lampens}, {Lombaert}, {Masseron}, {Menu}, {Neyskens}, {Raskin}, {Ringat}, {Rauch}, {Smolders}, \& {Tkachenko}}]{Vos+2012}
{Vos}, J., {{\O}stensen}, R.~H., {Degroote}, P., {et~al.} 2012, \aap, 548, A6

\bibitem[{{Vos} {et~al.}(2015){Vos}, {{\O}stensen}, {Marchant}, \& {Van Winckel}}]{Vos+2015}
{Vos}, J., {{\O}stensen}, R.~H., {Marchant}, P., \& {Van Winckel}, H. 2015, \aap, 579, A49

\bibitem[{{Vos} {et~al.}(2013){Vos}, {{\O}stensen}, {N{\'e}meth}, {Green}, {Heber}, \& {Van Winckel}}]{Vos+2013}
{Vos}, J., {{\O}stensen}, R.~H., {N{\'e}meth}, P., {et~al.} 2013, \aap, 559, A54

\bibitem[{{Vos} {et~al.}(2017){Vos}, {{\O}stensen}, {Vu{\v{c}}kovi{\'c}}, \& {Van Winckel}}]{Vos+2017}
{Vos}, J., {{\O}stensen}, R.~H., {Vu{\v{c}}kovi{\'c}}, M., \& {Van Winckel}, H. 2017, \aap, 605, A109

\bibitem[{Wagg \& Broekgaarden(2024)}]{software-citation-station-zenodo}
Wagg, T. \& Broekgaarden, F. 2024, The Software Citation Station

\bibitem[{{Wagg} \& {Broekgaarden}(2024)}]{software-citation-station-paper}
{Wagg}, T. \& {Broekgaarden}, F.~S. 2024, arXiv e-prints, arXiv:2406.04405

\bibitem[{{Wang} {et~al.}(2018){Wang}, {Gies}, \& {Peters}}]{Wang+2018}
{Wang}, L., {Gies}, D.~R., \& {Peters}, G.~J. 2018, \apj, 853, 156

\bibitem[{{Wang} {et~al.}(2021){Wang}, {Gies}, {Peters}, {G{\"o}tberg}, {Chojnowski}, {Lester}, \& {Howell}}]{Wang+2021}
{Wang}, L., {Gies}, D.~R., {Peters}, G.~J., {et~al.} 2021, \aj, 161, 248

\bibitem[{{Wang} {et~al.}(2023){Wang}, {Gies}, {Peters}, \& {Han}}]{Wang+2023}
{Wang}, L., {Gies}, D.~R., {Peters}, G.~J., \& {Han}, Z. 2023, \aj, 165, 203

\bibitem[{{Wright} {et~al.}(2010){Wright}, {Eisenhardt}, {Mainzer}, {Ressler}, {Cutri}, {Jarrett}, {Kirkpatrick}, {Padgett}, {McMillan}, {Skrutskie}, {Stanford}, {Cohen}, {Walker}, {Mather}, {Leisawitz}, {Gautier}, {McLean}, {Benford}, {Lonsdale}, {Blain}, {Mendez}, {Irace}, {Duval}, {Liu}, {Royer}, {Heinrichsen}, {Howard}, {Shannon}, {Kendall}, {Walsh}, {Larsen}, {Cardon}, {Schick}, {Schwalm}, {Abid}, {Fabinsky}, {Naes}, \& {Tsai}}]{Wright+2010}
{Wright}, E.~L., {Eisenhardt}, P. R.~M., {Mainzer}, A.~K., {et~al.} 2010, \aj, 140, 1868

\bibitem[{{Yamaguchi} {et~al.}(2024){Yamaguchi}, {El-Badry}, {Rees}, {Shahaf}, {Mazeh}, \& {Andrae}}]{Yamaguchi+2024}
{Yamaguchi}, N., {El-Badry}, K., {Rees}, N.~R., {et~al.} 2024, \pasp, 136, 084202

\bibitem[{{Zhang} \& {Green}(2025)}]{Zhang_Green2024}
{Zhang}, X. \& {Green}, G.~M. 2025, Science, 387, 1209

\bibitem[{{Zorec} {et~al.}(2016){Zorec}, {Fr{\'e}mat}, {Domiciano de Souza}, {Royer}, {Cidale}, {Hubert}, {Semaan}, {Martayan}, {Cochetti}, {Arias}, {Aidelman}, \& {Stee}}]{Zorec+2016}
{Zorec}, J., {Fr{\'e}mat}, Y., {Domiciano de Souza}, A., {et~al.} 2016, \aap, 595, A132

\end{thebibliography}

\begin{appendix}
\section{Observation overview}
\label{appendix:obs_overview}

In Table~\ref{tab:obs_overview} we provide an overview of observations including measured RVs of the narrow-lined stripped star and S/N estimates. Table~\ref{tab:phot_overview} lists the photometric flux measurements of the binary in different filter bands. These were collected with the help of the VOSA software package \citep{Bayo+2008} and used to fit the model SEDs (see Sect.~\ref{sec:sed}). 

\setlength{\extrarowheight}{3pt}
\begin{table}
\caption{Overview of observations.}\label{tab:obs_overview}
\centering
\begin{tabular}{c c c c c}
\hline
 Instrument & MJD [d] & RV [km/s] &$\sigma_\mathrm{RV}$ [km/s]& S/N \\
\hline
HERMES & 59828.194 & 67.3 & 2.9 & 62 \\
HERMES & 59861.158 & 39.0 & 3.5 & 63 \\
HERMES & 59879.167 & 32.6 & 5.5 & 38 \\
HERMES & 59899.053 & -0.9 & 3.3 & 51 \\
HERMES & 60001.897 & 103.1 & 2.9 & 67 \\
TRES & 59890.280 & 11.5 & 5.6 & 31 \\
TRES & 59900.356 & 5.6 & 2.6 & 21 \\
TRES & 59906.351 & 0.5 & 7.3 & 16 \\
TRES & 59925.188 & -9.4 & 1.8 & 23 \\
TRES & 59953.146 & 122.2 & 6.0 & 31 \\
TRES & 59970.144 & 134.5 & 2.2 & 19 \\
TRES & 59980.219 & 127.0 & 3.4 & 106 \\
TRES & 60190.401 & 138.0 & 2.9 & 47 \\
TRES & 60205.425 & 119.1 & 3.8 & 36 \\
TRES & 60217.398 & 103.3 & 2.3 & 37 \\
TRES & 60218.474 & 107.0 & 0.9 & 123 \\
TRES & 60219.412 & 103.9 & 2.3 & 151 \\
TRES & 60222.435 & 101.8 & 2.9 & 137 \\
TRES & 60226.338 & 98.5 & 1.3 & 120 \\
TRES & 60235.415 & 102.1 & 2.0 & 135 \\
TRES & 60240.324 & 92.2 & 0.7 & 125 \\
TRES & 60244.275 & 83.2 & 2.3 & 116 \\
TRES & 60253.241 & 79.6 & 3.6 & 136 \\
TRES & 60285.227 & 52.5 & 1.7 & 123 \\
TRES & 60299.262 & 43.1 & 7.2 & 110 \\
TRES & 60307.272 & 38.4 & 5.1 & 132 \\
TRES & 60327.153 & 26.5 & 5.8 & 128 \\
TRES & 60345.180 & -3.9 & 5.0 & 129 \\
TRES & 60388.105 & 91.0 & 2.0 & 105 \\
HIRES & 60274.406 & 60.2 & 4.2 & 200 \\
HIRES & 60275.386 & 60.4 & 4.0 & 129 \\
HIRES & 60283.541 & 51.4 & 5.0 & 159 \\
HIRES & 60293.523 & 45.6 & 7.0 & 172 \\
HIRES & 60310.488 & 29.6 & 6.5 & 192 \\
\hline
\end{tabular}
\tablefoot{Columns list observing instruments, dates, inferred RVs and uncertainties of the narrow-lined stripped star and estimated signal-to-noise ratios of the individual spectra.}
\end{table}
\setlength{\extrarowheight}{0pt}

\setlength{\extrarowheight}{3pt}
\begin{table}
\caption{Overview of literature photometric flux measurements of HIP~15429 used for the SED analysis.}\label{tab:phot_overview}
\centering
\begin{tabular}{c c c c}
\hline
        FilterID & $\lambda_\mathrm{eff}$ & Flux & Flux Error \\
                 & [\AA] &\multicolumn{2}{c}{[erg/cm$^2$/s/\AA]} \\
        \hline
        UVOT.UVM2\_trn & 2246.56 & 5.86e-14 & 2.0e-15 \\
        SDSS.u & 3608.04 & 4.831e-13 & 8.4e-15 \\
        SDSS.g & 4671.78 & 4.990e-13 & 1.6e-15 \\
        GAIA3.Gbp & 5035.75 & 4.534e-13 & 1.9e-15 \\
        GAIA3.G & 5822.39 & 3.785e-13 & 1.1e-15 \\
        SDSS.r & 6141.12 & 4.298e-13 & 1.1e-15 \\
        SDSS.i & 7457.89 & 3.399e-13 & 1.1e-15 \\
        GAIA3.Grp & 7619.96 & 3.120e-13 & 1.6e-15 \\
        SDSS.z & 8922.78 & 2.685e-13 & 0.9e-15 \\
        2MASS.J & 12350.0 & 1.278e-13 & 2.8e-15 \\
        2MASS.H & 16620.0 & 5.46e-14 & 1.1e-15 \\
        2MASS.Ks & 21590.0 & 2.363e-14 & 5.7e-16 \\
        WISE.W1 & 33526.0 & 5.39e-15 & 1.3e-16 \\
        WISE.W2 & 46028.0 & 1.784e-15 & 3.4e-17 \\
        WISE.W3 & 115608.0 & 6.87e-17 & 1.3e-18 \\
        WISE.W4 & 220883.0 & 8.74e-18 & 6.8e-19 \\
        \hline
\end{tabular}
\tablefoot{Columns list filter names, effective wavelengths, fluxes and flux uncertainties.}
\end{table}
\setlength{\extrarowheight}{0pt}
\setlength{\extrarowheight}{0pt}

\section{Spectral disentangling}
\label{appendix:disentangling}

Spectral disentangling aims to determine the two single-star spectra that can best reproduce the observed composite spectra of a binary at all observational epochs. For the shift-and-add method, the orbital parameters need to be known a priori to compute the relative RVs of the components. However, if only the RV semi-amplitudes are unknown, these can be included as additional parameters in the fit. 
For HIP~15429, the narrow-lined stripped star allowed us to trace the orbital parameters including the period and eccentricity of the binary and estimate $K_{\mathrm{B}}$, the RV semi-amplitude of the stripped B star. However, the $K_{\mathrm{Be}}$ value of the broad-lined companion star could not be estimated from the spectrum. 

Using the determined orbital parameters (see Sect.~\ref{sec:orbit_analysis}), we performed disentangling to infer the single-star spectra while simultaneously fitting for $K_{\mathrm{B}}$ and $K_{\mathrm{Be}}$. We allowed $K_{\mathrm{B}}$ to vary in the range of $\pm 2\sigma$ around the inferred value $74.1 \pm 2.2\, \mathrm{km \, s}^{-1}$ and for $K_{\mathrm{Be}}$, test values from 0 to $36\,\mathrm{km \, s}^{-1}$. The best-fit values $K_{\mathrm{B}}, \, K_{\mathrm{Be}}$ were determined by minimising the combined $\chi^2$ across all epochs. 

\begin{figure*}[t]
    \centering
    \begin{minipage}[t]{\columnwidth}
        \centering
        \includegraphics[width=\columnwidth]{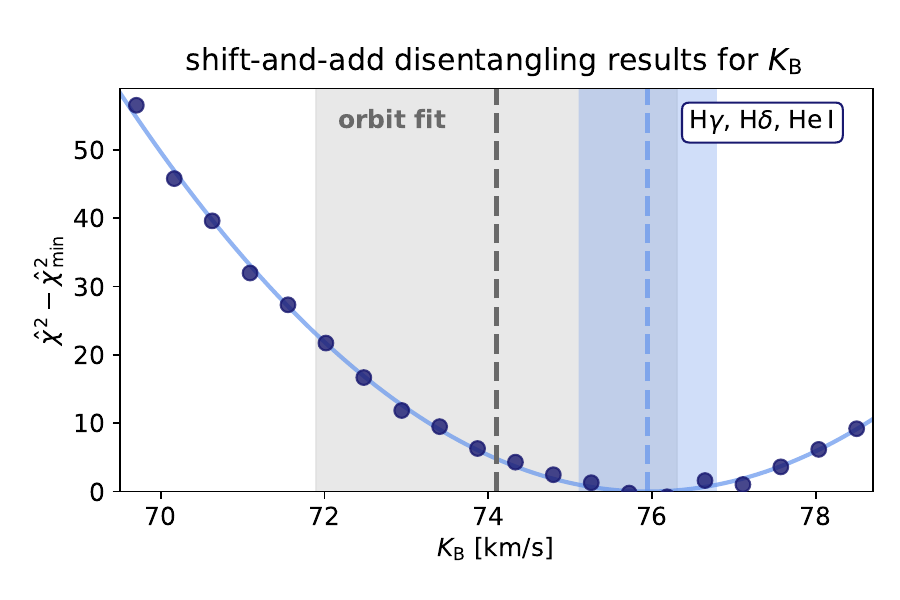}
    \end{minipage}
    \begin{minipage}[t]{\columnwidth}
        \centering
        \includegraphics[width=\columnwidth]{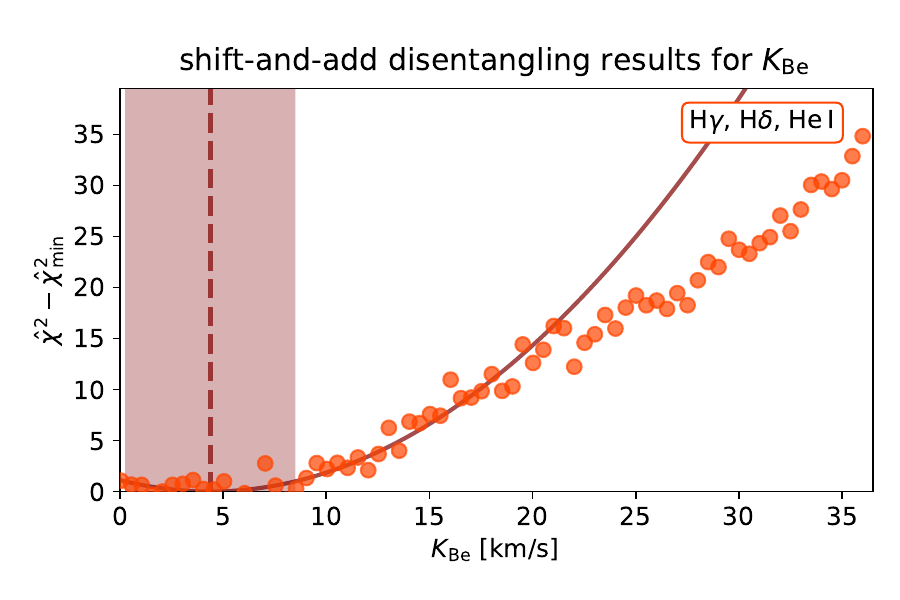}
    \end{minipage}
    \caption{Normalised $\hat \chi^2$ measurements for the RV semi-amplitudes $K_{\mathrm{B}}$ of the narrow-lined stripped star (left) and $K_{\mathrm{Be}}$ of the broad-lined Be companion (right). Values of $\chi^2$ were determined with the shift-and-add disentangling technique from the Balmer H$\delta$ and H$\gamma$, and the He~I $\lambda$4026, 4144, 4388 and 4472\,$\AA$ lines. Since the disentangling procedure is non-linear and introduces unrecognised sources of error, we opted to normalise the derived $\chi^2$ values by multiplication with a scaling factor such that the root mean squared error near the minimum is of order unity. Parabolic fits around the minimum yield best-fitting semi-amplitudes of $K_{\mathrm{B}} \simeq 76\pm0.9\,$km/s and $K_{\mathrm{Be}} \simeq 5\pm4 \,$km/s, indicated by vertical dashed lines and shaded regions. $1-\sigma$ uncertainties were determined as the intervals $\hat\chi^2 -\hat\chi^2_\mathrm{min} \leq 1$.}
    \label{fig:disentangling_results_chi2}
\end{figure*}

We used a series of photospheric helium absorption lines between 4000\,Å and 4500\,Å, as well as the Balmer lines H$_\gamma$ and H$_\delta$ for the initial disentangling and determination of $K_{\mathrm{B}}$ and $K_{\mathrm{Be}}$. Subsequently, we used the larger range of 4000\,Å to 4500\,Å to disentangle the entire spectrum with the fixed RV semi-amplitudes. The results of the initial disentangling are shown in Fig.~\ref{fig:disentangling_results_chi2} for both $K_{\mathrm{B}}$ and $K_{\mathrm{Be}}$.
For the RV amplitude of the stripped star, there is a clear minimum in the $\chi^2$ distribution for $K_{\mathrm{B}} = 76.0\pm0.9\,\mathrm{km \, s}^{-1}$, consistent within 1-$\sigma$ uncertainties with the results of the orbital analysis. 

\begin{figure*}[t]
    \centering
    \includegraphics[width=\textwidth]{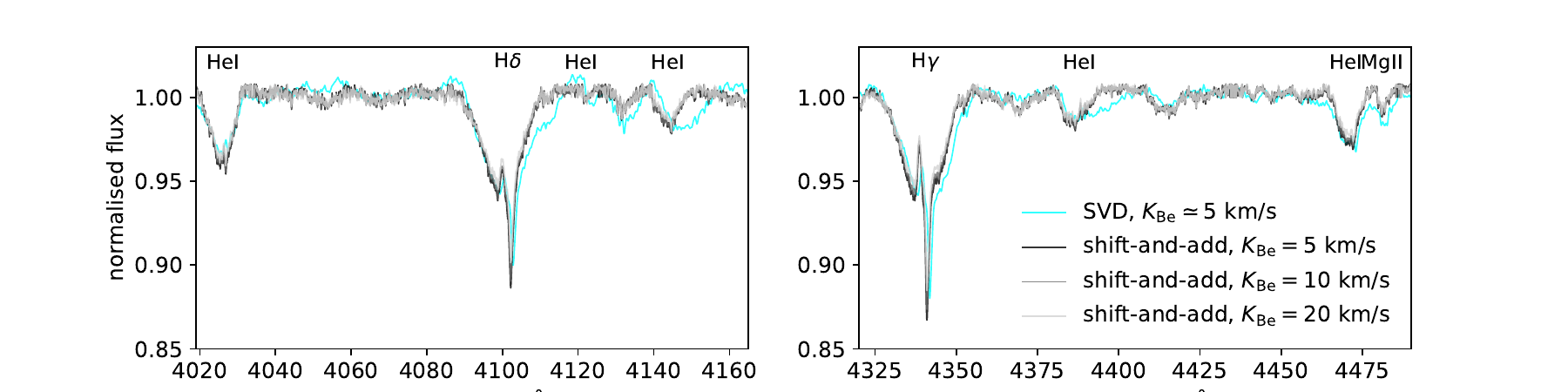}
    \caption{Disentangled spectra of the broad-lined, Be star companion. The figure compares spectra disentangled using the shift-and-add method for assumed Be star radial-velocity semi-amplitudes of 5, 10, and 20\,km/s shown in dark, medium and light grey, respectively. The cyan line shows the Be star spectrum obtained via the singular value decomposition (SVD) method of \cite{seeburger2024autonomous}. Unlike the shift-and-add method, this technique determines RVs as part of the disentangling process and does not require an input value of $K_\mathrm{Be}$. The inferred RVs imply a semi-amplitude of approximately $K_\mathrm{Be}\simeq 5\,$km/s.}
    \label{fig:disentangled_Be_star_K2_comparison}
\end{figure*}

For $K_{\mathrm{Be}}$, the $\chi^2$ distribution suggests a low semi-amplitude of $K_{\mathrm{Be}} = 5\pm 4\,\mathrm{km \, s}^{-1}$. However, the minimum is broad, and the determined value has larger uncertainties.
Due to the broad-lined nature of the companion star, the disentangled spectra are less sensitive to the choice of $K_{\mathrm{Be}}$. As shown in Fig.~\ref{fig:disentangled_Be_star_K2_comparison}, varying $K_{\mathrm{Be}} \lesssim 20$ km s$^{-1}$ produces nearly identical results, indicating that the substantial uncertainties in $K_{\mathrm{Be}}$ should not hinder further analysis.

We cross-checked our shift-and-add disentangling results with the method of \citet{seeburger2024autonomous}, which does not require orbital parameters as input. This technique provided disentangled component spectra, RVs for the narrow-lined B star, and an inferred mass ratio. The RVs of the donor star are consistent with those derived by cross-correlation, within 1–2$\sigma$. The inferred mass ratio, $q \approx 0.061$, implies a semi-amplitude for the Be star of $K_\mathrm{Be} \approx 5\,\mathrm{km/s}$, though both $q$ and $K_\mathrm{Be}$ have significant uncertainties. Overall, these results are in agreement with those obtained from the shift-and-add technique.

\begin{figure*}[t]
    \centering
    \includegraphics[width=\textwidth]{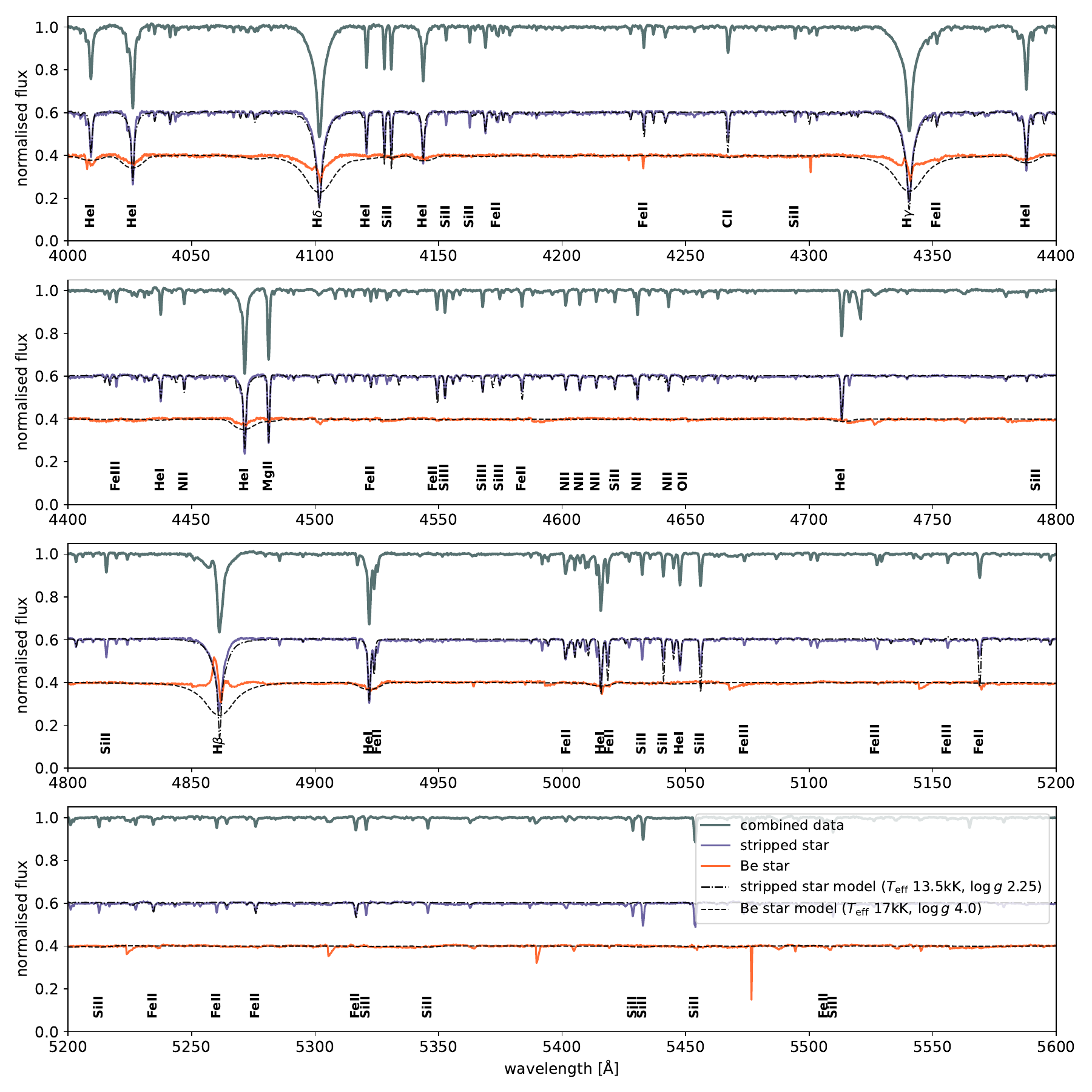}
    \caption{Full wavelength range of disentangled spectra and models.
    The disentangled spectra of the narrow-lined stripped star (dark blue) and the Be star (orange) are shown across the full wavelength range, scaled to their respective flux contributions. Overlaid are model spectra with the inferred parameters, plotted as dash-dotted and dashed lines, respectively. The observed spectra, co-added for improved S/N and shifted to the rest frame of the stripped star, are displayed in grey and normalised to a continuum level of one. Key spectral lines are labelled at the bottom. Narrow, unlabelled features in the Be star's disentangled spectrum are spurious, primarily resulting from telluric and interstellar absorption lines, as well as artefacts from the disentangling process.}
    \label{fig:full_range_disentangled_spectra_with_models_1}
\end{figure*}

The full wavelength range of the observed, disentangled, and modelled spectra is shown in Fig.~\ref{fig:full_range_disentangled_spectra_with_models_1}. The S/N of the disentangled spectra, based on the pixel-to-pixel scatter at $4500\,\AA$,  is approximately 370 for the stripped star and 280 for the Be star. However, this estimate does not account for systematic uncertainties introduced by the disentangling process, which dominate the overall uncertainty. By estimating the systematic uncertainties as the standard deviation of the flux residuals between the observed and disentangled spectra, we find that these are roughly 1.4 times larger than the statistical pixel-based noise.

\begin{figure*}
    \centering
    \includegraphics[width=\textwidth]{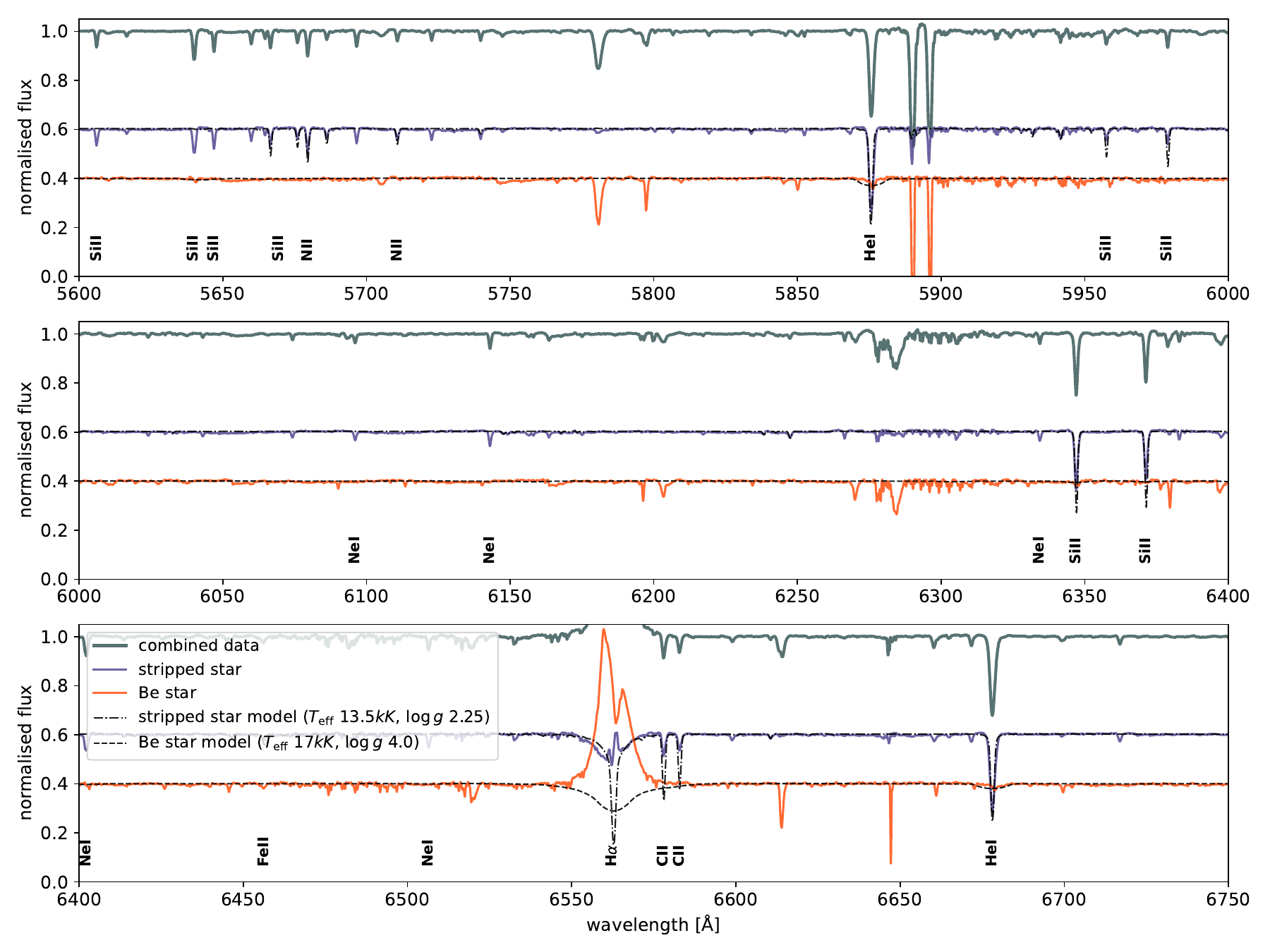}
    \caption*{Figure~\ref{fig:full_range_disentangled_spectra_with_models_1} continued.}
    \label{fig:full_range_disentangled_spectra_with_models_2}
\end{figure*}

\section{Stellar atmosphere models}
\label{appendix:model_abundances}

For the stripped star, we computed new model atmospheres with the non-LTE Potsdam Wolf-Rayet (\textsc{PoWR}) code \citep{Grafener+2002,Hamann_Grafener2003,Sander+2015}. The models were computed, with negligible mass-loss rates; $\dot{M} = 10^{-9}\, \mathrm{M}_\odot /$yr, microturbulent velocity of 10\,km/s, and a fixed terminal wind velocity 1000\,km/s. From these models, synthetic spectra were generated for stars of effective temperatures 10, 12.5, 13, 14, and 15\,kK and with surface gravity $\log g$ in the range 2.0 to 3.0 with step sizes of 0.25\,dex. 
Two sets of abundances were considered: one with solar composition following \citet{Asplund+2021}, and one representative of a stripped star, with nitrogen enrichment and carbon and oxygen depletion. 
The adopted mass fractions are listed in Table~\ref{tab:stripped_model_abundances}. To explore the effects of hydrogen depletion, we computed an additional series of models with fixed $T_{\rm eff} = 13.5$\,kK and $\log g = 2.25$, but with varying hydrogen mass fractions from $X = 0.001$ to $0.7$ in steps of 0.1, keeping the CNO-processed abundances fixed.

\setlength{\extrarowheight}{3pt}
\begin{table}[t]
\caption{Mass fractions adopted for the \textsc{PoWR} model stellar atmospheres.}\label{tab:stripped_model_abundances}
\centering
\begin{tabular}{c c c}
\hline
        Element & \multicolumn{2}{c}{Mass fraction} \\
         & solar \citep{Asplund+2021} & stripped \\
        \hline
        Hydrogen & 0.74  & 0.3 (0.001--0.7) \\
        Helium & 0.24 & 0.68 (1-X-Z) \\
        Nitrogen & 0.0007& 0.007 \\
        Carbon & 0.00255 & 0.00075 \\
        Oxygen & 0.0058 & 0.0050 \\
        Silicon & 0.00067& 0.00066 \\
        Magnesium & 0.00064& 0.00069 \\ 
        Iron group & 0.0012& 0.0012 \\
        \hline
\end{tabular}
\tablefoot{The solar abundances follow \cite{Asplund+2021}, the stripped star models are enriched in N and deficient in O and C compared to a solar composition. H and He are varied in the stripped-star grid, with He determined by $Y = 1-X-Z$, we list the adopted best-fit values.}
\end{table}
\setlength{\extrarowheight}{0pt}
\setlength{\extrarowheight}{0pt}

\section{Helium enrichment}
\label{appendix:helium_content}

\begin{figure*}[t]
    \centering
    \begin{minipage}[t]{\textwidth}
        \centering
        \includegraphics[width=\textwidth]{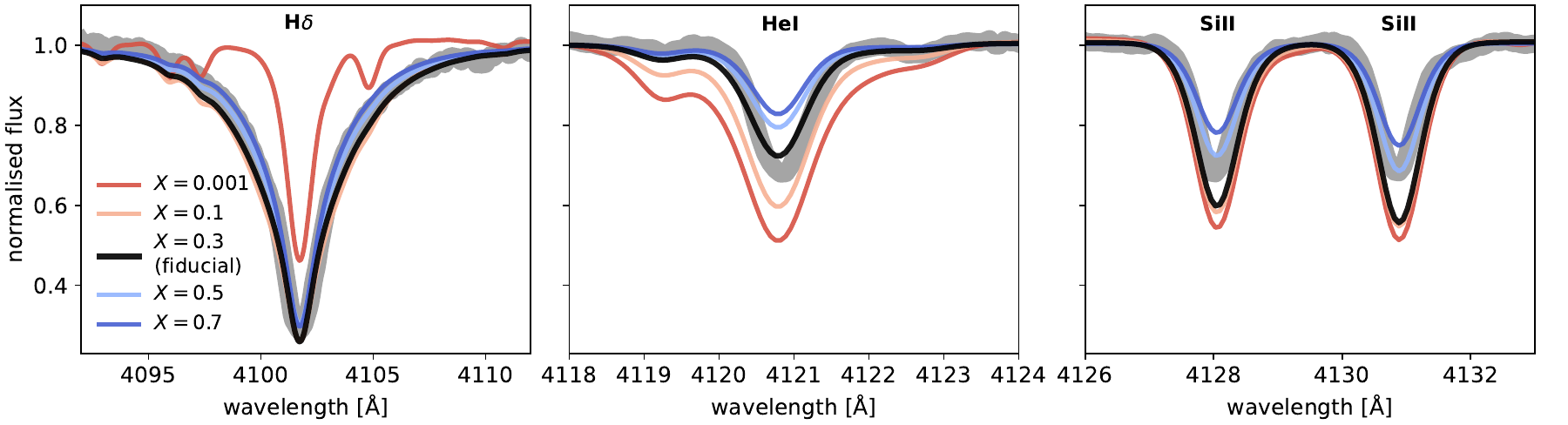}
    \end{minipage}
    \caption{Comparison of the disentangled spectrum of the stripped B star with stellar model atmospheres of varying hydrogen content. The panels show wavelength regions around the Balmer H$\delta$, He~I $4120\,\AA$ and Si~II $4128, 4131\,\AA$ lines (left to right). The disentangled spectrum is shown in grey, and the fiducial best-fit model ($T_\mathrm{eff, B} = 13.5\,$kK, $\log g_\mathrm{B} = 2.25$, flux ratio 60\%) with hydrogen mass fraction $X=0.3$ in black. The models vary $X$ while keeping other parameters fixed.}
    \label{fig:helium_content}
\end{figure*}

Initial comparisons between the disentangled spectrum of the stripped star and solar-composition models revealed that helium lines were systematically under-predicted (Fig.~\ref{fig:stripped_star_models}). This discrepancy points to helium enrichment, as expected for a stripped star that has lost much of its hydrogen envelope.
To quantify the effect, we computed additional \textsc{PoWR} models with fixed $T_{\rm eff} = 13.5$\,kK and $\log g = 2.25$, varying only the hydrogen (and hence helium) content while keeping the CNO-processed abundances fixed. As shown in Fig.~\ref{fig:helium_content}, increasing helium enhances the strength of the He~I lines. The model with $X = 0.3$ best reproduces the observed helium line equivalent widths and was adopted as our fiducial model. This value is consistent with predictions from the MESA evolutionary tracks (Sect.~\ref{sec:evolution}), which suggest $X \sim 0.25$ for the stripped star.
Notably, the Balmer line profiles remain largely unaffected for hydrogen mass fractions down to $X \sim 0.1$, supporting the robustness of our flux ratio and surface gravity estimates derived from these lines.
However, we note that helium enrichment can have an impact on metal line strengths. In particular, Si~II 4128, $4131\,\AA$ lines are somewhat over-predicted in the $X = 0.3$ model. A model with a slightly higher hydrogen content or a lower flux contribution would provide a better match. However, given the additional uncertainties related to microturbulence and metallicity, we adopt $X = 0.3$ as the best compromise and defer a detailed abundance analysis to future work.
Finally, we note that other bloated stripped stars, such as LB-1 \citep[$X \sim 0.55$,][]{Shenar+2020} and HR 6819 (solar composition per \citealt{Bodensteiner+2020} or $X \sim 0.55$ from \citealt{ElBadry_Quataert2020}), do not show strong evidence of significant hydrogen depletion. This suggests that helium enrichment alone is not a definitive indicator of a stripped star, possibly because of measurement challenges or because these stars retain a substantial hydrogen fraction.

\section{Rotation}
\label{appendix:rotation}

We determined the rotational velocities of the binary components by fitting rotationally broadened models to individual absorption lines in the disentangled spectra. For the narrow-lined stripped star, we used metal absorption lines while for the Be star companion we had to resort to He~I lines for lack of strong enough metal lines. The model spectra were convolved with a Gaussian kernel to account for instrumental broadening (using the resolution of the telescope with the lowest resolving power), a rotational kernel with varying values of $v_\mathrm{rot} \sin i$, and a radial-tangential kernel with the standard limb-darkening coefficient $\epsilon = 0.6$ \citep{Gray2005,Aerts+2014} and varying the macroturbulent velocity $v_\mathrm{mac}$. We adapted the Fortran code \texttt{rotin3} to Python for numerical implementation. For each spectral line, the best-fit parameters and their variances were determined using a least-squares approach implemented with the \texttt{curve\_fit} function from the \texttt{scipy.optimize} Python package. The reported values and errors for the rotation measurement are derived as the weighted mean and standard deviation of the weighted mean, calculated from the inverse-variance weighted results of all analysed spectral lines.

\begin{figure}
    \centering
    \includegraphics[width=\columnwidth]{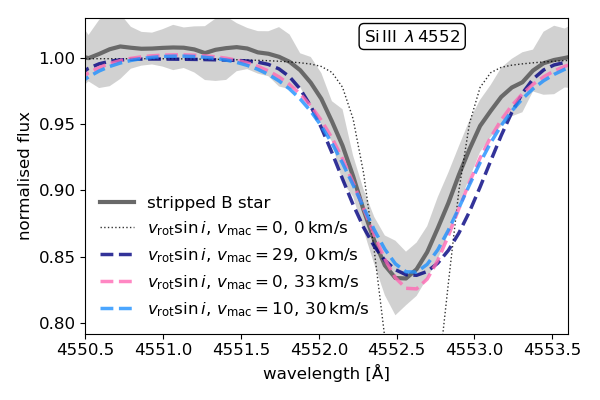}
    \caption{{Estimating rotational and macroturbulent line broadening of the stripped star.} Comparison between the disentangled spectrum (grey) and a \textsc{PoWR} model atmosphere around the Si~III $4552\,\AA$ line for the narrow-lined stripped star. Shaded regions indicate estimated uncertainties, calculated as the standard deviation of the flux residuals between the observed spectra and the disentangled spectrum.
    The model spectra were calculated with different values for rotation and macroturbulent velocities that are given in the legend. }
    \label{fig:stripped_star_rotation}
\end{figure}

Fig.~\ref{fig:stripped_star_rotation} shows the line profile of the Si~III $4552\,\AA$ line in the disentangled spectrum of the stripped star. Overplotted are the best-fit PoWR model spectra ($T_\mathrm{eff} = 13.5\,$kK, $\log g = 2.25$) with nothing but instrumental broadening, only applying rotational broadening, only macroturbulent broadening, and a combined fit of both broadening mechanisms. 
It is apparent that the triangular line profile shape is inconsistent with purely instrumental broadening and also not well-reproduced only using rotational broadening. However, the relative contributions of macroturbulent and rotational broadening cannot be reliably disentangled. We therefore only place an upper limit of $v_\mathrm{rot} \sin i \leq 30\,$km/s on the projected rotational velocity of the star.

\begin{figure}
    \centering
    \includegraphics[width=\columnwidth]{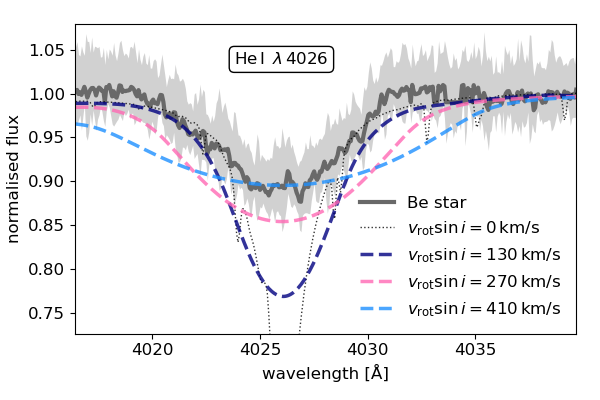}
    \caption{{Estimating rotational broadening of the broad-lined Be star.} The disentangled spectrum of the broad-lined companion star around the He~I $4026\,\AA$ line is shown in grey. Shaded regions indicate estimated uncertainties, calculated as the standard deviation of the flux residuals between the observed spectra and the disentangled spectrum. Overplotted are simulated spectra from the \textsc{Tlusty} library with $T_\mathrm{eff} = 17\,$kK and $\log g = 4.0$. The model spectra have been broadened using a fixed macroturbulent velocity of 50\,km/s and differing rotation velocities; the best-fit value of $v_\mathrm{rot,Be} = 270\pm70\,$km/s and $\pm 2-\sigma$ uncertainties.}
    \label{fig:Be_star_rotation}
\end{figure}

Fig.~\ref{fig:Be_star_rotation} displays the He~I 4026\,Å line of the disentangled Be star spectrum. Overplotted are \textsc{Tlusty} model spectra ($T_\mathrm{eff} = 17\,$kK, $\log g = 4.0$) with varying rotational broadening velocities $v_\mathrm{rot} \sin i$ and fixed macroturbulent broadening of $v_\mathrm{mac} = 50\,$km/s.

\section{Temperature determination}
\label{appendix:temperature_determination}

\begin{figure}
    \centering
    \includegraphics[width=\columnwidth]{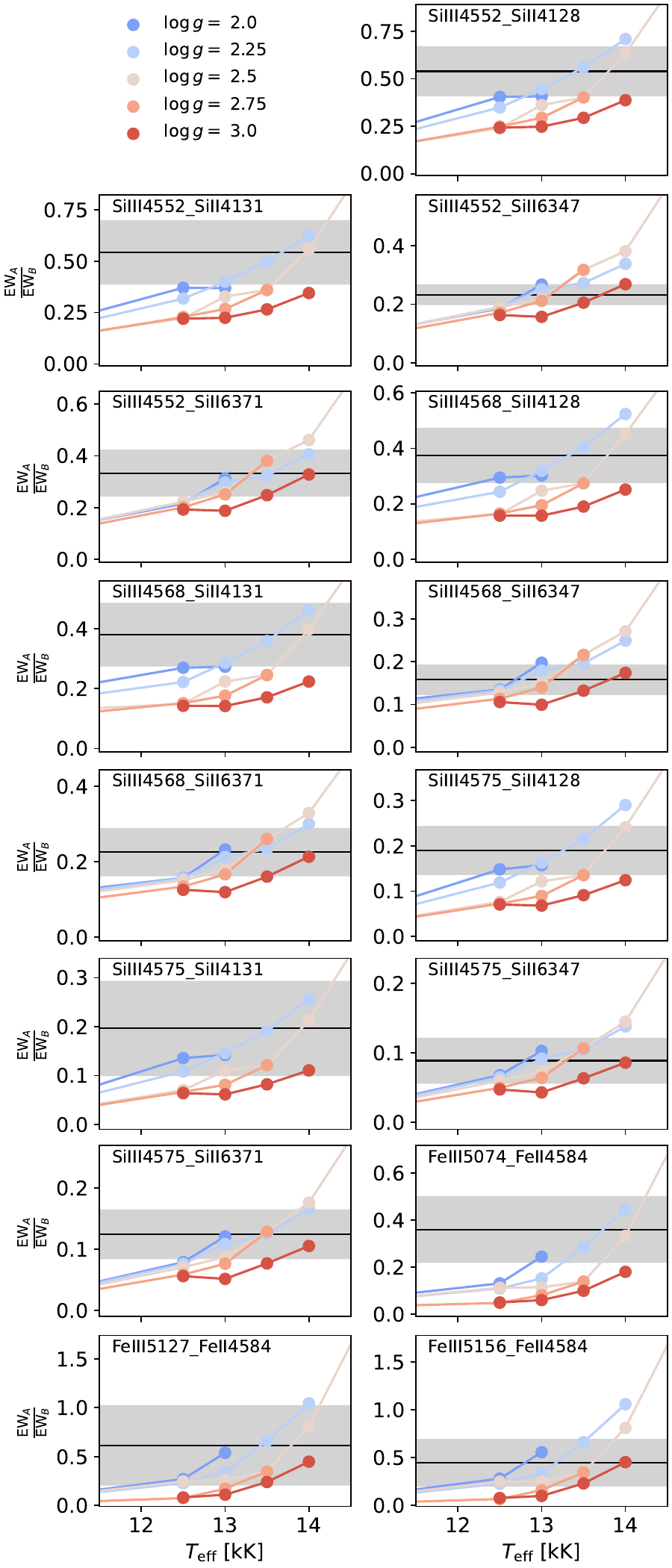}
    \caption{{Temperature determination of the B star from ionisation equilibrium}. The different panels are for different line combinations used to measure equivalent width ratios of Si III over Si II lines and Fe III over Fe II lines. The black line and shaded regions indicate the mean and uncertainties of the observed values. The coloured dots show the predicted ratios for the models of different $\log g$ as a function of $T_\mathrm{eff}$. For an estimated value of $\log g$ between $\approx 2.0$ and $2.5$, the best-fit $T_\mathrm{eff}$ lies between 13000 and 14000 K, respectively.}
    \label{fig:stripped_star_temperature}
\end{figure}

\begin{figure}
    \centering
    \includegraphics[width=\columnwidth]{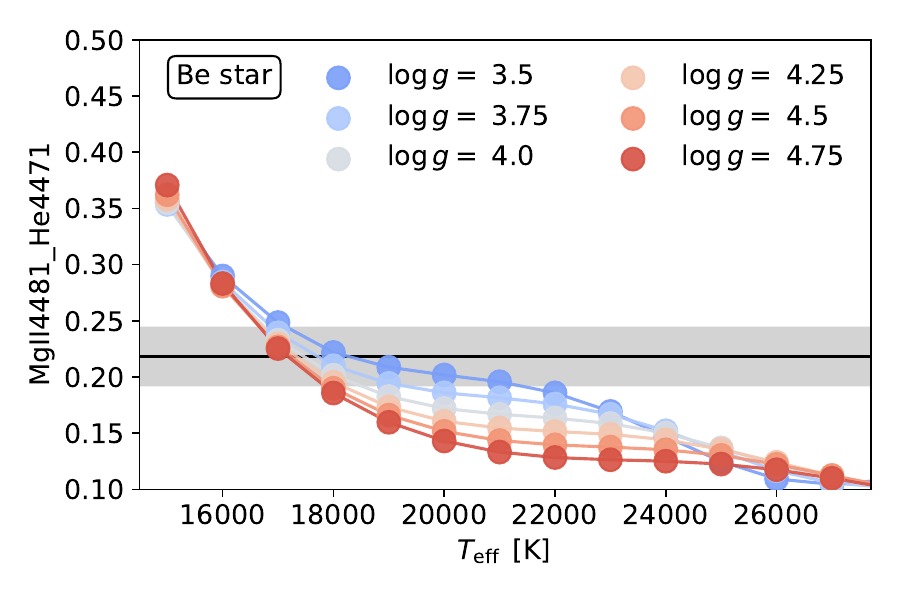}
    \caption{{Temperature determination of the Be star from EW ratios}. The ratio of the equivalent widths of Mg~II over He~I lines of the disentangled spectrum (black line) and model spectra of different temperatures and surface gravities (coloured dots, see legend). Shaded regions indicate 1-$\sigma$ uncertainties. For an estimated value of $\log g$ in the range $3.5$ to $4.5$, the best fit is found for effective temperatures between 16000 and 21000\,K.}
    \label{fig:Be_star_temperature}
\end{figure}

The effective temperatures of the binary components were determined on the basis of the
EW ratios of lines from different ionisation states. In Figures ~\ref{fig:stripped_star_temperature} and Fig.~\ref{fig:Be_star_temperature}, we show in detail the different EW ratios underlying the temperature determination of the narrow-lined, stripped star and the broad-lined Be star. We measured the EWs by integrating the area of the spectral lines under the continuum. For the Be star, the errors are estimated by propagating the flux uncertainty of the disentangled spectrum. For the B star, where the EWs are measured from the observed spectra, the error is determined as the standard deviation of the EW measurements obtained from all available spectra.

\section{MESA Set up}
\label{appendix:mesa_set-up}

\setlength{\extrarowheight}{3pt}
\begin{table*}
\caption{Overview of binary evolution models computed with MESA.} 
\label{tab:mesa_models}
\centering
\begin{tabularx}{\textwidth}{X X X X X X X X }
\toprule
&$M_{\mathrm{d,initial}}$  & $M_{\mathrm{a,initial}}$  & $P_\mathrm{initial}$ & $\beta$ &  $M_\mathrm{d, stripped}$ & $M_{\mathrm{a, Be}}$ & $P$   \\
\toprule
M1 & 4.0$\, \mathrm{M}_\odot$& 3.6$\, \mathrm{M}_\odot$& 12 d& 0.0 & 0.72$\, \mathrm{M}_\odot$& 6.9$\, \mathrm{M}_\odot$& 277 d \\
M2 & 5.0$\, \mathrm{M}_\odot$& 4.5$\, \mathrm{M}_\odot$& 8 d& 0.7 & 0.88$\, \mathrm{M}_\odot$& 5.7$\, \mathrm{M}_\odot$& 263 d \\
M3 & 5.5$\, \mathrm{M}_\odot$& 5.0$\, \mathrm{M}_\odot$& 8 d& 0.7 & 0.99$\, \mathrm{M}_\odot$& 6.3$\, \mathrm{M}_\odot$& 247 d \\
M4 & 6.0$\, \mathrm{M}_\odot$& 5.4$\, \mathrm{M}_\odot$& 8 d& 0.7 & 1.12$\, \mathrm{M}_\odot$& 6.8$\, \mathrm{M}_\odot$& 230 d \\
M5 & 7.0$\, \mathrm{M}_\odot$& 4.2$\, \mathrm{M}_\odot$& 8 d& 0.7 & 1.37$\, \mathrm{M}_\odot$& 8.0$\, \mathrm{M}_\odot$& 197 d \\
\hline
M3a & 5.5$\, \mathrm{M}_\odot$& 2.8$\, \mathrm{M}_\odot$& 30 d& 0.0 & 1.00$\, \mathrm{M}_\odot$& 7.3$\, \mathrm{M}_\odot$& 275 d \\
M3b & 5.5$\, \mathrm{M}_\odot$& 3.9$\, \mathrm{M}_\odot$& 12 d& 0.5 & 0.99$\, \mathrm{M}_\odot$& 6.1$\, \mathrm{M}_\odot$& 224 d \\
M3c & 5.5$\, \mathrm{M}_\odot$& 5.0$\, \mathrm{M}_\odot$& 8 d& 0.7 & 0.99$\, \mathrm{M}_\odot$& 6.3$\, \mathrm{M}_\odot$& 247 d \\
M3d & 5.5$\, \mathrm{M}_\odot$& 5.2$\, \mathrm{M}_\odot$& 5 d& 0.9 & 0.97$\, \mathrm{M}_\odot$& 5.6$\, \mathrm{M}_\odot$& 189 d \\
\bottomrule
\end{tabularx}
\tablefoot{Columns list the model index, initial stellar masses and periods, mass accretion efficiency, masses and periods during the stripped star phase. The first set of models is computed with varying primary star masses. For the second set of models with indices 3a - 3d, $M_{\mathrm{d,initial}}$ is fixed to $5.5\,\mathrm{M}_\odot$.}
\end{table*}
\setlength{\extrarowheight}{0pt}

The binary evolution models presented in Sect.~\ref{sec:evolution} were computed with version 24.03.1 of the Modules for Experiments in Stellar Astrophysics code \citep[MESA;][]{Paxton+2011, Paxton+2013, Paxton+2015, Paxton+2018, Paxton+2019, Jermyn+2023} and using \texttt{MESASDK} version 23.10.1 \citep{Townsend2013,Townsend2018,Townsend2019,ChristensenDalsgaard2008,Blinnikov2004,Blinnikov2006,Baklanov2005,Smolec2008}. The parameter values for all models are summarised in Table~\ref{tab:mesa_models}. The models were computed for initial donor masses in the range 4.0 to $7.0\,\mathrm{M}_\odot$. Simulations are initiated from prebuilt zero-age main-sequence (ZAMS) models with initial composition $X=0.7,\, Y=0.28$ and $Z= 0.02$ provided within MESA. We employ the standard \texttt{basic.net} nuclear network. The evolution of both binary components is computed, and the models are terminated when the stripped star reaches core-helium depletion, implemented as a central lower abundance limit of $X_\mathrm{He} = 0.01$. All of our models assume circular orbits and do not include rotation or related effects (e.g. rotational mixing, rotation-limited accretion), allowing for a free choice of accretion efficiency. This simplification is not expected to significantly impact the evolution of the stripped star -- the primary focus of our models \citep{Gotberg+2018} -- but it introduces substantial uncertainties in the envelope structure of the fast-rotating companion \citep[e.g.][]{Ramachandran+2023}. We use the wind mass-loss prescription from \cite{Reimers+1975} in the low temperature regime ($<8\,$kK), the 'Dutch' prescription \citep{Glebbeek+2009} in the high temperature regime ($>12\,$kK) and a ramp in between. 

Convective mixing is treated using the standard mixing length theory \citep{Bohm-Vitense1958} and the Ledoux criterion. We set the mixing length coefficient at $\alpha_\mathrm{MLT} = 1.73$ following \cite{ElBadry_Quataert2021} and \cite{Herwig2000} and employ an exponential overshoot scheme with the overshoot efficiency parameter $f_\mathrm{OV} = 0.014$ motivated by the results of \cite{Herwig2000}. As pointed out by \cite{ElBadry_Quataert2021}, the level of overshooting affects the final mass of the stripped star. For a fixed initial mass, a higher overshooting efficiency results in a larger convective core on the main sequence and in turn a larger He core. Semi-convective mixing is modelled with semi-convective coefficient $\alpha_\mathrm{SC} = 10$, following \cite{Schootemeijer+2019, Klencki+2020}. We include thermohaline mixing \citep{Kippenhahn+1980} with the standard thermohaline coefficient $\alpha_\mathrm{TH} = 1$ but we do not expect thermohaline mixing to have a significant impact on the stellar structure of the stripped stars \citep[e.g.][]{Gotberg+2018}.

Binary interaction and mass transfer during Roche-lobe overflow are handled using the implicit mass transfer approach proposed by \cite{Kolb_Ritter1990}. The parameters $\alpha,\, \beta,\, \delta$ and $\gamma$ as described in \cite{Tauris_vandenHeuvel2006} control the efficiency of mass transfer. We initially ran our simulations at a coarser resolution and then tested numerical convergence by increasing the mesh and time resolution for the final models, ensuring that the results remained qualitatively consistent.

The MESA EOS is a blend of the EOSs OPAL \citep{Rogers2002}, SCVH
\citep{Saumon1995}, FreeEOS \citep{Irwin2004}, HELM \citep{Timmes2000},
PC \citep{Potekhin2010}, and Skye \citep{Jermyn2021}.

Radiative opacities are primarily from OPAL \citep{Iglesias1993,
Iglesias1996}, with low-temperature data from \citet{Ferguson2005}
and the high-temperature, Compton-scattering-dominated regime from
\citet{Poutanen2017}.  The electron conduction opacities are from
\citet{Cassisi2007} and \citet{Blouin2020}.

Nuclear reaction rates are from JINA REACLIB \citep{Cyburt2010}, NACRE \citep{Angulo1999}, and
additional tabulated weak reaction rates \citet{Fuller1985, Oda1994,
Langanke2000}.  Screening is included with the prescription of \citet{Chugunov2007}.
The thermal neutrino loss rates are from \citet{Itoh1996}.

Roche-lobe radii in binary systems are computed using the fit of
\citet{Eggleton1983}.  Mass transfer rates in Roche-lobe
overflowing binary systems are determined following the
prescription of \citet{Ritter1988}.

\section{Spectral energy distribution fit parameters}
\label{appendix:sed_fit}
\begin{figure*}
    \centering
    \includegraphics[width=0.9\textwidth]{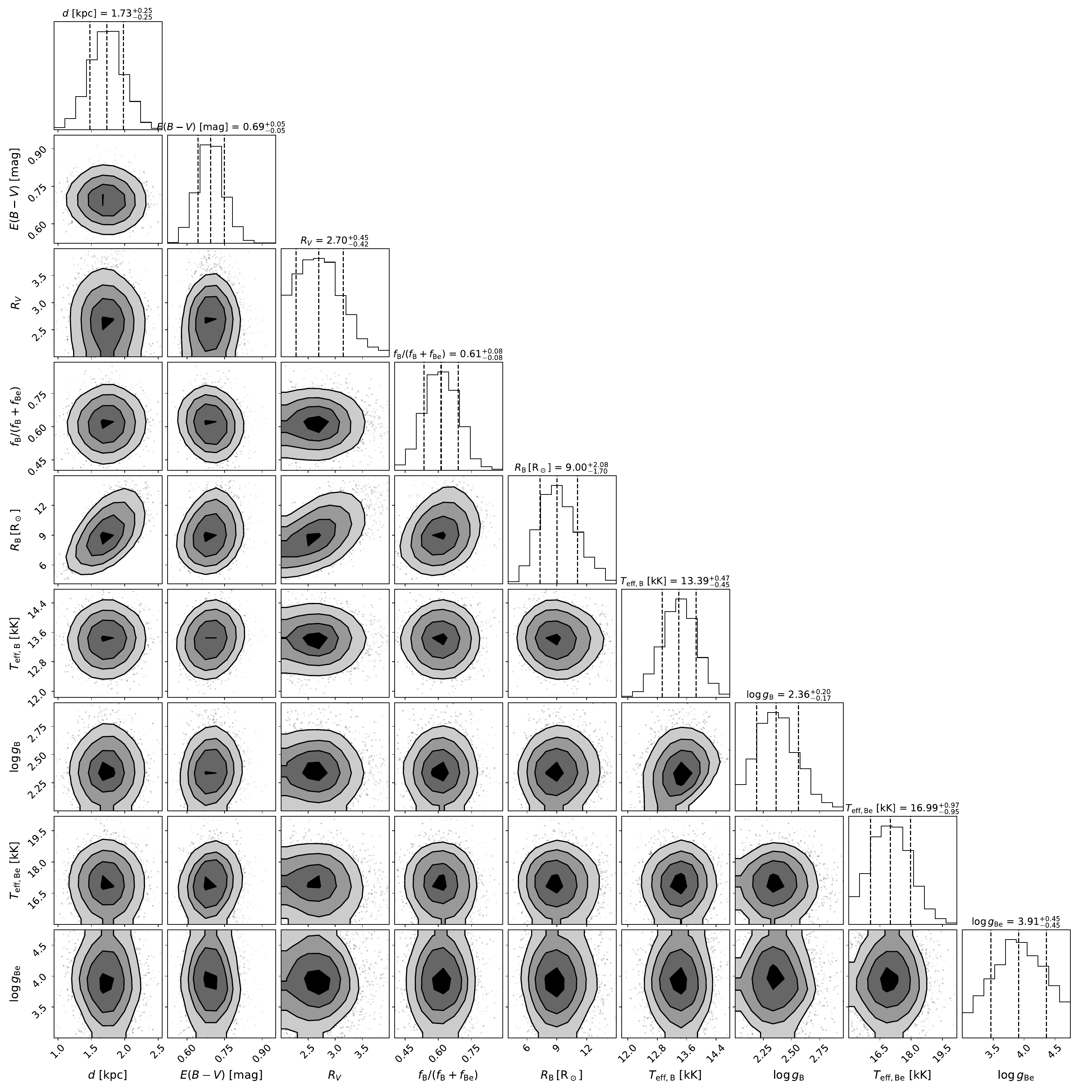}
    \caption{Corner plot visualising the binary parameter values and correlations inferred based on the SED photometry (see Sect.~\ref{sec:sed}). Top panels in each column show histograms of posterior samples for individual parameters. Here, dashed lines mark median values and 1-$\sigma$  confidence intervals. Central panels show the correlations of posterior samples for parameter pairs. } 
    \label{fig:sed_corner_plot}
\end{figure*}

The corner plot in Fig.~\ref{fig:sed_corner_plot} shows the detailed results of the SED fit. Parameter distributions and correlations were inferred using a nested sampling scheme. Inferred parameters include the distance, extinction, flux ratio at $4300\,\AA$, radius, effective temperature and surface gravity of the stripped star, and effective temperature and surface gravity of the companion star. 

\end{appendix}

\end{document}